%% file: main.tex
\documentclass[11pt]{article}

\newcommand{\hiredraceaware}{hired_race_aware_comb_seed1666583033}

\newcommand{\offerraceaware}{offered_race_aware_jan2023_seed1666577807}

\input{preambles/header}

\input{preambles/notation}
\pgfplotsset{compat=1.17}
\hypersetup{
  pdfauthor={Li, Raymond, Bergman},
  pdftitle={Hiring as Exploration},
  pdfdisplaydoctitle=true,
  pdfpagelayout=OneColumn,
  pdfstartview=FitH,
  pdfpagemode=UseOutlines,
  pdfpagemode=UseNone,
  bookmarksnumbered=false,
  verbose=false,
  colorlinks=true,
  citecolor=blue,
  linkcolor=blue,
  urlcolor=blue
}

\graphicspath{{../Empirics/}{./}}

\usepackage{pgfplotstable}

\usepackage{geometry}
\usepackage{caption}
\begin{document}
\thispagestyle{empty}

\vspace{20pt}
\begin{center}
\textcolor{white}{fill}\\
\vspace{20pt}
\LARGE{Hiring as Exploration$^{*}$}\\\mbox{ }
\end{center}

\begin{singlespace}
\begin{center}
\begin{tabular}{ccccccc}
\large{Danielle Li} & \mbox{ } & \mbox{ } & \large{Lindsey Raymond} & \mbox{ } & \mbox{ } & \large{Peter Bergman} \\
\large{MIT \& NBER} & \mbox{ } & \mbox{ } & \large{MIT} & \mbox{ } & \mbox{ } & \large{UT Austin \& NBER} \\
\end{tabular}
\end{center}

\end{singlespace}

\begin{center}
\vspace{.20in}
\normalsize{\today}

\vspace{10pt}
First Draft: August 2020\\
\end{center}
\vspace{10pt}

\begin{abstract}
\vspace{-11pt}
\singlespacing
\noindent
This paper views hiring as a contextual bandit problem: to find the best workers over time, firms must balance ``exploitation'' (selecting from groups with proven track records) with ``exploration'' (selecting from under-represented groups to learn about quality).  Yet modern hiring algorithms, based on ``supervised learning'' approaches, are designed solely for exploitation.  Instead, we build a resume screening algorithm that values exploration by evaluating candidates according to their statistical upside potential.  Using data from professional services recruiting within a Fortune 500 firm, we show that this approach improves the quality (as measured by eventual hiring rates) of candidates selected for an interview, while also increasing demographic diversity, relative to the firm's existing practices.  The same is not true for traditional supervised learning based algorithms, which improve hiring rates but select far fewer Black and Hispanic applicants. 
 Together, our results highlight the importance of incorporating exploration in developing decision-making algorithms that are potentially both more efficient and equitable.
\end{abstract}



\bigskip

\hspace{-25pt}\rule[0.1ex]{0.33\textwidth}{0.2mm}\vspace{0.05in}\\\indent
\begin{singlespace}
\vspace{-25pt}
\noindent \scriptsize{\mbox{  }\mbox{  }\mbox{  }\mbox{  }\mbox{  }$^{*}$Correspondence to d\_li@mit.edu, lraymond@mit.edu, and peterbergman@utexas.edu. We are grateful to David Autor, Pierre Azoulay, Dan Bjorkegren, Emma Brunskill, Max Cytrynbaum, Eleanor Dillon, Alex Frankel, Bob Gibbons, Nathan Hendren, Max Kasy, Pat Kline, Jin Li, Fiona Murray, Anja Sautmann, Scott Stern, John Van Reenen, Kathryn Shaw, and various seminar participants for helpful comments and suggestions. The content is solely the responsibility of the authors and does not necessarily represent the official views of MIT, Columbia University, or the NBER.}
\end{singlespace}

\thispagestyle{empty}


 \setcounter{page}{0}

\clearpage

\newpage

Increasing access to job opportunity for minorities and women is crucial for reducing well-documented race, ethnicity, and gender gaps in the economy.  While a proliferation of initiatives related to diversity, equity, and inclusion speak to firms' interest in these issues, a persistent doubt remains: how can firms increase diversity without sacrificing quality?  

Concerns about ``equity--efficiency'' tradeoffs in hiring are predicated on the assumption that firms are able to perfectly predict the quality of the applicants they encounter.  In such a case, any deviation from the predicted ranking---whether to select more minority or majority group members---would result in a decline in worker quality.  In practice, however, an extensive literature has documented that firms, and the recruiters that they employ, are often inaccurate or biased in their predictions \citep{bensonlishue2021, klinerosewalters}.  Given that firms appear to be far from perfect in their ability to forecast quality, there may be significant scope for improved evaluation tools to expand opportunities for a broader range of candidates while maintaining or even improving worker quality.

In this paper, we examine the role of algorithms in the hiring process. Resume screening algorithms have become increasingly prevalent in recent years, and have been used to assess job candidates across various industries and occupations.\footnote{Accurate adoption rates are elusive, but a 2020 survey of human resource executives found that 39\% reported using predictive analytics in their hiring processes, a significant increase from just 10\% in 2016 \citep{mercer2020}. Furthermore, a survey of technology companies indicates that 60\% plan to invest in AI-powered recruiting software in 2018, and over 75\% of recruiters believe that artificial intelligence will revolutionize hiring practices \citep{bogen2018help}. We discuss evidence on algorithmic adoption further in Section \ref{sec:adopt}
Throughout this paper, we use the terms ``hiring algorithm,'' ``hiring ML,'' and ``resume screening algorithm'' interchangeably to refer to algorithms that assist in making initial interview recommendations. It remains rare for algorithms to make final hiring decisions \citep{raghavan2019}.}
 \footnote{For example, see \citet{barzilay2019, mckinney2020A, mullainathanobermeyer2019, schrittwieser2019mastering, Russakovsky_2015, ajunwaParadox2020}.}  Amazon, for instance, was widely criticized for using a resume screening algorithm that penalized the presence of the term ``women'' (for example, ``captain of women's crew team'') on resumes.\footnote{See https://www.reuters.com/article/us-amazon-com-jobs-automation-insight/amazon-scraps-secret-ai-recruiting-tool-that-showed-bias-against-women-idUSKCN1MK08G.} 


Our paper uses data from a large Fortune 500 firm to study the decision to grant first-round interviews for high-skill positions in consulting, financial analysis, and data science---sectors which offer lucrative jobs with opportunities for career advancement, but which have also been criticized for their lack of diversity.  We study the impact of two types of algorithmic approaches: a ``supervised learning'' model that selects the best candidates as predicted based on its current training data and a ``contextual bandit'' model that seeks to expand its training data in order to learn about the best candidates over time.  Our findings demonstrate that while both algorithmic approaches improve the quality of applicants selected by the firm, they differ in their ability to select diverse candidates. The traditional supervised learning approach leads to a significant reduction in the number of Black and Hispanic workers receiving interviews compared to human hiring practices.  In contrast, the contextual bandit approach increases the representation of underrepresented minorities.  To our knowledge, this study provides the first empirical evidence that algorithmic design can lead to Pareto improvements in both representation and worker quality.

Modern hiring algorithms typically model the relationship between applicant covariates and outcomes in a given training dataset, and then apply this model to predict outcomes for subsequent applicants.  By systematically analyzing historical examples, this supervised learning approach can unearth predictive relationships that may be overlooked by human recruiters. Yet because this approach implicitly assumes that past examples extend to future applicants, firms that rely on this approach may favor groups with proven track records, to the detriment of non-traditional applicants. 
Indeed, because algorithms are most frequently used at the very top of the hiring funnel, this may prevent such applicants from accessing even initial interviews.

We develop and evaluate an alternative algorithm that explicitly values exploration.  Our approach begins with the idea that the hiring process can be thought of as a contextual bandit problem: in looking for the best applicants over time, a firm must balance ``exploitation'' with ``exploration'' as it seeks to learn the predictive relationship between applicant covariates (the ``context'') and applicant quality (the ``reward''). Whereas the optimal solution to bandit problems is widely known to incorporate some exploration, supervised learning based algorithms engage only in exploitation because they are designed to solve static prediction problems.  By contrast, bandit models are designed to solve dynamic prediction problems that involve learning from sequential actions: in the case of hiring, these algorithms value exploration because learning improves future choices.

Our supervised learning model (hereafter, ``SL'') is based on a logit LASSO that is trained to predict an applicant's underlying ``hiring potential,'' e.g. whether they would receive and accept and offer if interviewed.\footnote{In Section \ref{sec:offer}, we consider models which maximize offer likelihood.}  
Our model is dynamic in the sense that we update its training data throughout our analysis period with the offer and hiring outcomes of the applicants it chooses to interview.\footnote{In practice, we can only update the model with hiring outcomes for applicants it selects who are also actually interviewed in practice.  See Section \ref{sec:feasible} for a more detailed discussion of how this impacts our analysis.}  This updating allows the SL model to learn about the quality of the applicants it selects, but the model remains myopic in the sense that it does not incorporate the value of this learning into its selection decisions ex-ante. 

Our contextual bandit approach implements an Upper Confidence Bound (hereafter, ``UCB'') algorithm.  In contrast to the SL model, which evaluates candidates based on their point estimates of hiring potential, a UCB contextual bandit selects applicants based on the most optimistic assessment of their hiring potential.  That is, among applicants with the same predicted hiring potential, the UCB model would prefer the one for whom its estimate is most uncertain.  Once candidates are selected, we incorporate their realized offer and hiring outcomes into the training data and update the algorithm for the next period.


In terms of demographics, we show that a traditional SL model would interview substantially fewer Black and Hispanic applicants, relative to the firm's current hiring practices: a reduction from 9.4 percent under the status quo to 4.2 percent with the SL model.  In contrast, implementing a UCB model would more than double the share of interviewed applicants who are Black or Hispanic, from 9.4 percent to 24.3 percent.  Both models increase the share of women relative to human recruiters.  These results suggest that exploration in the bandit sense---selecting candidates with covariates for which there is more uncertainty---can lead firms to give more opportunities to workers from groups that are under-represented in their training data, even if diversity goals are not a distinct part of the algorithm's mandate.

A key question, however, is what would happen to worker quality.  Bandit algorithms may increase demographic representation by exploring, but this exploration could come at the expense of worker quality.  To assess this, we must overcome a missing data problem: we only observe hiring outcomes for candidates who were interviewed in reality.\footnote{This is also referred to as a ``selective labels'' problem.  See, for instance, \citet{lakkaraju2017,mullainathan2018, adh_bias2020}.} 
We take three complementary approaches, each based on different assumptions, all of which show that algorithms outperform human recruiters in terms of identifying applicants who are more likely to be hired by the firm.  




First, we focus on the sample of interviewed candidates for whom we directly observe hiring outcomes. Within this sample, we ask whether applicants preferred by our ML models have a higher likelihood of being hired than applicants preferred by a human recruiter. We find that, for both ML models, applicants with high scores are much more likely to be hired than those with low scores.  In contrast, there is almost no relationship between an applicant's propensity to be selected by a human, and their eventual hiring outcome; if anything, this relationship is negative.  

Our second approach uses inverse propensity score weighting to recover an estimate of mean hiring likelihood among applicants selected from our full applicant sample.  This approach infers hiring outcomes for applicants who are not interviewed using observed outcomes among interviewed applicants with similar covariates. The propensity score weighting approach is consistent as long as there is no selection on unobservables.  In our setting, this assumption is realistic because human recruiters have access to largely the same resume information we do prior to making an interview decision and do not interact with applicants. We continue to find that ML models improve hiring yield (that is, average hire rates among interviewed applicants): 32 and 27 percent of applicants selected by the SL and UCB models are eventually hired, respectively, compared with only 10 percent among those selected by human recruiters.   

Our third approach uses an instrumental variables strategy to address concerns about the potential for selection on unobservables.  In our setting, applicants are randomly assigned to initial resume screeners, who vary in their leniency in granting an interview.  We show that applicants selected by stringent screeners (e.g. those subject to a higher bar) have no better outcomes than those selected by more lax screeners: this suggests that humans are not positively screening candidates based on their unobservables.  We use this same variation to identify the returns to following ML recommendations on the margin by looking at instrument compliers.  We find that marginal candidates with high UCB scores have better hiring outcomes and are also more likely to be Black or Hispanic.  Such a finding suggests that following UCB recommendations on the margin would increase both the hiring yield and the demographic diversity of selected interviewees.  In contrast, following SL recommendations on the margin would generate similar increases in hiring yield but decrease minority representation. 

We also provide some evidence relating hiring yield to other measures of applicant quality.  We observe job performance ratings and promotion outcomes for a small subset of workers hired in our sample.  Among this selected group, we show that our ML models (trained to maximize hiring likelihood) appear more positively correlated with on-the-job performance ratings and future promotion outcomes than a model trained to mimic the choices of human recruiters.  This provides suggestive evidence that following ML recommendations designed to maximize hiring yield does not come at the expense of on-the-job performance, relative to following human recommendations.  

Finally, we also show that our results are broadly robust to focusing on whether an applicant receives an offer, rather than whether they are hired (e.g. receive and accept an offer).  We repeat much of our initial analysis using models designed to maximize offer likelihood rather than hiring likelihood.  Again, we find similar results: relative to human practices, UCB models select a more diverse set of candidates who are also more likely to receive offers.  This same is not true for the SL-based offer model which improves offer likelihood but selects fewer minority applicants.  

Together, our main findings show that there need not be an equity-efficiency tradeoff when it comes to expanding diversity in the workplace.  Specifically, firms' \textit{current} recruiting practices appear to be far from the Pareto frontier, leaving substantial scope for new ML tools to improve both hiring rates and demographic representation.  Incorporating exploration in our setting would lead our firm to interview twice as many under-represented minorities while more than doubling its predicted hiring yield.   This logic is consistent with a growing number of studies showing that firms may hold persistently inaccurate beliefs about the quality of minority applicants, and may benefit from nudges (algorithmic or otherwise) that generate additional signals of their quality.\footnote{For instance, see \citet{conradjmp, bohren1, bohren2, lepage, lepageexp}.}

At the same time, our SL model leads to similar increases in hiring yield, but at the cost of drastically reducing the number of Black and Hispanic applicants who are interviewed.  This divergence in demographic representation between our SL and UCB results demonstrates the importance of algorithmic design for shaping access to labor market opportunities.

In extensions, we consider several alternative screening policies.  We show that blinding our algorithms to race, ethnicity, and gender variables still generates increases in the share of Black, Hispanic, and female applicants who are selected relative to human hiring.  The main difference between our blinded and unblinded UCB models is that the share of selected Asian applicants increases while the share of selected White applicants decreases.  We also show that our UCB model performs better on quality when compared to a supervised learning model in which we implement group-specific quotas.  Our model further has the advantage of achieving increases in diversity without requiring firms to explicitly specify interview slots by sensitive categories such as race, ethnicity, and gender, a practice that often faces legal challenges or requires information that firms may not always be able to collect.

\section{Bias in Hiring Practices} \label{setting}

\subsection{Human Hiring}\label{sec:humans}

To be successful, firms must identify and hire the right workers. In most firms, this task falls to human workers, who screen initial applications for further consideration, conduct interviews, and make final hiring decisions.  Because resumes, interviews, and other assessment tools are limited in their ability to reveal an applicant's potential, firms ultimately have to rely on the personal judgment of their recruiters.

A longstanding social sciences literature shows that human evaluators perform their jobs imperfectly.  Human decision-makers may be simultaneously cognitively limited in their ability to process data \citep{TREISMAN198097, GABAIX2019261, benjamin}, overconfident in their assessments \citep{svenson, fischhof, kausel}, and update both too little and too much in response to feedback \citep{mobius}.  In addition to these behavioral biases, evaluators may have social preferences for particular applicants.  For example, in an ethnographic study, \citet{Rivera2012} documents how recruiters at elite professional services firms favor applicants who share the same hobbies (``She plays squash. Anyone who plays squash I love'').  Reviewers' biases may further be exacerbated by time pressure, which may lead them to lean more heavily on unreliable heuristics.  In a study of hiring for software engineering roles, \citet{alinerrecruiters2024} found that recruiters spent a median of 31 seconds per resume.  

Such behaviors may contribute to already well-documented race, ethnicity, and gender gaps in the labor market \citep{bertrandduflo, blaukahn, pager}.  For example, role congruity theory suggests that managers may find it more difficult to imagine women succeeding in high-level roles because of a mismatch between the qualities stereotypically associated with effective leaders and with women \citep{eagly2002role}. \citet{bensonlishue2021} find, indeed, that managers incorrectly assess women as having lower ``potential'' within the firm.  In a large scale correspondence study, \citet{klinerosewalters} find evidence that recruiters discriminate against Black applicants across a range of firms and industries.  In a study in Eastern Europe, \citet{bartos} shows that discrimination in outcomes may be presaged by discrimination in attention: hiring managers pay less attention when evaluating resumes with Roma-sounding names.  Recent work by \citet{bohren1} suggests that some of these differences may be due to managers having incorrect biased beliefs. 
 
A variety of studies consider ways to mitigate these biases, with mixed results.  For example, a common suggestion is that women and minorities may benefit by being evaluated by other women and minorities.  This solution, however, is often not supported in the data: 
\citet{baguesev}, for example, finds that the presence of women on recruiting committees can, in fact, hurt female applicants.  Another suggestion is to require decision-makers to undertake anti-bias trainings.  Yet, while lab studies have shown that de-biasing exercises (perspective taking, counter-stereotyping) can reduce biases, there is less evidence about their efficacy in real organizations \citep{paluck}.  Rather, evidence on durable changes in attitudes seems to come from prolonged cross-group exposure (e.g. shared living, schooling, or service) that is difficult for firms to implement as a policy \citep{baguesroth2021, gautamjmp}. Finally, affirmative-action approaches to redressing discrimination face increasing legal scrutiny \citep{SFFA_v_Harvard_2022}. 


Rather than mitigating the biases that evaluators may hold, another strand of research considers the impact of limiting their ability to exercise unconstrained judgment.  \citet{dawesbook} surveys studies examining the predictive accuracy of human evaluators across a range of settings and concludes that ``expert judgments are rarely impressively accurate and virtually never better than a mechanical judgment rule.''  
Less is known about how constraining human judgment may impact diversity outcomes.  Proponents of holistic review have argued that minority groups can benefit from an evaluator's ability to account for assessments of adversity.  At the same time, allowing for discretion may introduce opportunities for decisions to be clouded by an evaluator's implicit biases or personal preference \citep{prendergast1993discretion, bertrand2005implicit}. 

Rules-based assessments, in essence, suggest that decisions can be improved if humans behaved more like machines.  Our paper takes this idea to its conclusion and examines how the growing adoption of algorithms may impact both the quality and equity of firms' hiring practices.

\subsection{Algorithmic approaches} \label{sec:adopt}

Firms are increasingly turning toward data-driven tools to improve their hiring practices. 
The most ubiquitous hiring technology is an Applicant Tracking System (ATS).  While the baseline versions of these tools simply keep track of applicants, they often offer additional functionality, such as allowing recruiters to filter applicants based on whether their resumes meet the requirements listed on-the-job application.  A 2021 survey of employers in the US, UK, and Germany found that 94\% of respondents used some automated tools to filter or rank candidates \citep{fullerHiddenWorkersUntapped}. 

In recent years, a growing number of firms have begun offering more powerful ML-based tools that predict a candidate's suitability for a role based on historical data on hiring and performance in that role.\footnote{\citet{raghavan2019} provides an overview of such vendors.}  These types of algorithms are most commonly used at the ``top of the funnel,'' to prioritize applicants for initial interviews. Recruiters at this stage often face the task of sifting through thousands of applications for just a handful of open positions.\footnote{\citet{fullerHiddenWorkersUntapped} reports that the average job opening for a corporate position posted in 2020 receives 200 applications, up from 100 in 2010.}  Organizations surveyed about their use of such algorithms frequently express the hope that these tools will enable them to efficiently identify qualified candidates and fill vacancies more quickly \citep{upturn}. A 2020 industry survey found that 55\% of US firms use predictive analytics at some point in their human resource decision-making process, while 41\% use algorithms to make predictions about worker fit \citep{mercer2020}.\footnote{Many well-known companies, such as Intel, Johnson and Johnson, Dominos, JP Morgan, United Parcel Service, Mastercard, LinkedIn, Unilever, and Accenture, have openly acknowledged using algorithmic hiring tools for a variety of job roles \citep{qz}.  Indeed, a recent 2023 study found that 60\% of Fortune 100 firms work with a single hiring analytics company, HireVue \citep{nawratHireVue2023}.}  Algorithms are also commonly used in the public sector; \citet{nawratHireVue2023} reports that eight of the ten largest US federal agencies use algorithmic screening for some roles.

The use of algorithms, moreover, is not restricted to organizations who haven specifically chosen to buy a customized algorithmic solution.  Employers, including small employers, often post their job openings on third party job search platforms such as LinkedIn or ZipRecruiter, all of which use ML-based tools to decide which applicants to recommend for an open position.  As a result, algorithms play a role in screening applicants even for organizations that do not actively choose to employ algorithmic tools.  Further, as AI tools become increasingly integrated with widely used ATS systems, the use of algorithms is likely to grow.\footnote{For example, the ATS vendor Workday recently acquired the hiring prediction firm HiredScore in order to build more AI tools into their platform: \href{https://investor.workday.com/2024-02-26-Workday-Announces-Intent-to-Acquire-HiredScore}{acquisition annnouncement here}.}


Algorithms may not suffer from some of the key limitations that human recruiters face.  Whereas individual recruiters are likely to base their judgments on their own narrow experience, algorithms are trained on much larger datasets of applicants.  For any given applicant, algorithms are able to form predictions using many variables, while human attention is more limited.\footnote{\citet{mullainathanobermeyer2019}, for instance, provides evidence that the optimal number of variables that predict patient outcomes is greater than the number that doctors can attend to.}  Algorithms also assess applicants instantaneously, consistently, and without fatigue, in contrast with research showing that human evaluators are inconsistent and suffer from cognitive fatigue \citep{GABAIX2019261, HIRSHLEIFER201983}.  

Consistent with these advantages, existing evidence suggests that algorithms may improve the quality of hiring decisions \citep{hoffman2018, cowgill2020}.
Crucially, however, a growing literature has raised questions about how the increasing adoption of algorithms may impact equity and access to job opportunities.\footnote{For surveys of algorithmic fairness, see \citet{bakalar2021fairness, barocas_big_2016, corbett-davies_measure_2018, cowgilltucker2019}.  For a discussion of broader notions of algorithmic fairness, see \citet{kasyabebe, kleinberg_inherent_2016}.}  
A key concern is that algorithms may be trained on data that reflects historical inequities and, in turn, replicate these biases \citep{obermeyer2019, lambrecht} 
Anecdotal accounts of algorithmic bias in hiring have also been widely reported in the popular press: an audit of one resume screening model, for instance, found that the two variables it most strongly favored were being named ``Jared'' and playing high school lacrosse.\footnote{See https://qz.com/1427621/companies-are-on-the-hook-if-their-hiring-algorithms-are-biased.}  

Much of this criticism has implicitly focused on algorithms based on ``supervised learning.''  Supervised learning relies on the existence of labeled datasets to train models to predict a given outcome.  In the context of hiring, these datasets tend to be based on applicants that a firm has seen and hired in the past.  A supervised learning model may then favor applicants who play lacrosse because socioeconomic status or cultural fit has historically been predictive of success in the hiring process. To the best of our knowledge, most commercially available hiring algorithms are based on this type of approach.\footnote{In general, most firms do not provide information on the specifics of their proprietary algorithms.  However, several industry sources have indicated that this is true of their own algorithms.  Further, most discussions of hiring ML implicitly assume that this is the case.  For example, in a survey of firm approaches, \citet{raghavan2019} discuss many different ways in which firms may implement supervised learning approaches (e.g. what outcomes to train on or what historical data to use), but there is no discussion of any alternative algorithmic approaches that firms may take.  We were also unable to find any reports of firms using bandit approaches in our review of various industry surveys, e.g. \citet{mercer2020, upturn}.}  

In this paper, we highlight an alternative class of algorithms that have thus far not been applied or studied in the context of hiring: contextual bandit algorithms. Whereas supervised learning models focus solely on selecting applicants with high predicted quality, bandit algorithms also seek out candidates in order to learn about their quality.  A small empirical literature has shown that firms can benefit from non-algorithmic policies that push them to adopt more exploratory practices: \citet{conradjmp} shows that temporary affirmative action policies can generate persistent gains in minority representation, while \citet{whatley_1990} documents a similar finding by examining the racial integration of firms following World War I.  We take this idea and ask whether algorithms can implement exploration in a more efficient way.  

These ideas have broader implications for various settings in which decision-makers need to assess the quality of applicants. While we focus on hiring, these same selection problems arise in promotion, credit scoring, loan approval, university admissions, investing, and allocating research funding. In all these cases, decision-makers strive to gain insights into an applicant's quality, often in situations where historical data records may be incomplete or biased. 
In such scenarios, supervised learning algorithms designed to solve static prediction problems may not be the most suitable tools.\footnote{In many settings, such as college admissions, there has been less algorithmic adoption compared to hiring.  However, we note that the absence of explicit algorithms does not necessarily mean the absence of (potentially biased) algorithm-like thinking.  For example, academics frequently rely on institutional affiliation when assessing the quality of a piece of research.}  Instead, our paper proposes that these problems could be reframed as dynamic learning problems, for which exploration-based algorithms can be particularly useful. By embracing a more adaptive and exploratory approach, decision-makers can potentially overcome the limitations of historical data and make more informed, forward-looking assessments of applicant quality.



\section{Our Setting} \label{sec:setting}

We focus on recruiting for high-skilled, professional services positions, a sector that has seen substantial wage and employment growth in the past two decades \citep{bls2019}.  At the same time, this sector has attracted criticism for its perceived lack of diversity: female, Black, and Hispanic applicants are substantially under-represented relative to their overall shares of the workforce \citep{pew2018}.  This concern is acute enough that companies such as Microsoft, Oracle, Allstate, Dell, JP Morgan Chase, and Citigroup offer scholarships and internship opportunities targeted toward increasing the recruiting, retention, and promotion of those from low-income and historically under-represented groups.\footnote{For instance, see \href{https://www.Blacknews.com/news/top-internship-programs-for-minority-and-Black-students-for-2018/}{here} for a list of internship opportunities focused on minority applicants. JP Morgan Chase created Launching Leaders and Citigroup offers the HSF/Citigroup Fellows Award.}  However, despite these efforts, organizations routinely struggle to expand the demographic diversity of their workforce---and to retain and promote those workers---particularly in technical positions \citep{jackson2020, castilla2008, atheymentoring}.   

The hiring process at our firm works as follows.  Applicants submit their resumes to posted jobs using the firm's online portal.  On average, our firm receives approximately 200 applications per opening, which is in line with a 2021 industry study showing that corporate job postings receive an average of 250 applications \citep{fullerHiddenWorkersUntapped}.  Applicants are first screened by human recruiters, who are full-time HR professionals employed by the firm.  Recruiters do not meet or interact with applicants and their job is to decide which applicants to interview based only on the information submitted via the online portal.  


Once a candidate passes the initial resume screening, they are invited to participate in an initial interview. 
This is often a case-style interview designed to assess job skills.\footnote{For example, if the applicant was applying for a business consultant role, the interviewer may describe a hypothetical business problem and ask how the applicant would approach it.  If the candidate were applying for a data science role, the candidate may be asked to demonstrate knowledge of statistical principles and describe the types of analyses they might run to answer a particular question.}
In our data, we will use the term ``interviewed'' to refer to candidates who reach this initial interview stage.  Candidates who do well in this initial interview are then invited for a visit in which they complete additional interviews, this time conducted by employees whom the candidate is likely to be working under if hired.  These interviews include additional case interviews, as well as assessments of job fit, problem solving, and leadership.  Finally, after the interviews, the hiring team meets to discuss each candidate's strengths and weaknesses, grading them on a pre-defined rubric that includes assessments of their technical skills, communication skills, and cultural fit.  The group then ultimately votes on whether or not to extend an offer.  

We focus on the resume review stage.  Due to the need to divert current employees from other productive tasks to conduct interviews, firms are highly selective when choosing which applicants to interview: our firm rejects 95\% of applicants during its initial resume review.  Because of the volume of applicants, recruiters must eliminate many applications quickly, based on limited information. For instance, \citet{alineresumes2024} found that the median time recruiters spent reviewing a resume was only 31 seconds. Mistakes in screening not only impact firm productivity but may also perpetuate inequalities if recruiters inadvertently rely on heuristics that disadvantage qualified individuals who do not fit traditional models of success \citep{classceiling, Rivera2012}.  In light of these issues, we believe that it is particularly important to understand whether algorithmic tools can be used to improve decisions at the critical initial screening stage.   


\subsection{Data}

Our data come from a Fortune 500 company in the United States that hires workers in several job families spanning business and data analytics.  All of these positions require a bachelor's degree, with a preference for candidates graduating with a STEM major, a master's degree, and, often, experience with programming in Python, R or SQL.  Like other firms in its sector, our data provider faces challenges in identifying and hiring applicants from under-represented groups.  We have data on 88,666 job applications from January 2016 to April 2019, as described in Table \ref{sumstat_app}.   Most applicants in our data are male (68\%), Asian (58\%), or White (29\%). Black and Hispanic candidates comprise 13\% of all applications, but under 5\% of hires.  Women, meanwhile, make up 33\% of applicants and 34\% of hires. We describe our sample and variables in more detail in our Data Appendix, Sections \ref{asec:sample_details} and \ref{asec:key_vars}. 

\subsubsection{Applicant covariates}

We have information on applicants' educational background, work experience, referral status, basic demographics, as well as the type of position to which they applied.  Appendix Table \ref{atab:features} provides a list of these raw variables, as well as some summary statistics.  We have self-reported race/ethnicity (White, Asian, Hispanic, Black, not disclosed and other), gender, veteran status, community college experience, associate, bachelor, PhD, JD or other advanced degree, number of unique degrees, quantitative background (defined having a degree in a science/social science field), business background, internship experience, service sector experience, work history at a Fortune 500 company, and education at elite (Top 50 ranked) US or non-US educational institution. We record the geographic location of education experience at an aggregated level (India, China, Europe). We also track the job family each candidate applied to, the number of applications submitted, and the time between the first and most recent application. 

\subsubsection{Quality measures}

A key challenge our firm faces is being able to hire qualified workers to meet its labor demands; even after rejecting 95\% of candidates in deciding whom to interview, 90\% of interviews do not result in a hire.  These interviews are costly because they divert high-skill current employees from other productive tasks \citep{kuhn2019}.  In our paper, we therefore measure an applicant's quality as their likelihood of actually being hired by the firm.  By this definition, a high quality applicant is one that meets the firm's own hiring criteria (whatever that may be) and who accepts the firm's offer of employment.  In Section \ref{sec:offer}, we consider quality defined only by whether the firm chooses to extend an offer. 

Of course, in deciding whom to interview, firms may also care about other objectives: they may look for applicants who have the potential to become superstars---either as individuals, or in their ability to manage and work in teams---or they may avoid applicants who are more likely to become toxic employees \citep{benson2019, demingsocialskills, minortoxic, reagans2001, schumann2019making}.  Unfortunately, we observe little information on applicants' post-hire performance.  For the small set of workers for which we observe this data, we provide noisy evidence that ML models trained to maximize hiring rates are also positively related to performance ratings and promotion rates (see Section \ref{sec:onthejob}).

\section{Empirical Strategy} \label{sec:empirics}

The goal of our paper is to understand how implementing an exploration-based resume screening algorithm would impact firms' interview outcomes, relative to it's existing practices, and relative to traditional supervised learning approaches.   An ideal comparison would involve randomizing screening technologies (human, supervised ML, or exploration ML) through an experiment.  
Our analysis, however, relies on archival data.  While we observe demographics for all applicants regardless of whether they are interviewed, we observe quality measures---hiring and offer outcomes---when an applicant is interviewed.  This means that we face a ``selective labels'' problem: if an algorithm selects a candidate who is not interviewed in practice, we will not observe that candidate's interview outcome.  

In this section, we describe our framework for addressing this inference challenge.  For this discussion, it is sufficient to consider a generic ML-based interview policy.  In Section \ref{sec:algoconstruction}, we will describe the details of the specific algorithms we implement.  

\subsection{Baseline Framework} \label{sec:framework}

We consider a firm that makes interview decisions over time.  In each period $t$, the firm sees a set of job applicants indexed by $i$, and must choose which candidates to interview $I_{it} \in \{0, 1\}$.  The firm would only like to interview candidates that meet its hiring criterion, so a measure of an applicant's quality is her ``hiring likelihood'': $Y_{it} \in \{0,1\}$.  $Y_{it}$ should be thought of as a potential outcome: would applicant $i$ applying at time $t$ be hired by the firm \textit{if} they were granted an interview?  Empirically, $Y_{it}$ is an indicator for whether an applicant receives an offer from the firm or is actually hired (receives and accepts an offer).  Regardless of outcomes, the firm pays a cost, $c_t$, per interview, which can vary exogenously with time to reflect the number of interview slots or other constraints in a given period.  

The firm's payoff for interviewing worker $i$ is given by:
$$
\textit{Payoff}_{it}=
\begin{cases}
Y_{it}-c_t \text{ if } I_{it}=1\\
0 \text{ if } I_{it}=0
\end{cases}
$$
For each applicant $i$ in period $t$, the firm also observes a vector of demographic, education, and work history information, denoted by $X'_{it}$.  These variables provide ``context'' that can inform the expected returns to interviewing a candidate.  We write $E[Y_{it} | X'_{it}]= \mu(X'_{it}\theta_t^*)$, where $\mu: \mathbb{R} \to \mathbb{R}$ is a link function and $\theta_t^*$ is an unobserved vector describing the true predictive relationship between covariates $X'_{it}$ and hiring potential $Y_{it}$.\footnote{In practice, when estimating contextual bandit models, most algorithms make functional form assumptions about the underlying true relationships. Of course, in practice firms do not know the true relationship, preventing firms from implementing the ideal decision rule. } We allow $X'_{it}$ to include components that are observed and unobserved by the econometrician.  After each period $t$, the firm observes the payoffs associated with its chosen actions. 

Given this information, we can think of a firm's interview decision for applicant $i$ at time $t$ as given by:
\begin{equation} \label{interviewpolicy}
I_{it}=\mathbb{I}(s_t(X'_{it}) > c_t)
\end{equation}
Here, $s_t(X'_{it})$ can be thought of as a score measuring the value the firm places on a candidate with covariates $X'_{it}$ at time $t$. This score is indexed by $t$ to reflect the fact that the value of a given applicant can change over time if the firm's beliefs about their quality change or if the firm's priorities do.  The firm's goal is to identify a scoring function $s_t(X'_{it})$ that leads it to identify and interview applicants with $Y_{it}=1$ as often as possible.  

Our model mirrors a standard contextual multi-arm bandit (MAB) problem.\footnote{In a generic contextual MAB, an agent receives information on ``context'' before deciding which bandit ``arm'' to pull in order to receive different ``rewards''.  In our case, the context information is an applicant's resume and demographics $X_{it}$; the arms are the decision of whether to interview or not $I_{it} \in \{0, 1\}$; and the rewards are the associated payoffs $Y_{it}-c_t \text{ if } I_{it}$ if interview or 0 if not.} Leaving aside the optimal choice of scoring function (which we discuss later in Section \ref{sec:algoconstruction}), we can think of \textit{any} interview policy as being described by some scoring function $s_t(\cdot)$ and its associated interview decision $I_{it}$.  In particular, we write $s_t^H$ and $I_{it}^H$ to refer to the (H)uman interview policy that is used by the firm and $s_t^{ML}$ and $I_{it}^{ML}$ to refer to any counterfactual machine-learning (ML) based interview policy.  For notational simplicity, we suppress the subscripts for applicant $i$ at time $t$ for the remainder of the paper, unless we are discussing specific regressions or details of algorithm construction.  

\subsection{Addressing sample selection}


We are interested in understanding how the quality and demographics of the interviewed candidates change under different interview policies. We see demographics for all applicants and therefore do not face sample selection problems in comparing the demographics of counterfactual interview policies. However, we would also like to examine quality by comparing $E[Y|I^H=1]$ versus $E[Y|I^{ML}=1]$ for traditional and exploration-based ML approaches.  $E[Y|I^H=1]$ is readily computable because we directly observe the hiring potential $Y$ for all workers chosen to be interviewed by human recruiters.  $E[Y|I^{ML}=1]$, however, is only partially observable because we only see hiring outcomes for the subset of ML-selected applicants who are actually interviewed (e.g. selected by human recruiters): $E[Y|I^{ML}=1 \cap I^H=1]$.  In our analysis, we address potential biases in assessing the counterfactuals associated with sample selection in three complementary ways. In Section \ref{mainresults}, we show that we obtain similar results under each approach.  

\subsubsection{Interviewed sample only} \label{sec:intonly}

Our first approach examines the predictive relationship between algorithm scores $s^{ML}$ and applicant quality among the subset of applicants who are actually interviewed, e.g. those for whom we observe realized hiring potential.  
To compare our ML model's preferences to that of human recruiters, we construct a measure of $s^H$, the implicit ``score'' that humans assign to applicants by training a model to predict an applicant's likelihood of being selected for an interview $E[I^H|X]$, as described in Appendix \ref{asec:m_humandetails}.  We then compare the predictive power of $\hat{s}^H$ with that of $s^{ML}$ among the interviewed sample. Our findings, discussed in Section \ref{sec:intonlyres}, will show that the average quality among candidates preferred by ML-based screening approaches is higher than that of candidates preferred by human recruiters. 

\subsubsection{Full sample, assuming no selection on unobservables} \label{sec:ipw}

A concern with our above approach is that human recruiters may add value by screening out particularly poor candidates so that they are never observed in the interview sample to begin with.  In this case, there may be no correlation between human preferences and hiring potential among those who are interviewed, even if human preferences are highly predictive of quality in the full sample.

Our next approach addresses this by estimating the average quality of \textit{all} ML-selected applicants, $E[Y|I^{ML}=1]$. We infer hiring likelihoods for ML-selected applicants who were not interviewed using observed hiring outcomes from applicants with similar covariates who were interviewed, assuming no selection on unobservables: $E[Y|I^{ML}=1, X] =  E[Y|I^{ML}=1, I=1, X]$.  

In our setting, this is a plausible assumption because recruiters have very little additional information relative to what we also observe. Screeners never interact with applicants and make interview decisions on the basis of applicant resumes.  Because the hiring software used by our data firm further standardizes this information into a fixed set of variables, they generally do not observe cover letters or even resume formatting.  Given this, the types of applicant information that are observable to recruiters but not to the econometrician are predominately related to resume information that we do not code into our feature set.  For example, we convert education information into indicator variables for college major designations, institutional ranks, and types of degrees.  A recruiter, by contrast, will see whether someone attended the University of Wisconsin or the University of Michigan.\footnote{Adding additional granularity in terms of our existing variables into our model does not improve its AUC.}  In addition to worker characteristics, our models also include characteristics of the job search itself to account for factors that influence hiring demand independent of applicant characteristics. For more details on the construction of our key variables, see Appendix \ref{asec:datamodelsappendix}.

Following \citet{hirano2003} and assuming no selection on unobservables, we can write the inverse propensity weighted estimate of ML-selected workers' hiring likelihood as:
\begin{eqnarray}
	E[Y|I^{ML}=1] &=& \frac{p(I=1)}{p(I^{ML}=1)}E\left[\frac{p(I^{ML}=1 | X)}{p(I=1|X)}Y|I=1\right] \label{decompeqn}
\end{eqnarray} 

Equation \eqref{decompeqn} says that we can recover the mean quality of ML-selected applicants by reweighting outcomes among the human-selected interview sample, using the ratio of ML and human-interview propensity scores.\footnote{See Appendix \ref{asec:ipw} for the full derivation of Equation \eqref{decompeqn}.}  The ML decision rule is a deterministic function of covariates $X$, meaning that the term $p(I^{ML}=1 | X)$ is an indicator function equal to one if the ML rule would interview the applicant, and zero if not.  The term $p(I^{H}=1 | X)$ is just the human selection propensity which we estimate as $\hat{s}^H$, described previously.  Finally, because we always select the same number of applicants as are actually interviewed in practice, the term $\frac{p(I=1)}{p(I^{ML}=1)}$ is equal to one by construction.  


\subsubsection{Marginally promoted sample, IV analysis} \label{sec:iv}


We continue to be concerned about the possibility of selection on unobservables, particularly if human recruiters screen out unobservably bad applicants.  In that scenario, our previous approach would overstate hiring outcomes for ML-selected applicants who do not receive an interview. 

To address this, our next approach compares the quality of human and ML selected candidates, using exogenous variation in interview propensity arising from the random assignment of applicants to resume screeners, following the methodology pioneered by \citet{kling2006}. For each applicant, we form the jackknife mean interview rate of their assigned screener and use the leniency of the screener as an instrument, $Z$, for whether the applicant is interviewed.

Appendix Figure \ref{hist_iv} plots the distribution of jackknife interview pass rates in our data, restricting to the 54 screeners (two thirds of the sample) who evaluate more than 50 applications (the mean in the sample overall is 156).  After controlling for job family, job level, and work location fixed effects, the 75th percentile screener has a 50\% higher pass rate than the 25th percentile screener.  Table \ref{iv_validity} shows that this variation is predictive of whether a given applicant is interviewed (Column 1), but is not related to any of the applicant's covariates (Columns 2-8).   In Appendix Figure \ref{I_ML_byleniency}, we plot the relationship between ML scores (both UCB and SL) and interview likelihood separately for applicants assigned to strict or lenient screeners.  We show that those assigned to lenient screeners are more likely to be interviewed across all ML scores, and that this advantage is essentially constant. In Appendix \ref{asec:ivvalidity}, we show that strict and lenient screeners appear to have similar preference orderings of applicants, reducing concerns about violations of monotonicity.

Given this instrument, the goal of our analysis in this section is to ask whether firms can improve both hiring yield by following ML recommendations ``on the margin,'' that is, in cases when the human recruiter appears to be on-the-fence about whether or not to interview a candidate.  Specifically, consider the following counterfactual interview policy, given our recruiter leniency instrument $Z$ and algorithmic score $s^{ML}$:
$$
\tilde{I}=
\begin{cases}
I^{Z=1} \text{ if } s^{ML} \geq \tau,\\
I^{Z=0} \text{ if } s^{ML} < \tau.
\end{cases}
$$
$\tilde{I}$ takes the firm's existing interview policy, $I$, and modifies it at the margin. The new policy $\tilde{I}$ favors applicants with high ML scores by asking the firm to make interview decisions $I$ as if these applicants were randomly assigned to a generous initial screener ($Z=1$).\footnote{For simplicity in exposition, we let $Z$ be a binary instrument in this example (whether an applicant is assigned to an above or below median stringency screener) although in practice we will use a continuous variable.}  That is, $I^{Z=1}$ refers to the counterfactual interview outcome that would be obtained, if an applicant were evaluated by a lenient screener.  Similarly, $\tilde{I}$ penalizes applicants with low ML scores by making interview decisions for them as though they were assigned to a stringent screener ($Z=0$).  

By construction, the interview policy $\tilde{I}$ differs from the status quo policy $I$ only in its treatment of instrument compliers.  Compliers with high ML scores will be selected under $\tilde{I}$ because they are treated as if they are assigned to lenient recruiters.  Conversely, compliers with low ML scores are rejected because they are treated as if they are assigned to strict reviewers.  As such, $\tilde{I}$ provides a concrete example of an alternative policy that improves hiring yield by following ML recommendations more if compliers with high ML scores have greater hiring potential than compliers with low ML scores, $E[Y| I^{Z=1} > I^{Z=0}, s^{ML} \geq \tau]$ vs. $E[Y| I^{Z=1} > I^{Z=0}, s^{ML} < \tau]$.  


\section{Algorithm Design} \label{sec:algoconstruction}

Having discussed our general empirical strategy, we now provide an overview of the specific algorithms we consider.  In Appendix \ref{asec:datamodelsappendix}, we provide additional information regarding our sample and feature construction, as well as model training, fitting, and updating procedures.

\subsection{Preliminaries}

We begin by clarifying some relevant issues for all models we consider.

We divide our data into two periods, the first consisting of the 48,719 applicants arriving before 2018 (2,617 of whom are interviewed), and the second consisting of the 39,947 applications that arrive in 2018-2019 (2,275 of whom are interviewed).  We think of the 2016-2017 period as our ``training'' data and the 2018-2019 period as our ``analysis'' period.  This approach to defining a training dataset (rather than taking a random sample of our entire data) most closely approximates a real world setting in which firms would likely use historical data to train a model that is then applied prospectively.  We also continue to update our models during the 2018-2019 analysis sample by adding the outcomes of the applicants it selects back into its training data.


Our goal is to predict applicants' hiring potential, $Y$, as defined in Section \ref{sec:framework}. Because hiring potential is only directly observed for applicants who are interviewed, we train our models using data from interviewed applicants only. We note that not all screening algorithms are trained in this way.  One common approach is to predict the human recruiter's interview decision, $I$, rather than the applicant's hiring potential $Y$.  In this case, vendors are able to train their data on all applicants.  Another common approach is to focus on predicting hiring likelihood $Y$, but to set $Y=0$ for all applicants who are not interviewed.  This essentially assumes that candidates who were not interviewed had low hiring potential.  We choose not to follow either of these approaches because they conflate a recruiter's decision $I$ with the quality of that decision $Y$. \citet{rambachan2019bias}, in addition, show that such approaches tend to be more biased against racial minorities.  

Next, we acknowledge that our measures of quality---hire and offer outcomes---are based on the discretion of managers and potentially subject to various types of evaluation biases \citep{Quadlin2018, castilla2011}.  Indeed, many on-the-job performance metrics, such as performance evaluations, would also be subject to this concern.  Without a truly ``objective" measure of quality, we interpret our results as asking whether ML tools can improve firm decisions, as measured by its own revealed preference metrics. 

Finally, our models may generate inaccurate predictions if the relationship between covariates $X$ and hiring likelihood $Y$ differs between the full applicant sample and the interviewed subset.  While there are a growing set of advanced ML tools that seek to correct for training-sample selection,\footnote{See, for example, \citet{dimakopoulou_balanced_2018}, \citet{dimakopoulou_estimation_2018} which discuss doubly robust estimators to remove sample selection and \citet{si2020distributional}.} testing these approaches is outside of the scope of this paper and we are not aware of any commercially available algorithms that employ sample selection correction.  We provide evidence that selection on unobservables does not appear to be a large concern in Section \ref{sec:ipwres}.

In general, we emphasize that the ML models we build should not be thought of as ``optimal'' in either their design or performance, but as an example of what could be feasibly achieved by most firms that are able to organize their administrative records into a modest training dataset, with a standard set of resume-level input features, using a technically accessible ML toolkit.

	
\subsubsection{Supervised Learning ( ``SL'')} \label{sl_details}

Our first model uses a standard supervised learning approach to predict an applicant's likelihood of being hired, conditional on being interviewed.  We begin with an initial training dataset, $D_0$ and use it to form an estimate of applicant quality $\hat{E}[Y_{it} | X'_{it}; D_0]$ using a L1-regularized logistic regression (LASSO).  Appendix Figure \ref{roc} plots the receiver operating characteristic (ROC) curve and its associated AUC, or area under the curve.  This model has an AUC of 0.64, meaning that it will rank an interviewed applicant who is hired ahead of an interviewed but not hired applicant 64 percent of the time.  We also plot the confusion matrix in Appendix Figure \ref{cm}, which provides more information on the model's classification performance.

Having trained this initial model on 2016-17 data, we use it to make interview decisions for future applicants.  It is common for firms to train once and then continue to use the same static model.  In our paper, however, we estimate a dynamic SL model that updates the firm's training data with the outcomes of applicants it selects later on.  Specifically, we divide our analysis sample (2018-2019) into ``rounds'' of 100 applicants.  After each round, we take the applicants the model has selected and update its training data.  We then retrain the model and use its updated predictions to make selection decisions in the next round.  At any given point $t$, the SL model's interview policy is as follows, based on Equation \eqref{interviewpolicy} of our conceptual framework:
\begin{equation} \label{SLrule}
I_{it}^{SL}=\mathbb{I}(s_t^{SL}(X'_{it})>c_t), \text{ where } s_t^{SL}(X'_{it})=\hat{E}[Y_{it}|X'_{it}; D^{SL}_{t}].
\end{equation}
Here, $D^{SL}_{t}$ is the training data available to the algorithm at time $t$.  

It is important to emphasize that we can only update the model's training data with \textit{observed} outcomes for the set of applicants selected in the previous period: that is, $D^{SL}_{t+1} = D^{SL}_{t} \cup (I^{SL}_{t} \cap I_{t})$.  Because we cannot observe hiring outcomes for applicants who are not interviewed in practice, we can only update our data with outcomes for applicants selected by both the model and actual human recruiters.  This may impact the degree to which the SL model can learn about the quality of the applicants selected, relative to a world in which hiring potential is fully observed for all applicants. We discuss this in more detail shortly, in Section \ref{sec:feasible}. 

\subsubsection{Upper Confidence Bound (``UCB'')} \label{ucb_details}

While there is, in general, no generic optimal strategy for the contextual bandit model described in Section \ref{sec:framework}, it is widely known that exploitation-only approaches---such as the SL model described above---are inefficient solutions because they do not factor the ex-post value of learning into their ex-ante selection decisions \citep{dimakopoulou_estimation_2018, bastani2019}.   An emerging literature in computer science has therefore focused on developing a range of computationally tractable algorithms that incorporate exploration.\footnote{The best choice of algorithm for a given situation will depend on the number of possible actions and contexts, as well as on assumptions regarding the parametric form relating context to reward.  For example, recently proposed contextual bandit algorithms include UCB (\citet{auer_using_2002}), Thompson Sampling (\citet{agrawal2013}), and LinUCB (\citet{li_contextual-bandit_2010}).  In addition, see \citet{agrawal2013}, and \citet{bastani_online_2019}.  Furthermore, the existing literature has provided regret bounds---e.g., the general bounds of \citet{russo2015informationtheoretic}, as well as the bounds of \citet{rigollet2010nonparametric} and \citet{slivkins_contextual_2014} in the case of non-parametric function of arm rewards---and has demonstrated several successful applications areas of application---e.g., news article recommendations (\citet{li_contextual-bandit_2010}) or mobile health (\citet{lei2017actorcritic}). For more general scenarios with partially observed feedback, see \citet{rejwan2019topk} and \citet{bechavod2020equal}. } 


The particular exploration-based implementation we use is an Upper Confidence Bound Generalized Linear Model (UCB-GLM) described in \citet{li2017}.  We choose this approach because it best fits our setting. UCB-GLM works well when the relationship between ``context'' variables (covariates $X_{it}$) and ``reward'' (hiring potential, $Y_{it}$) follows a generalized linear functional form ($E[Y_{it} | X'_{it}] = \mu(X'\theta_t^*)$): we measure $Y_{it}$ as a binary hiring outcome and estimate $E[Y_{it} | X'_{it}]$ using a logistic regression.  Under these circumstances, \citet{li2017} provides the algorithm implementation we follow and shows that it is asymptotically regret-minimizing.\footnote{See Equation 6 and Theorem 2 of their paper.}

Specifically, our UCB algorithm scores applicant $i$ in period $t$ as follows:
\begin{equation} \label{UCBrule}
I_{it}^{UCB}=\mathbb{I}(s_t^{UCB}(X'_{it})>c_t), \text{ where } s_t^{UCB}(X'_{it})=\hat{E}[Y_{it}|X'_{it}; D^{UCB}_{t}] + \alpha B(X'_{it}; D^{UCB}_{t}).
\end{equation}
In Equation \eqref{UCBrule}, the scoring function $s_t^{UCB}(X'_{it})$ is a combination of the algorithm's expectations of an applicant's quality based on its training data and an ``exploration bonus'' given by:
\begin{equation} \label{UCBBonus} 
B(X'_{it}; D^{UCB}_{t}) = \sqrt{(X_{it}-\bar{X_t})'V_t^{-1}(X_{it}-\bar{X_t})}, \text{ where } V_t  = \sum_{j \in D^{UCB}_{t}} (X_{jt}-\bar{X_t})(X_{jt}-\bar{X_t})'.
\end{equation}

Intuitively, Equation \eqref{UCBrule} breaks down the value of an action into an exploitation component and an exploration component.  In any given period, a strategy that purely focuses on exploitation would choose to interview a candidate on the basis of her expected hiring potential: this is encapsulated in the first term, $\hat{E}[Y_{it}|X'_{it}; D_t^{UCB}]$. Indeed, this is essentially the scoring function for the SL model, described in Equation \eqref{SLrule}.  Meanwhile, a strategy that purely focuses on exploration would choose to interview a candidate on the basis of the distinctiveness of her covariates: this is encapsulated in the second term, $B(X'_{it}; D_t^{UCB})$, which shows that applicants receive higher bonuses if their covariates deviate from the mean in the population ($X_{it}-\bar{X_t}$), especially for variables $X'_{it}$ that generally have little variance in the training data (e.g. weighted by the precision matrix $V_t^{-1}$).  To balance exploitation and exploration, Equation \eqref{UCBrule} combines these two terms.  As a result, candidates are judged on their mean expected quality \textit{plus} their distinctiveness from the existing training data.  The term $\alpha$ captures the weight that we put on the exploration component relative to the exploitation component.  Taken together, \citet{li2017} shows that this provides an upper bound on the confidence interval associated with an applicant's true quality, given the training data $D_t^{UCB}$, hence the term UCB.  In our model, we follow the approach described in \citet{li2017} and choose an $\alpha$ of 1.96 so that we are, in fact, using the upper 95th percentile bound. 
In essence, UCB approaches are based on the principle of ``optimism in the face of uncertainty,'' favoring candidates with the highest statistical upside potential.\footnote{The basic UCB approach for non-contextual bandits was introduced by \citet{lai1985} and, since then, various versions of this approach have been developed for different types of contextual bandit settings, and shown to be regret minimizing.}

At time $t=0$ of our analysis sample, our UCB and SL models share the same predicted quality estimate, which is based on the baseline model trained on the 2016-2017 sample.  As with the SL model, we update the UCB model's training data with the outcomes of applicants it has selected during the 2018-2019 analysis period.  Based on these new training data, the UCB algorithm updates both its beliefs about hiring potential and the bonuses it assigns.  As was the case with the SL model, we can only add applicants who are selected by the model and also interviewed in practice.   

\subsubsection{Model comparisons} 

A large theoretical literature shows that exploration-based model such as UCB will outperform exploitation-only based approaches in the long run \citep{dimakopoulou_estimation_2018}.  \citet{li2017} proves that the specific model we adopt, UCB-GLM will asymptotically minimize regret via more efficient learning: that is, it will select applicants with greater hiring potential. Yet while a UCB based approach is expected to out-perform SL models in the long run, the quality differences we would observe in practice capture both the long term benefits of learning and the short term costs of exploration. This tradeoff will also depend on the specifics of our empirical setting.  In particular, if quality is not evolving and there is relatively rich initial training data, SL models may perform as well as, if not better than, UCB models because the value of exploration will be limited. If, however, the training data were sparse or if the predictive relation between context and rewards evolves over time, then the value of exploration is likely to be greater.  

In terms of diversity, our UCB algorithm favors candidates with distinctive covariates because this helps the algorithm learn more about the relationship between applicant covariates and hiring outcomes.  This suggests that a UCB model would, at least in the short run, select more applicants from demographic groups that are under-represented in its training data, relative to an SL model.\footnote{We note, however, that we calculate bonuses over all covariates.  If White and Asian applicants are more heterogeneous along other dimensions such as education and work history, then they may nonetheless receive high exploration bonuses.} Over time, however, exploration bonuses will decline as the model receives more information about applicants of all types.  As a result, long run differences between SL and UCB models will be primarily driven by differences in their beliefs about applicant quality.  Gains in diversity driven by exploration bonuses will not be sustainable if minority applicants are actually weaker.  

Our main results will come from a 16 month period from January 2018 to April 2019.  Because most organizations cannot afford to care only about the long run impacts of new hiring policies, it becomes important to empirically examine how exploration-based algorithms behave over medium-run time scales. 
 
%

\subsection{Feasible versus Live Model Implementation} \label{sec:feasible}

In a live implementation, each algorithm would select which applicants to interview, and the model would be updated with the outcomes of all selected candidates.   In our retrospective analysis, we are only able to update our models with outcomes for ML selected candidates who were actually interviewed in practice.  Here, we discuss how the actual implementation of our models---which we term ``feasible'' SL or UCB---may differ from a live implementation.

For concreteness, suppose that the UCB model wants to select 50 theater majors but, in practice, only 5 such applicants were actually interviewed.  In our feasible implementation, we would only be able to update the UCB's training data with the outcomes of these five applicants, whereas in a live implementation, we would be able to update with outcomes for all 50 UCB-selected candidates.

If humans are not selecting on unobservables, the feasible UCB's estimate of the quality of theater majors would be the same as the live UCB's estimate but, because it observes five rather than 50 instances, its estimates would be less precise.  This would impact the exploration in the next period: even though it has the same beliefs about quality, the feasible UCB would select more theater majors in the next period because its uncertainty about these applicants is higher.  This would slow down the learning of the feasible UCB model relative to the live model but, with a large enough sample, both should converge to the same beliefs and actions.\footnote{Formally, the distinction between the feasible and live versions of our ML models is related to regression in which outcomes are missing at random conditional on unobservables.  Under the assumption of no selection on unobservables, common support, and well-specification of the regression function (in our case, the logit), the feasible and live versions of our models should both be consistent estimators of the underlying parameter $\theta^*$ linking covariates with hiring outcomes: $E[Y_{it} | X'_{it}]= \mu(X'_{it}\theta_t^*)$ \citep{wangetal2010, robinsetal1995}.  In a finite sample, of course, the point estimates of the feasible and live models may differ.}

This analysis changes if human screeners select based on unobservables.  Suppose the 5 theater majors who are interviewed are unobservably better than the 45 theater majors who are not interviewed. A feasible UCB model would then be too optimistic about the quality of this population relative to a live UCB model that correctly learns the quality of all 50 applicants.  In the next period, the feasible UCB model would select more theater majors both because uncertainty remains higher and because positive selection on unobservables induces upwardly biased beliefs.  This latter bias can lead our approach to select too many applicants from groups whose weaker members are screened out of the model's training data by human recruiters.  In Section \ref{sec:ipwres} we provide IV-based evidence that human recruiters do not appear to be selecting on unobservables. In addition, Section \ref{sec:simulations} shows simulation results that more closely approximate a live implementation in which we can update outcomes for all selected candidates.

\section{Main Results} \label{mainresults}

\subsection{UCB and SL versus Human Recruiters: Diversity of Selected Applicants} \label{resdiversity}

We begin by assessing the impact of each policy on the diversity of candidates selected for an interview in our analysis sample.  This is done by comparing $E[X| I=1]$, $E[X | I^{SL}=1]$, and $E[X | I^{UCB}=1]$, for various demographic measures $X$, where we choose to interview the same number of people as the actual recruiters in a given year-month.  We observe demographic covariates such as race or ethnicity and gender for all applicants, regardless of their interview status, and do not face a selective labels problem for comparisons of demographics.

We focus on the racial and ethnic composition of selected applicants.  Panel A of Figure \ref{pie_race} shows that, at baseline, 58\% of applicants in our analysis sample are Asian, 29\% are White, 9\% are Black, and 4\% are Hispanic.  Panel B shows that human recruiters select a similar proportion of Asian and Hispanic applicants (57\% and 5\%, respectively), but relatively more White and fewer Black applicants (34\% and 5\%, respectively).  In Panel C, we show that the SL model reduces the share of Black and Hispanic applicants from 10\% to under 5\%, White representation increases more modestly from 34\% to 40\%, and Asian representation stays largely constant.  In contrast, Panel D shows that the UCB model increases the Black share of selected applicants from 5\% to 16\%, and the Hispanic share from 4\% to 9\%.  The White share stays constant, while the Asian share falls from 57\% to 43\%.

Appendix Figure \ref{pie_gender} plots the same set of results for gender.  Panel B shows that 66\% of interviewed applicants are men and 34\% are women; this is largely similar to the gender composition of the overall applicant pool.  Unlike the case of race or ethnicity, both our ML models are aligned in selecting more women than human recruiters, increasing their representation to 41\% (SL) or 38\% (UCB).


Next, we explore why our UCB model selects more Black and Hispanic applicants.  Appendix Figure \ref{ucb_bonuses_static} shows that Black and Hispanic applicants receive slightly larger exploration bonuses on average. This reflects both direct differences in population size by race or ethnicity, as well as indirect differences arising from the correlation between race or ethnicity and other variables that also factor into bonus calculations.  

A crucial question raised by this analysis is whether these differences in diversity are associated with differences in applicant quality.  We will discuss this extensively in the next section and provide evidence that, despite their demographic differences, hiring outcomes for applicants selected by our SL and UCB models are comparable to each other, and much better than those selected by human recruiters.

\subsection{UCB and SL versus Human Recruiters: Quality of Selected Applicants} \label{results:quality}

While we observe demographics for all applicants, we only observe hiring potential $H$ for applicants who are actually interviewed.  We therefore cannot directly observe hiring potential for applicants selected by an algorithm, but not by the human reviewer.  To address this, we take three complementary approaches, described previously in Section \ref{sec:empirics}.  Across all three approaches, we find evidence that both SL and UCB models would select applicants with greater hiring potential, relative to human screening.  

\subsubsection{Interviewed sample} \label{sec:intonlyres}
Our first approach restricts to the sample of applicants who are interviewed, for whom we directly observe hiring outcomes.  Among this set, we directly observe ML scores $s^{SL}$ and $s^{UCB}$.  We do not, however, directly observe the implicit score that human recruiters give each candidate.  Before continuing, we therefore need to generate an estimate of ``$s^H$,'' an applicant's propensity to be selected for an interview by a human recruiter.  To do this, we simply generate a model of $E[I|X]$ where $I \in \{0,1\}$ are realized human interview outcomes, using same logistic LASSO approach described in Section \ref{sl_details}.\footnote{The only methodological difference between this model and our SL model is that, because we are trying to predict interview outcomes as opposed to hiring outcomes conditional on interview, our training sample consists of all applicants in the training period, rather than only those who are interviewed.} We describe this model construction and the training and procedure in Appendix Section \ref{asec:m_humandetails}. Appendix Figure \ref{roc_human} plots the ROC associated with this model.  Our model ranks a randomly chosen interviewed applicant ahead of a randomly chosen applicant who is not interviewed 77\% of the time.\footnote{Although a ``good" AUC number is heavily context specific, a general rule of thumb is that models with an AUC in the range of $0.75-0.85$ have acceptable discriminative properties, depending on the specific context and shape of the curve \citep{fischer2013}.}

Figure \ref{corr_quality} plots a binned scatterplot depicting the relationship between algorithm scores and hiring outcomes among the set of interviewed applicants; each dot represents the average hiring outcome for applicants in a given scoring bucket.  Among those who are interviewed, applicants' human scores are uninformative about their hiring likelihood; if anything this relationship is slightly negative.\footnote{This weak relation between human preferences and outcomes is consistent with existing work documenting that humans often have incorrect perceptions of worker quality.  For instance, \citet{hoffman2018} find that firms see worse hiring outcomes when humans make exceptions to algorithmic suggestions.  In a study of personnel assessment, \citet{yu2020} finds that the scores of expert human resource managers were weakly related to on-the-job performance. Similarly, in a study of recruiters for software engineering positions, \citet{alineresumes2024} found a weak correlation between recruiter ranking and eventual hiring likelihood.}

In contrast, all ML scores have a statistically significant, positive relation between algorithmic priority selection scores and an applicant's (out of sample) likelihood of being hired.\footnote{ Appendix Table \ref{allalgoscorr} shows these results as regressions to test whether the relationships are statistically significant.} 
%

Table \ref{disagreement} examines how these differences in scores translate into differences in interview policies.  To do so, we consider ``interview'' strategies that select the top 25, 50, or 75\% of applicants as ranked by each model; we then examine how often these policies agree on whom to select, and which policy performs better when they disagree.  Panel A compares the SL model to the human interview model and shows that the human model performs substantially worse in terms of predicting hiring likelihood when the models disagree: only 5-8\% of candidates favored by the human model are eventually hired, compared with 17-20\% of candidates favored by the SL model.  Panel B finds similar results when comparing the human model to the UCB model.  Finally, Panel C shows that, despite their demographic differences, the SL and UCB models agree on a greater share of candidates relative to the human model, and there do not appear to be significant differences in overall hiring likelihoods when they disagree: if anything, the UCB model performs slightly better.  

For consistency, Appendix Figure \ref{pie_diversity_interviewed} revisits our analysis of diversity using the same type of selection rule described in this section: specifically, picking the top 50\% of candidates among the set of interviewed.  Again, we find that UCB selects a substantially more diverse set of candidates than the SL model.  

\subsubsection{Full sample} \label{sec:ipwres}

A concern with our analysis of the $I=1$ sample is that human recruiters may add value by screening out particularly poor candidates so that they are never observed in the interview sample to begin with.  In this case, then we may see little relation between human preferences and hiring potential among those who are interviewed, even though human preferences are highly predictive of quality in the full sample.

To address this, we compute estimates of the hiring likelihood of all ML selected applicants, using the inverse propensity weighting (IPW) approach described earlier in Section \ref{sec:ipw}.  Our results are presented in Figure \ref{bar_quality}.  Among those selected by human recruiters, the average observed hiring likelihood is 10 percent.  In contrast, our calculations show that ML models select applicants with almost 3 times higher predicted hiring potential.  In particular, the average expected hiring likelihood for applicants selected by the UCB model is 27 percent and 32 percent for the SL model.  The slightly weaker performance of the UCB model may be explained by the fact that an emphasis on exploration means that the UCB algorithm may select weaker candidates, particularly in earlier periods.  Together, this set of results is consistent with our findings from the interviewed-only subsample: the hiring yield of ML algorithms are similar to each other and, in all cases, better than the human decision-maker.  We find no evidence that the gains in diversity that we document in Section \ref{resdiversity} come at the cost of substantially reducing hiring rates among selected applicants.  \\

\noindent\textbf{Testing IPW assumptions}: This analysis relies on two assumptions---no selection on unobservables and a common support between ML and human preferences.

First, we test for the presence of selection on unobservables using variation from random assignment to lenient and strict reviewers, as described in Section \ref{sec:iv}.  The logic is as follows: if humans are, on average, positively selecting candidates on observables, then it should be the case that applicants selected by more stringent reviewers---e.g. those who are subjected to a higher human threshold---should be more likely to be hired conditional on being interviewed than those selected by more lenient reviewers. That is, if there is a positive (negative) relationship between human selection propensities and hiring outcomes, then going further down the distribution by selecting more candidates should decrease (increase) average quality. 

\begin{figure}[ht!]
\begin{center}
\captionsetup{justification=centering}
\caption{\textsc{Figure \ref{selectiononunobs}: Testing for Positive Selection}}
\makebox[\linewidth]{
\begin{tabular}{c}
\textsc{\footnotesize{A. Full Sample, No Controls}} \\
\includegraphics[scale=0.7]{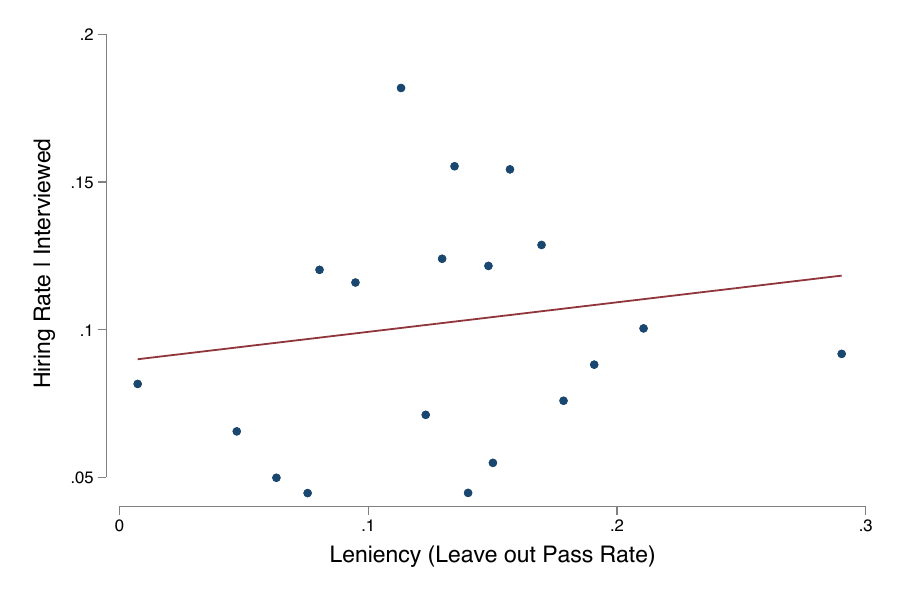}\\
\textsc{\footnotesize{B. Full Sample, Controls}} \\ 
\includegraphics[scale=0.7]{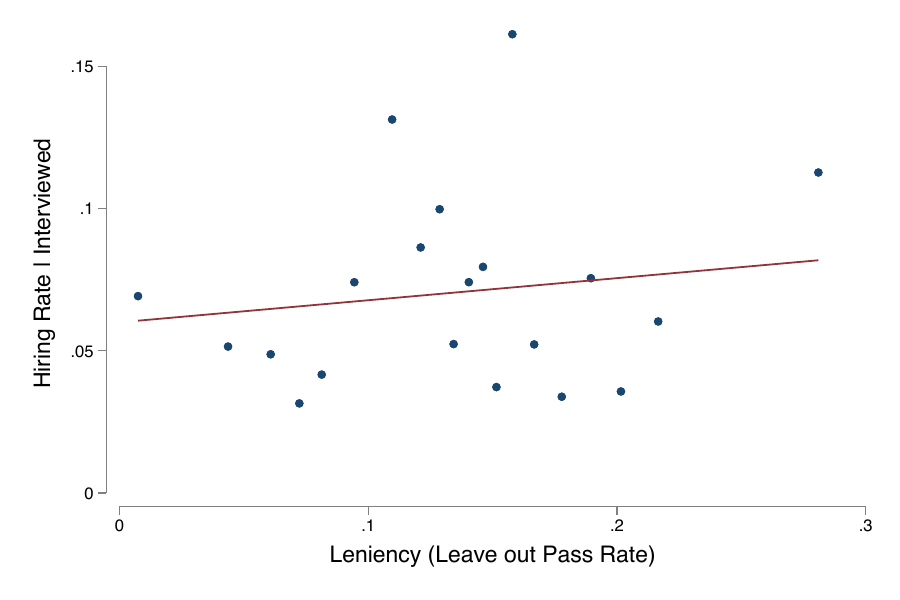}
\end{tabular}
}
\label{selectiononunobs}
\end{center}
\end{figure}

Figure \ref{selectiononunobs} plots the relationship between screener leniency and hiring outcomes.  Panel A of Figure \ref{selectiononunobs} does not control for applicant observables while Panel B does.  In both cases, there appears to be little relationship between leniency and hiring outcomes, suggesting that strict reviewers do not appear to select a stronger set of candidates.  If anything, there appears to be a slightly positive relation, suggesting that lenient reviewers may weakly select a stronger group of stronger candidates.  The possibility that human recruiters are actively bad at selecting candidates is consistent with our results in Panel A of Figure \ref{corr_quality}, which shows that workers preferred by humans have somewhat lower hiring rates among those who were interviewed.  We note that both these figures include fixed effects for job family, job level, work location, and application year so that this slightly positive association is not being driven by confounding differences in hiring demand across positions or times.
Interestingly, we also note that the pattern is similar regardless of whether we control for applicant covariates: this suggests that strict screeners are not better at selecting applicants based on either observed or unobserved covariates.

Next, we test for common support among human and ML preferences.  Intuitively, the IPW approach infers the quality of ML-selected candidates using actual hiring outcomes from candidates with similar covariates who were actually selected to be interviewed by a human recruiter.  We therefore require that candidates selected by the ML algorithm have some non-zero probability of also being selected by human recruiters.  

Appendix Figure \ref{common_support} plots the distribution of a candidate's estimated propensity to be selected by a human recruiter for the set of applicants chosen by our SL and UCB models.  In both cases, we find that all ML-selected applicants have a human selection propensity strictly between 0 and 1; we see no mass at or near zero.

\subsubsection{Marginally interviewed sample} \label{sec:ivres}

Our final approach asks the firm to modify its current human interview decisions by following ML recommendations when evaluating ``marginal'' candidates.  Section \ref{sec:iv} shows that we can assess the impact of this alternative interview policy $\tilde{I}$ by comparing the characteristics of instrument compliers with high and low ML scores.  Compliers can be thought of as ``marginal'' in that they are interviewed only because they were randomly assigned to a lenient recruiter.  

\begin{figure}[ht!]
\begin{center}
\captionsetup{justification=centering}
\caption{\textsc{Figure \ref{iv_marginal_hired}: Characteristics of marginal interviewees}}
\makebox[\linewidth]{
\begin{tabular}{cc}
\textsc{\footnotesize{A. Hiring Likelihood, SL}} & \textsc{\footnotesize{B.  Hiring Likelihood, UCB}} \\
\includegraphics[scale=0.5]{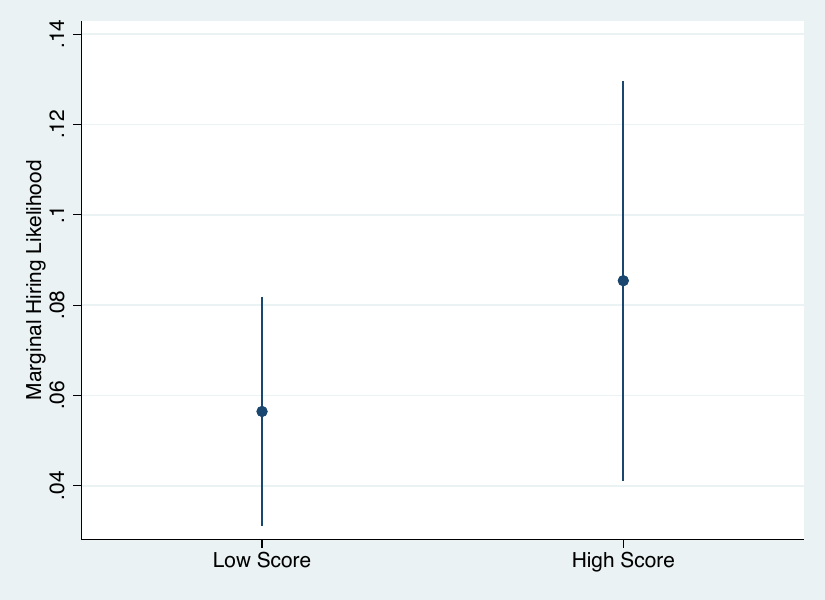} & \includegraphics[scale=0.5]{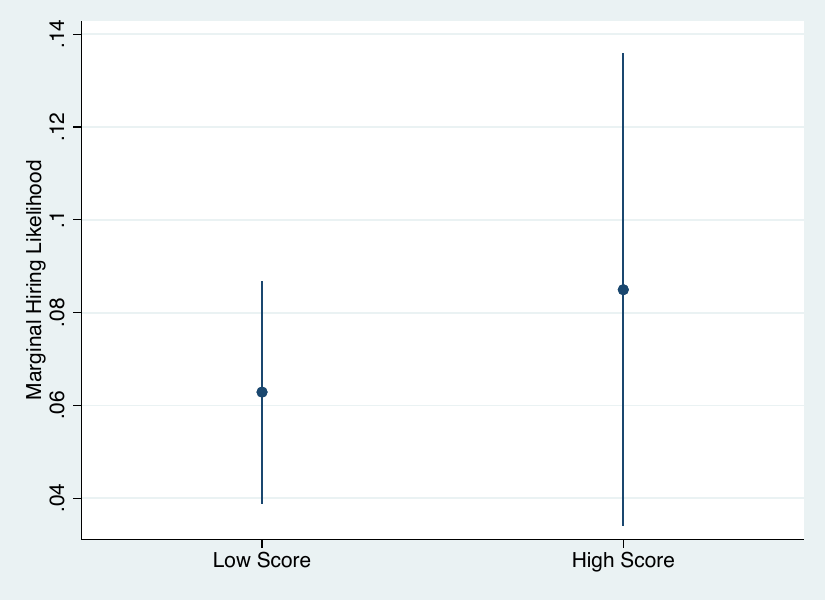}\\
\textsc{\footnotesize{C. Black/Hispanic, SL}} & \textsc{\footnotesize{D.  Black/Hispanic, UCB}} \\
\includegraphics[scale=0.5]{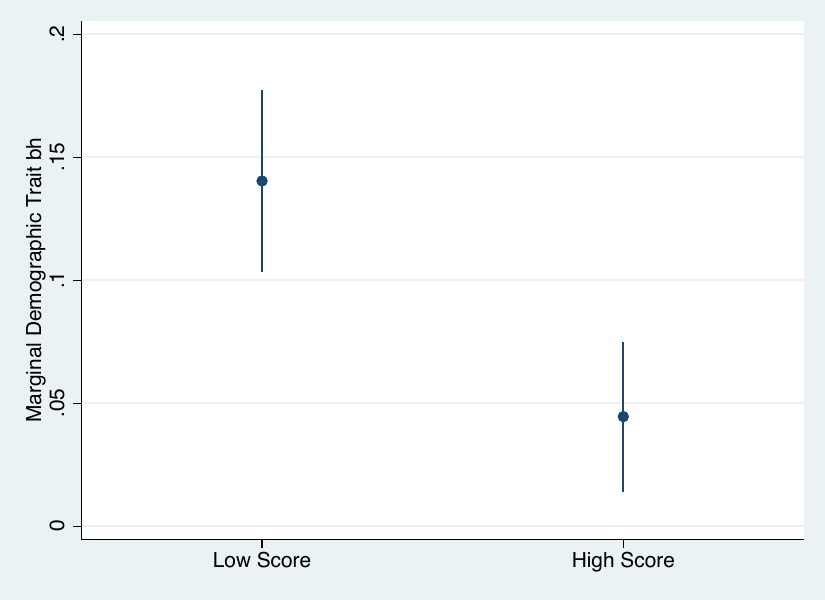} & \includegraphics[scale=0.5]{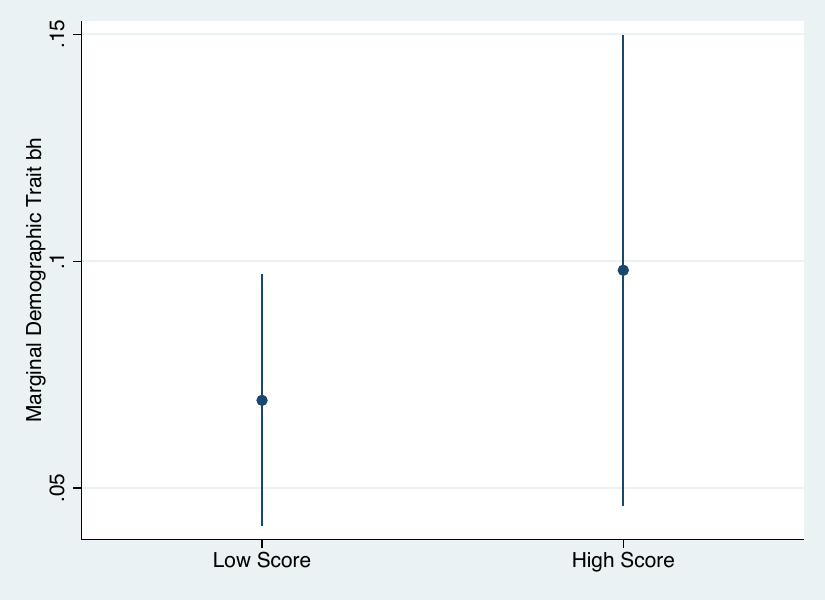}\\
\textsc{\footnotesize{E. Female, SL}} & \textsc{\footnotesize{F.  Female, UCB}} \\
\includegraphics[scale=0.5]{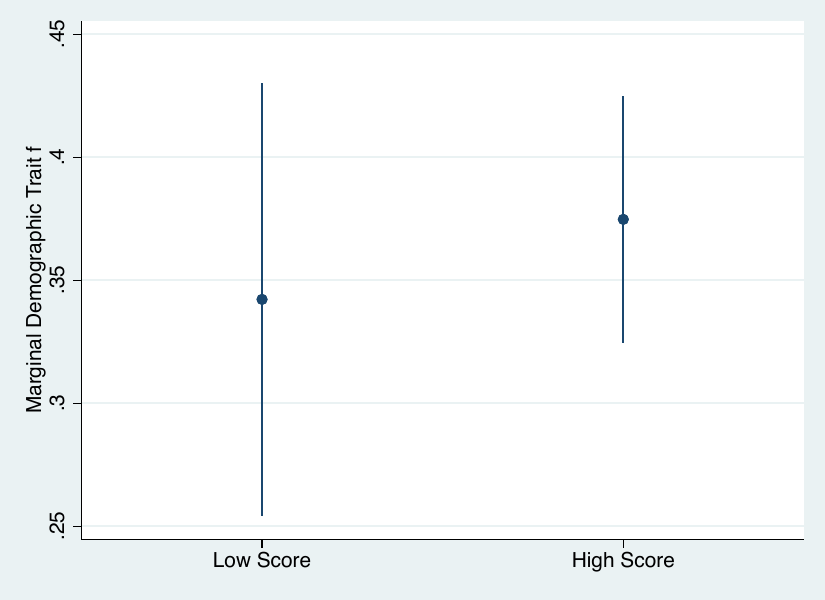} & \includegraphics[scale=0.5]{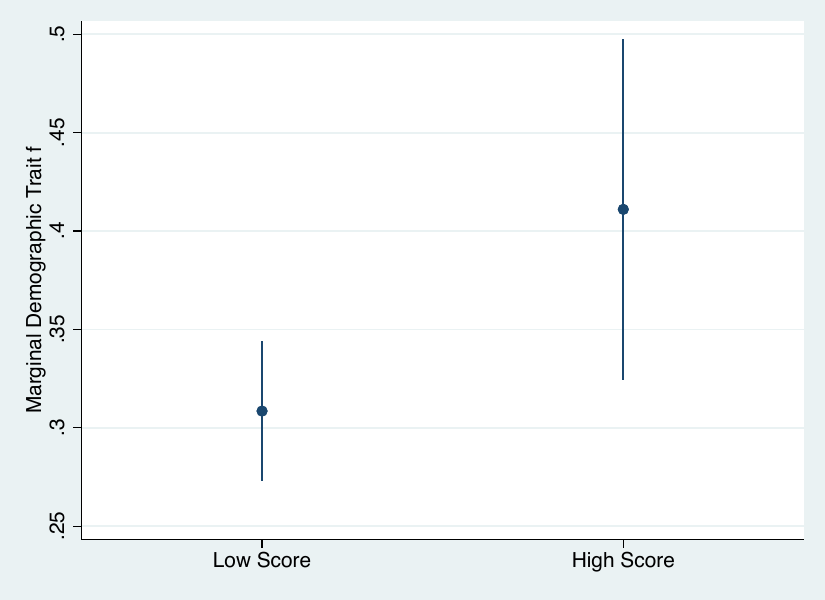}\\
\end{tabular}
}
\label{iv_marginal_hired}
\end{center}
\end{figure}

Figure \ref{iv_marginal_hired} presents our results.  Panels A, C, and E focus on applicants who are marginally selected based on SL model scores while Panels B, D, and F focus on marginal applicants as defined by UCB scores.  In Panels A and B, we see that compliers with high SL and UCB scores are both more likely to be hired than those with low scores.  This indicates that, on the margin, nudging human interview decisions toward either UCB or SL preferences would increase the expected hiring yield. 

In the remaining panels, we consider how following ML recommendations on the margin would change the demographics of selected candidates.  In Panel C, we see that marginally selected applicants with high SL scores are substantially less likely to identify as Black or Hispanic.  As such, nudging toward SL scores would tend to decrease the racial and ethnic diversity of selected applicants, relative to existing human decisions.  In contrast, Panel D shows the opposite for the UCB model. Here, we find that compliers with high UCB scores are more likely to be Black or Hispanic.  As such, the interview policy defined by $\tilde{I}$ would increase quality and diversity on the margin, relative to the firm's current practices.  In Panels E and F, we show that both the SL and UCB models would tend to increase the representation of women. 

These results are again consistent with our earlier results.  In both cases, following UCB recommendations can increase hiring yield and diversity relative to the firm's present policies, while following traditional SL recommendations increases quality but decreases racial and ethnic diversity.

\section{Alternative measures of quality} 
 
\subsection{Maximizing Offer Rates} \label{sec:offer}

In our main analyses, we focus on screening models that are designed to maximize hiring yield.  This is our preferred specification as it captures the key reason why firms turn to algorithms in the first place: the desire to fill vacancies with qualified workers.  

Hiring requires that a worker both receive and accept a job offer.  To isolate an algorithm's ability to identify applicants a firm would \textit{like} to hire, we build an alternative set of UCB and SL models that maximize the likelihood that an applicant is extended an offer, regardless of whether they accept.  These models are trained in the same way as our main models, except using receiving an offer as the outcome variable of interest.  Appendix Figure \ref{roc_offer} shows that we correctly predict offer outcomes in our baseline training data approximately 68 percent of the time. Appendix Sections \ref{asec:m_sldetails} and \ref{asec:m_trainingdetails} provide additional details on the training and out-of-sample accuracy of our offer model. 

\begin{figure}[ht!]
	\begin{center}
		\caption{\textsc{Figure \ref{bar_quality}:  Average Hiring Likelihood}}
		\vspace{20pt}
		\makebox[\linewidth]{\includegraphics[scale=.6]{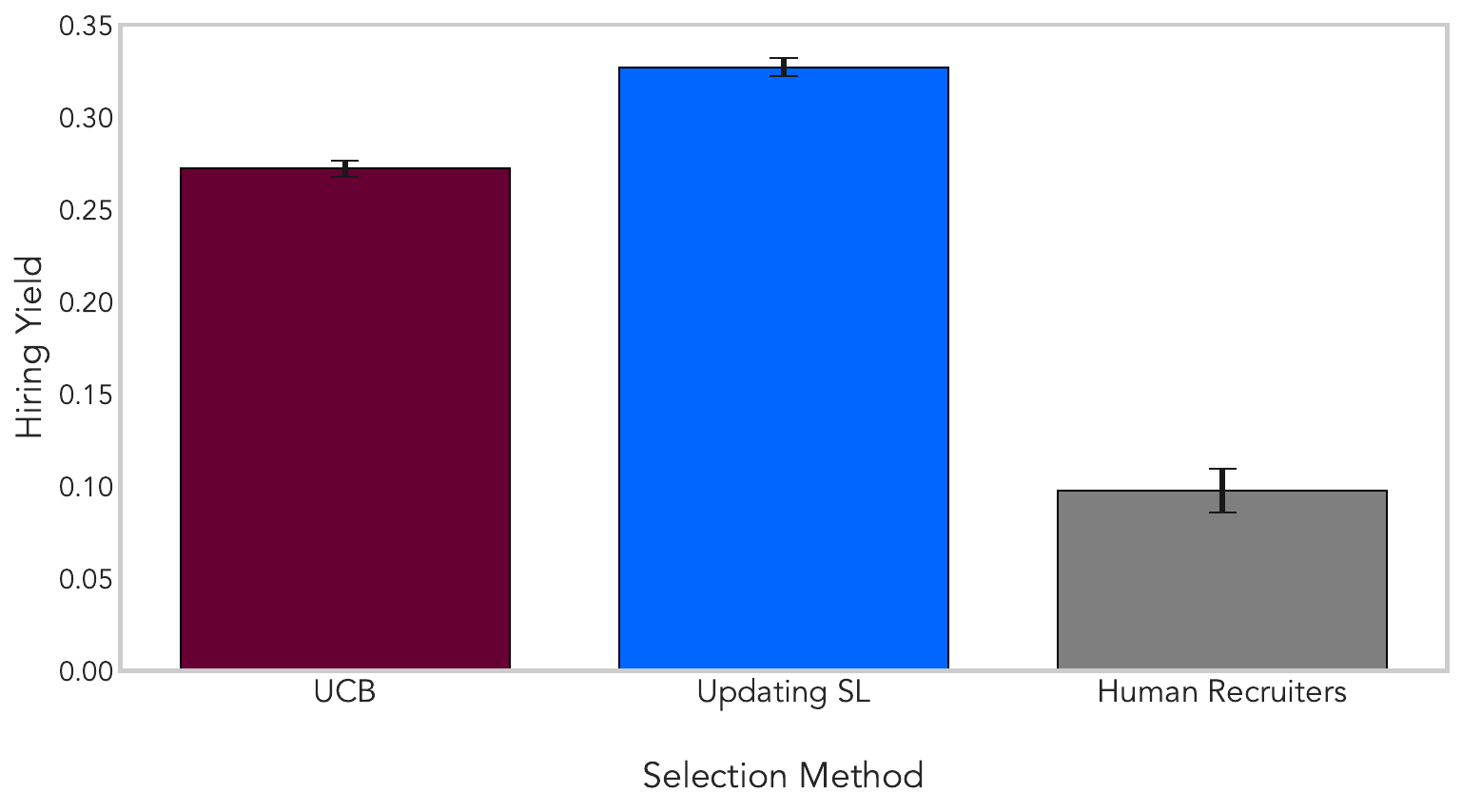}}\\
		\label{bar_quality}
	\end{center}
\end{figure}

\begin{figure}[ht!]
\begin{center}
\captionsetup{justification=centering}
\caption{\textsc{Figure \ref{offer_main}: Racial Composition---Offer Model}}
\makebox[\linewidth]{
\begin{tabular}{cc}
\textsc{\footnotesize{A. SL Offer Model}} & \textsc{\footnotesize{B. UCB Offer Model}} \\
\includegraphics[scale=0.5]{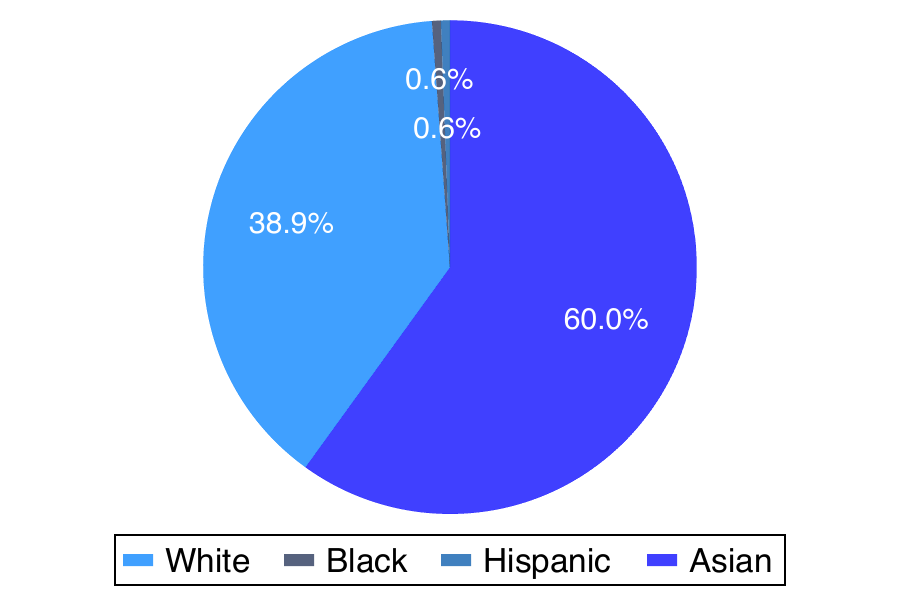} & \includegraphics[scale=0.5]{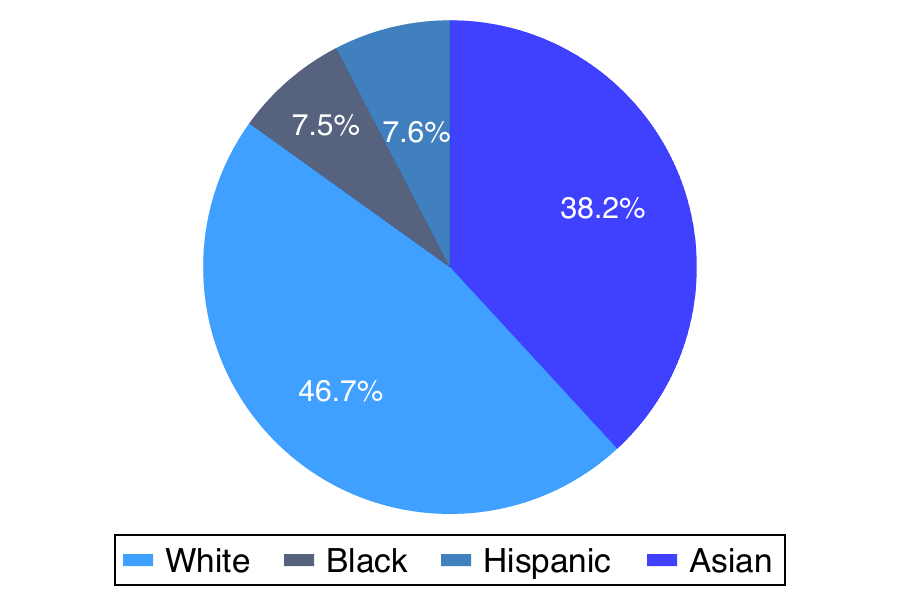}\\
\end{tabular}
}
\makebox[\linewidth]{
\begin{tabular}{c}
	\textsc{\footnotesize{C. Average Offer Rate}}  \\
	\makebox[\linewidth]{\includegraphics[scale=.6]{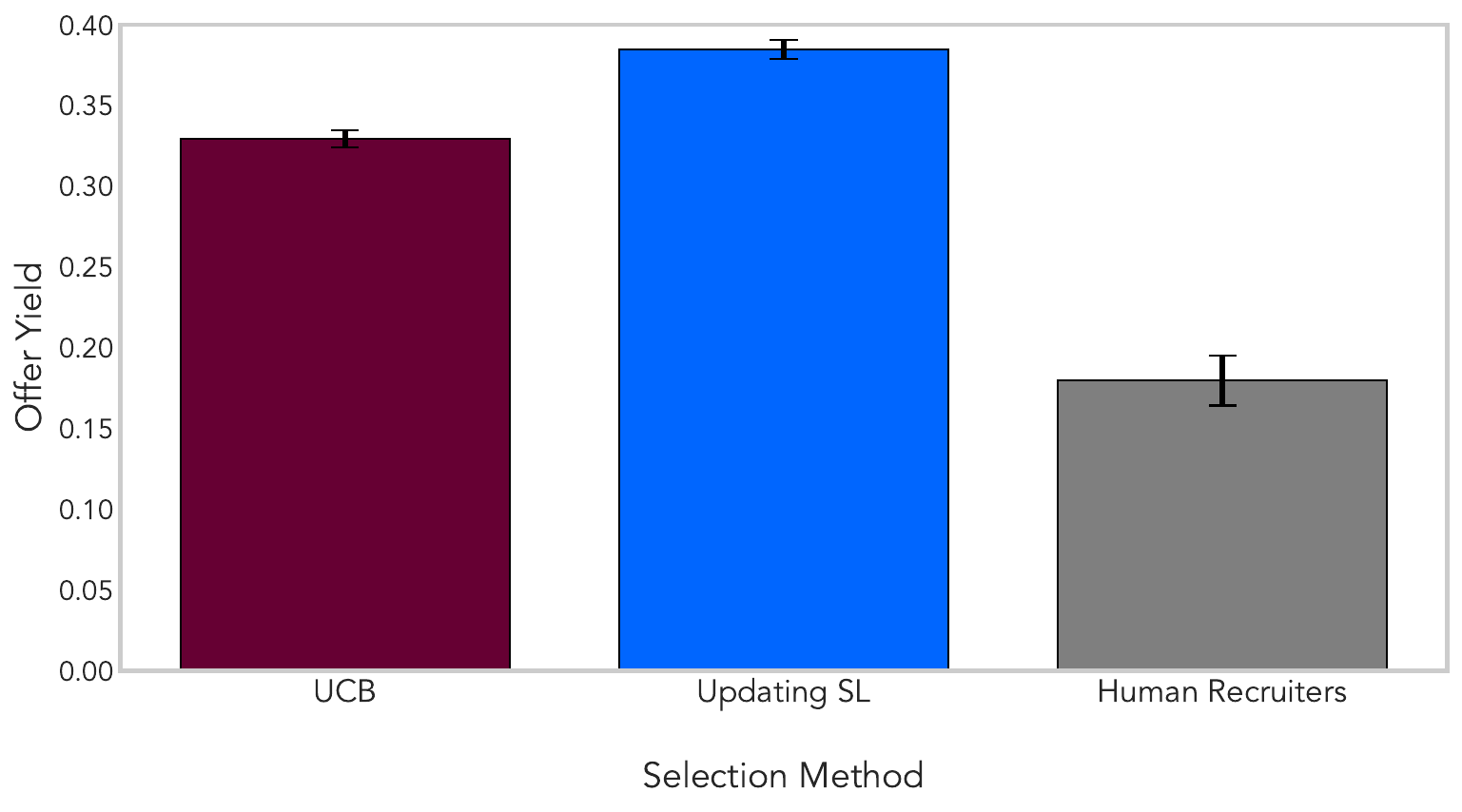}}\\
\end{tabular}
}
\label{offer_main}
\end{center}

\medskip
\begin{footnotesize} 
\begin{singlespace}
\justifying \noindent \textsc{Notes}: Panel A shows the race/ethnicity composition of an interview pool selected by a supervised learning algorithm described in Equation \eqref{SLrule} predicting offer potential. Panel B shows the composition of applicants who would be selected for an interview by the UCB algorithm described in Equation \eqref{UCBrule} predicting offer. By construction, all methods are constrained to match the number of applicants interviewed by human recruiters. Applicants' demographic information is collected by our firm during the application process. Panel C compares the quality (measured as the percentage of selected applicants who receive an offer using inverse propensity reweighting method described in Section \ref{sec:ipw}) for the two ML models, as well as the true offer yield from human interview decisions. All data come from the firm's application and hiring records. 
\end{singlespace}
\end{footnotesize}
\end{figure}

In Figure \ref{offer_main}, we show that offer-based SL and UCB models behave similarly to our hire-based models.  Panels A and B compare the demographics of applicants selected under SL and UCB models.  Similar to our main results in Figure \ref{pie_race}, we find that the SL model dramatically reduces the share of Black and Hispanic applicants who are selected for an interview (to less than 2\% from a human recruiter baseline of just under 10\%) while the UCB model increases this share to approximately 15\%).   In Panel C, we compare the average offer rate of UCB, SL and human selected applicants, using our inverse propensity weighting estimates discussed in Section \ref{sec:ipwres}.\footnote{Appendix Figure \ref{common_support_offer} shows that all offer model selected applicants have a human selection propensity strictly between 0 and 1 with no mass at or near zero.}  Consistent with Figure \ref{bar_quality}, we find that both UCB and SL models outperform human recruiters, with the SL model somewhat outperforming UCB over the 18 months of our analysis period.  

\begin{figure}[ht!]
\begin{center}
\captionsetup{justification=centering}
\caption{\textsc{Figure \ref{corr_quality}: Correlations between algorithm scores and hiring likelihood}}
\makebox[\linewidth]{
\begin{tabular}{c}
\textsc{\footnotesize{A. Human}}  \\
\includegraphics[scale=0.55]{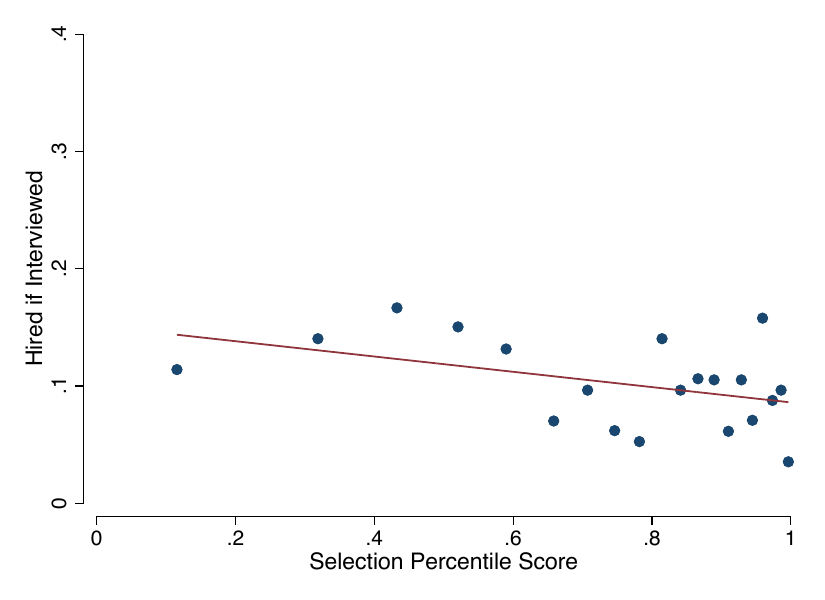} \\
\textsc{\footnotesize{B. SL Model}} \\
\includegraphics[scale=.55]{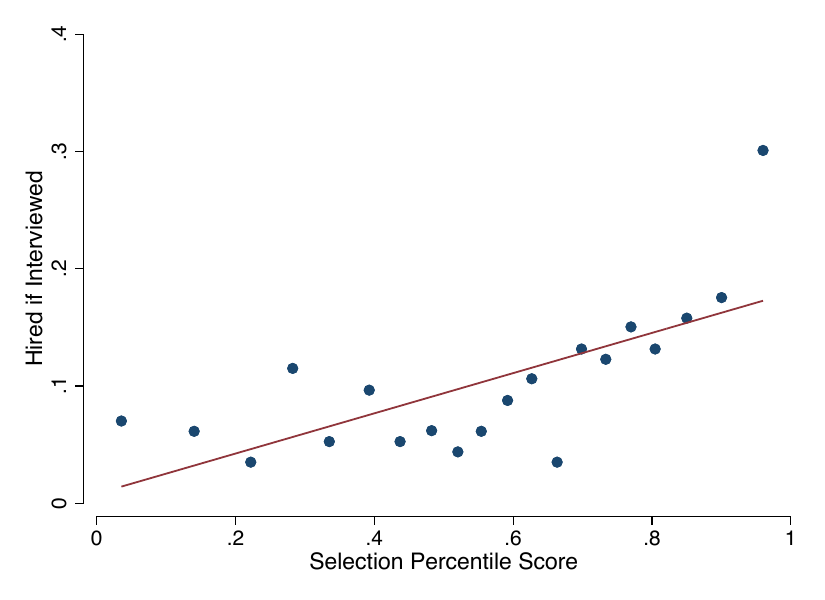} \\
\textsc{\footnotesize{C. UCB Model}} \\
\includegraphics[scale=.55]{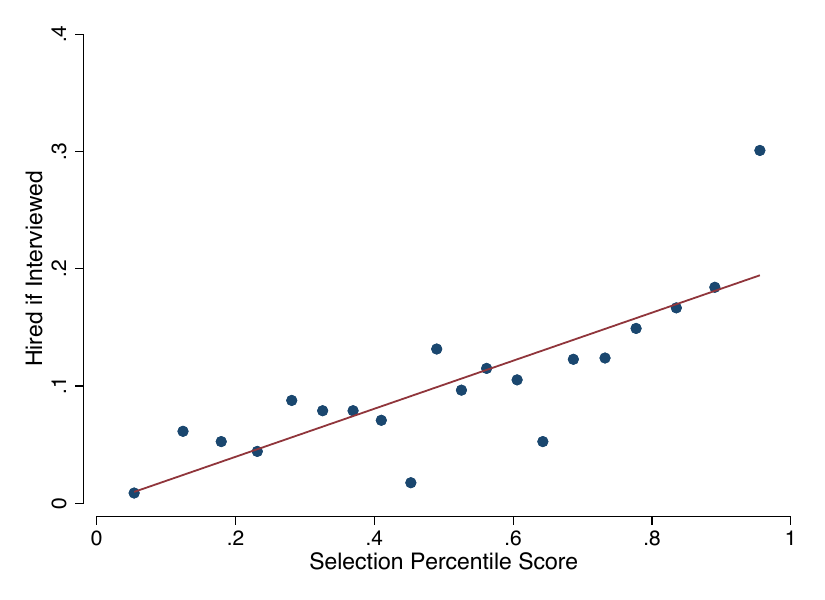} \\
\end{tabular}
}
\label{corr_quality}
\end{center}
\begin{singlespace}
\begin{footnotesize} 
\justifying \noindent \textsc{Notes}: Each panel of this figure plots algorithm selection scores on the $x$-axis and the likelihood of an applicant being hired if interviewed on the $y$-axis. Panel A shows the selection scores from an algorithm that predicts human recruiters interview selection choices. Panel B shows the selection scores from the supervised learning algorithm described by Equation \eqref{SLrule}. Panel C shows the selection scores from the UCB algorithm described in Equation \eqref{UCBrule}. All data come from the firm's application and hiring records. 
\end{footnotesize}
\end{singlespace}
\end{figure}

Appendix Figure \ref{corr_quality_offer} plots the correlation between UCB and SL scores and offer rates, among the set of applicants who are interviewed, analogous to the results presented for the hire model in Figure \ref{corr_quality}.  Similarly, Appendix Figure \ref{iv_marginal_offered} repeats our marginal sample IV analysis for the offered models.  Our results are very similar.  Appendix Figure \ref{corr_quality_offer} shows that candidates with higher UCB or SL scores are more likely to receive an offer, whereas applicants preferred by human recruiters tend to have, if anything, worse offer outcomes. Appendix Figure \ref{iv_marginal_offered} shows that firms can improve offer rates by following the recommendations of either ML model on the margin, but that SL recommendations decrease the share of under-represented minorities while UCB recommendations increase representation.  


\subsection{On-the-job Performance} \label{sec:onthejob}
 
A concern with our analysis is that both hiring and offer outcomes may not be the measure of quality that firms are seeking to maximize.  If firms ultimately care about on-the-job performance metrics, then they may prefer that it's recruiters pass up candidates who are likely to be hired in order to look for candidates that have a better chance of performing well, if hired.  

Our ability to assess this possibility is limited by a lack of data: of the nearly 49,000 applicants in our training data, only 296 have data on job performance ratings, making it difficult to accurately build such a model.  As a result, we take an alternative approach and correlate measures of on-the-job performance with ML and human preferences.  If humans were trading off hiring likelihood for job performance, then our human SL score, $s^H$, should be positively predictive of job performance relative to $s^{SL}$ and $s^{UCB}$.

Table \ref{tab_jobperformance} presents results using two measures: mid-year performance ratings and whether a worker was promoted.  Performance ratings are given on a scale of 1 (below), 2 (at), or 3 (above) average performance and 13\% of workers receive an above average rating.  Eight percent of hires in our sample are promoted during the analysis period. 
Panel A examines the correlation between an applicant's likelihood of being selected by a human recruiter and their likelihood of receiving a top performance rating (Column 1) and or a promotion (Column 2).  In both cases, we observe a negatively signed and sometimes statistically significant relationship: if anything, human recruiters are less likely to interview candidates who turn out to do well on-the-job.  In contrast, Panels B and C conduct the same exercise for each of our ML models.  For our SL hired model, these correlations are positively signed but statistically insignificant.  For the SL offered model, we see a positive and statistically significant correlation between scores and top performance ratings, and a zero correlation for promotions. We find a similar pattern for the UCB scores: we see a positive and sometimes statistically significant relationships between the UCB hired model score and on-the-job performance.   For the offered model, we again see a positive and statistically significant correlation between scores and top performance ratings, and no  correlation with promotions.

We caution that these data are potentially subject to strong sample selection due to the small proportion of workers for whom we have data.  That said, our results provide no evidence to support the hypothesis that human recruiters are successfully trading off hiring likelihood in order to improve expected on-the-job performance among the set of applicants they choose to interview.

\section{Alternative Policies}\label{sec:altpolicies}

So far, we have given our algorithms access to applicant's demographics and have made no restrictions on which applicants it can select.  Here, we consider two alternative approaches that treat demographic information differently. The first regulates algorithmic \textit{inputs}: it restricts the model's ability to access information on race, gender, and ethnicity.  The second approach regulates algorithmic \textit{outputs}: it maintains access to demographic information, but imposes a quota on which applicants the model can select. 

These policies correspond to principles of discrimination law.  In the US, the Equal Opportunity Employment Commission (EEOC) looks for ``disparate treatment'' (treating applicants differently on the basis of their demographics) or ``disparate outcomes'' (success rates that are substantially different by demographic group).  This is similar to the European Union's Equal Treatment Directive, which prohibits ``direct discrimination'' and ``indirect discrimination.''  Blinding algorithms are a way of preventing disparate treatment by regulating algorithmic inputs while quotas are a way of preventing disparate outcomes by regulating algorithmic outputs.

\subsection{Demographics Blinding} \label{sec:blinding}

Our main algorithms are trained on a variety of applicant characteristics, including explicit information on race, ethnicity, and gender.  As a result, these models can treat applicants differently on the basis of protected categories, a legal area \citep{kleinberg2018discrimination}.  It therefore natural to ask how our results would change if we eliminated the use of race, ethnicity, and gender as model inputs.\footnote{A number of recent papers have considered the impacts of anonymizing applicant information on employment outcomes \citep{goldin2000,aaslund2012,behaghel2015,agan2018,alston2019,doleac2020,craigie2020, murraygender}.}  Demographics-blind algorithms can also be useful in settings where firms do not have access to these data, either because applicants choose not to provide it or where collecting data on demographics is restricted.  

The impact of blinding is difficult to predict because demographic information enters the UCB model in two ways: as features of the model that are used to predict quality and as inputs in calculating exploration bonuses.  Eliminating this information can therefore shifts the model's predictive abilities as well as its exploration behavior.  
To examine what occurs in our setting, we re-estimate the UCB model without applicants' race, gender, and ethnicity in either prediction or bonus provision.  As a practical matter, we continue to allow the inclusion of other variables, such as geography, which may be correlated with race and ethnicity.  

\begin{figure}[ht!]
\begin{center}
\captionsetup{justification=centering}
\caption{\textsc{Figure \ref{pie_diversity_norg_RL}: Demographics Blinding}}

\makebox[\linewidth]{
\begin{tabular}{cc}
\textsc{\footnotesize{A. Race/Ethnicity, UCB}} & \textsc{\footnotesize{B. Race/Ethnicity, Blinded UCB}}  \\
\includegraphics[scale=0.45]{Output/May2024/pie_selAI_Rlogucb_byrace.pdf}  &  \includegraphics[scale=0.45]{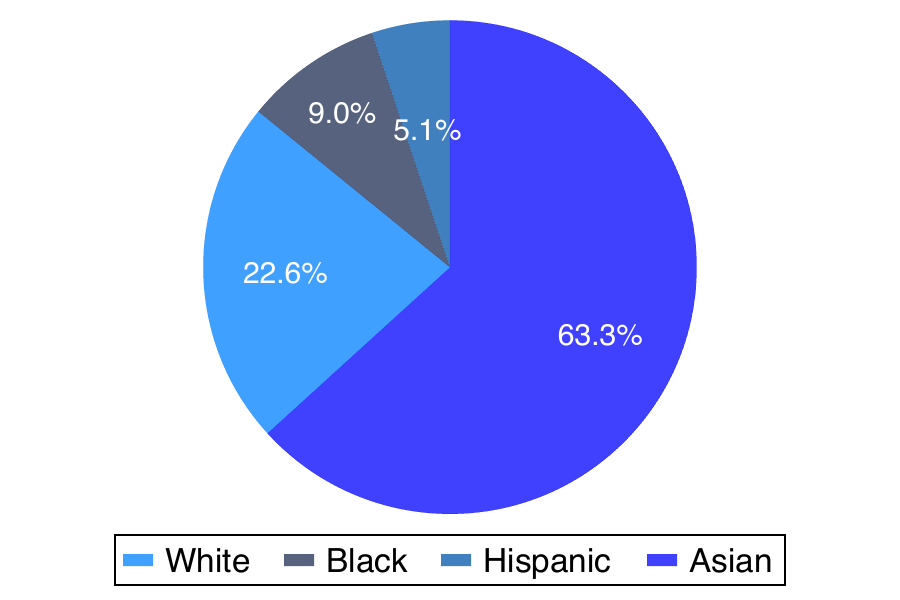}\\
\end{tabular}
}

\vspace{15pt}
\textsc{\footnotesize{C.  Average Hiring Likelihood}}  \\
\makebox[\linewidth]{\includegraphics[scale=.6]{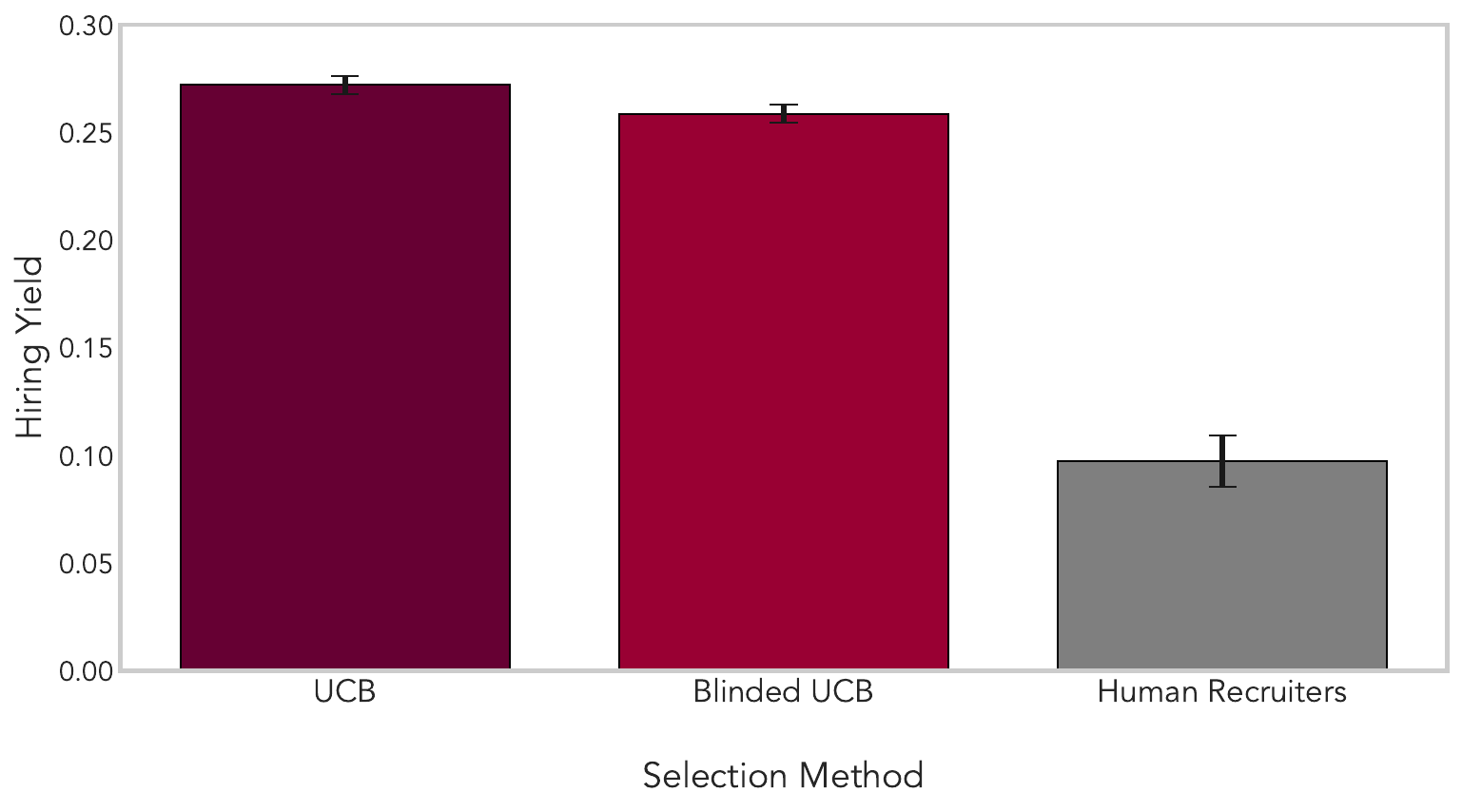}}\
\label{pie_diversity_norg_RL}

\end{center}
\end{figure}

Figure \ref{pie_diversity_norg_RL} shows how this blinding impacts diversity.  Panel A reproduces the composition of applicants selected by the unblinded UCB model and Panel B displays the blinded results.  Blinding reduces the share of selected applicants who are Black or Hispanic, from 24\% to 14\%, although there is still greater representation relative to human hiring (10\%).  The most stark differences come in the treatment of White and Asian applicants.  In the non-blinded model, White and Asian applicants make up a similar share of interviewed applicants (33\% and 43\%, respectively), even though there are substantially more Asian applicants in the overall pool.  When the algorithm is blinded, however, many more Asian applicants are selected relative to White applicants (63\% vs. 23\%, recalling that Asian and White applicants make up 57\% and 30\% of the applicant pool at large, respectively).  

Appendix Figure \ref{afig:blinding} provides additional analysis of how blinding impacts UCB model scores.  Panel A shows that blinding decreases exploration bonuses for Black and Hispanic applicants while increasing bonuses for Asian applicants.  
In the demographics-aware model, Asian applicants received smaller bonuses because they share a covariate---being Asian---that is very common in the sample.  When the algorithm is no longer able to observe this common trait, Asian applicants appear more distinctive because of their less common work and educational backgrounds. Panel B of Appendix Figure \ref{afig:blinding} plots the correlation between an important covariate---attending a highly-ranked school---and UCB beliefs about an applicant's quality.  Under the demographics-aware model, there is substantial heterogeneity in the ``returns'' to school rank across demographic groups, with Asian applicants being the least rewarded for having attended a top school.  Blinding, however, prevents the model from assigning race-specific returns; in our data, this increases the relative returns to elite education among Asian applicants.  Taken together, these findings provide intuition for the large increase in Asian representation under blinding.

Panel C of Figure \ref{pie_diversity_norg_RL} examines the accuracy of blinded vs. unblinded UCB, using the reweighting approach described in Section \ref{sec:ipw}.  We find that blinding leads to a small, modest decline in in the quality of algorithmically selected candidates; both models continue to substantially outperform human evaluators.  In our setting, the small difference in outcomes between the blinded and unblinded UCB models likely combines two distinct impacts.  First, blinding reduces the predictive ability of our models.  At the same time, Asian applicants tend to have relatively higher hire rates in our data so that, in our case, blinding shifts exploration toward a higher yield group.

%

\subsection{Supervised Learning with Quota} \label{sec:quota}

An alternative approach to achieving greater representation is to introduce diversity as an explicit constraint.  In this section, we consider a policy in which applicants are scored by a supervised learning model with access to demographic information, but where the composition of selected applicants must reflect that of the applicant pool.

\begin{figure}[ht!]
	\begin{center}
		\captionsetup{justification=centering}
		\caption{\textsc{Figure \ref{pie_diversity_SLq}: Supervised Learning with Quota}}
		
		\makebox[\linewidth]{
			\begin{tabular}{cc}
\textsc{\footnotesize{A. Race/Ethnicity, UCB}} & \textsc{\footnotesize{B. Race/Ethnicity, SL with Quota}}  \\
\includegraphics[scale=0.45]{Output/May2024/pie_selAI_Rlogucb_byrace.pdf}  & \includegraphics[scale=0.45]{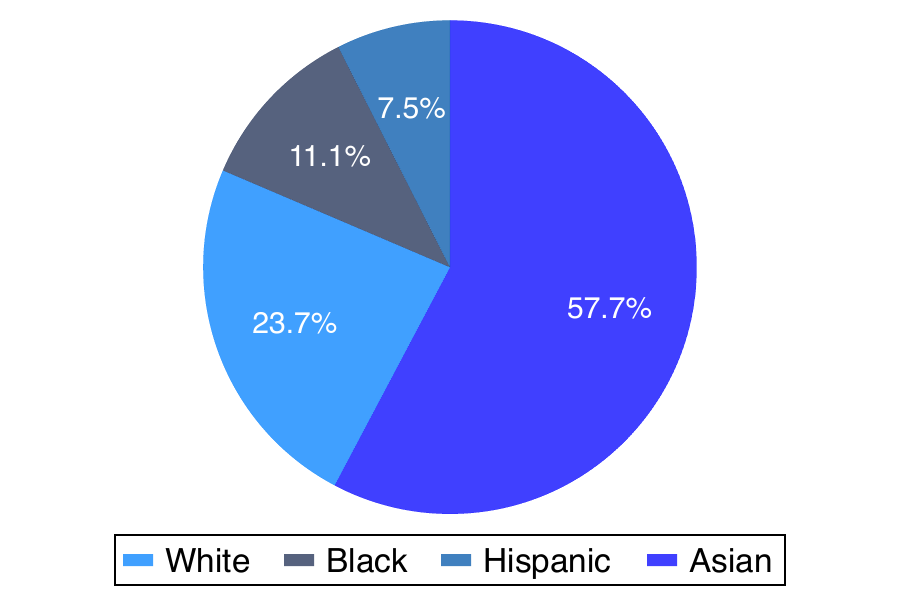}  
			\end{tabular}
		}
		
		\textsc{\footnotesize{C. Average Hiring Likelihood}}  \\
		\makebox[\linewidth]{\includegraphics[scale=.5]{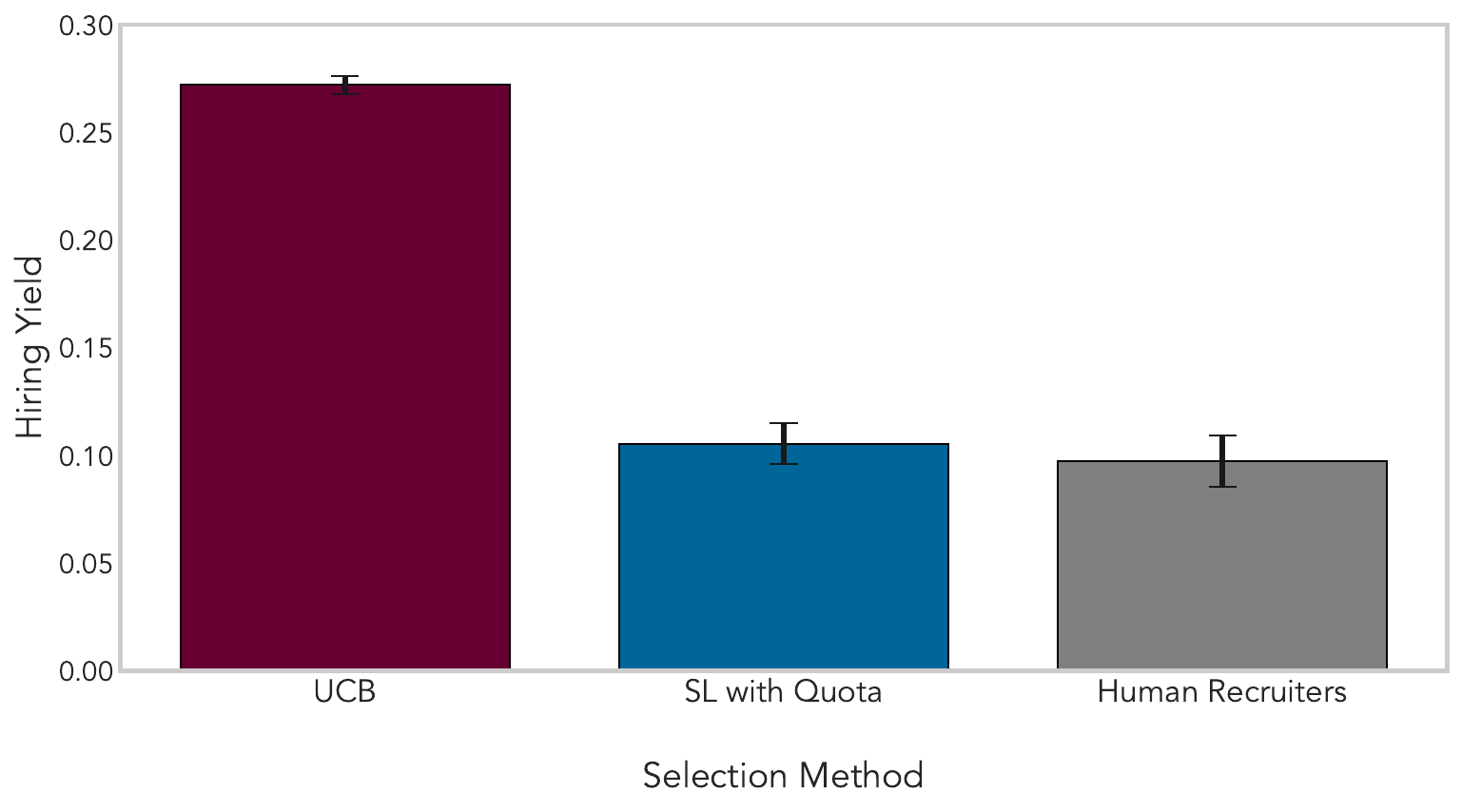}}\\
		\label{pie_diversity_SLq}
		
	\end{center}
\end{figure}

Panels A and B of Figure \ref{pie_diversity_SLq} compare the demographics of candidates selected under our baseline UCB model to those selected with our SL with quota policy.  By construction, the composition of applicants selected under our quota model is similar to that of the overall applicant pool (Panel A of Figure \ref{pie_race}).  We note that our percentages are not exact because we are working with small discrete numbers so it is not always possible for the share of selected applicants to equal the population share.
In Panel C, we show that the quality of workers hired under the quota model is \textit{substantially} worse: about 10 percent of selected applicants are predicted to be hired, compared with close to 30 percent for the unconstrained SL and UCB models.  

We believe that this is due to the fact that a quota model substantially constrains the firm in terms of \textit{when} it must select minority candidates.  In settings where applicants are selected as part of a defined cohort (say, college admissions), it is straightforward to define the applicant pool over which the quota must be enforced.  In most hiring settings, however, applicants are selected on a rolling basis, making it conceptually challenging to specify how many members of each group to select over a given period.  Any ex-ante constraint could reduce quality by selecting too many minority candidates when their quality is low, and too few when their quality is high. 

Appendix Figure \ref{afig_quotatiming} shows that this is a real constraint in our setting. 
Specifically, we compare the average number of Black or Hispanic applicants who are selected by our UCB model (dotted blue line) and the SL with quota over our analysis period (solid red line).  The UCB model selects more Black or Hispanic applicants on average but varies significantly in the number it selects each period.  By contrast, the quota model is restricted to selecting, on average, one such applicant each period---no more, no less. These results come from using a window of 100 applicants over which to define our quota, but the nature of this challenge is general.

\section{Additional Results: Time Dynamics and Learning} \label{sec:learning}

Our main results show that our UCB algorithm increases the hiring yield of selected applicants, while also increasing demographic diversity.  A key question relates to how these patterns evolve over time: are gains in diversity transient, and does exploration generate greater losses in efficiency in the short run?  In this section, we explore how the hiring yield and demographics associated with selected candidates evolve over time.  

\subsection{Time dynamics in analysis data} \label{sec:dynamics}

\begin{figure}[ht!]
\begin{center}
\captionsetup{justification=centering}
\caption{\textsc{Figure \ref{quality_over_time}: Hiring Yield Over Time}}
\makebox[\linewidth]{
			\begin{tabular}{c}
				\textsc{\footnotesize{A. Hiring Yield (IPW estimates)}}  \\
				\includegraphics[scale=0.5]{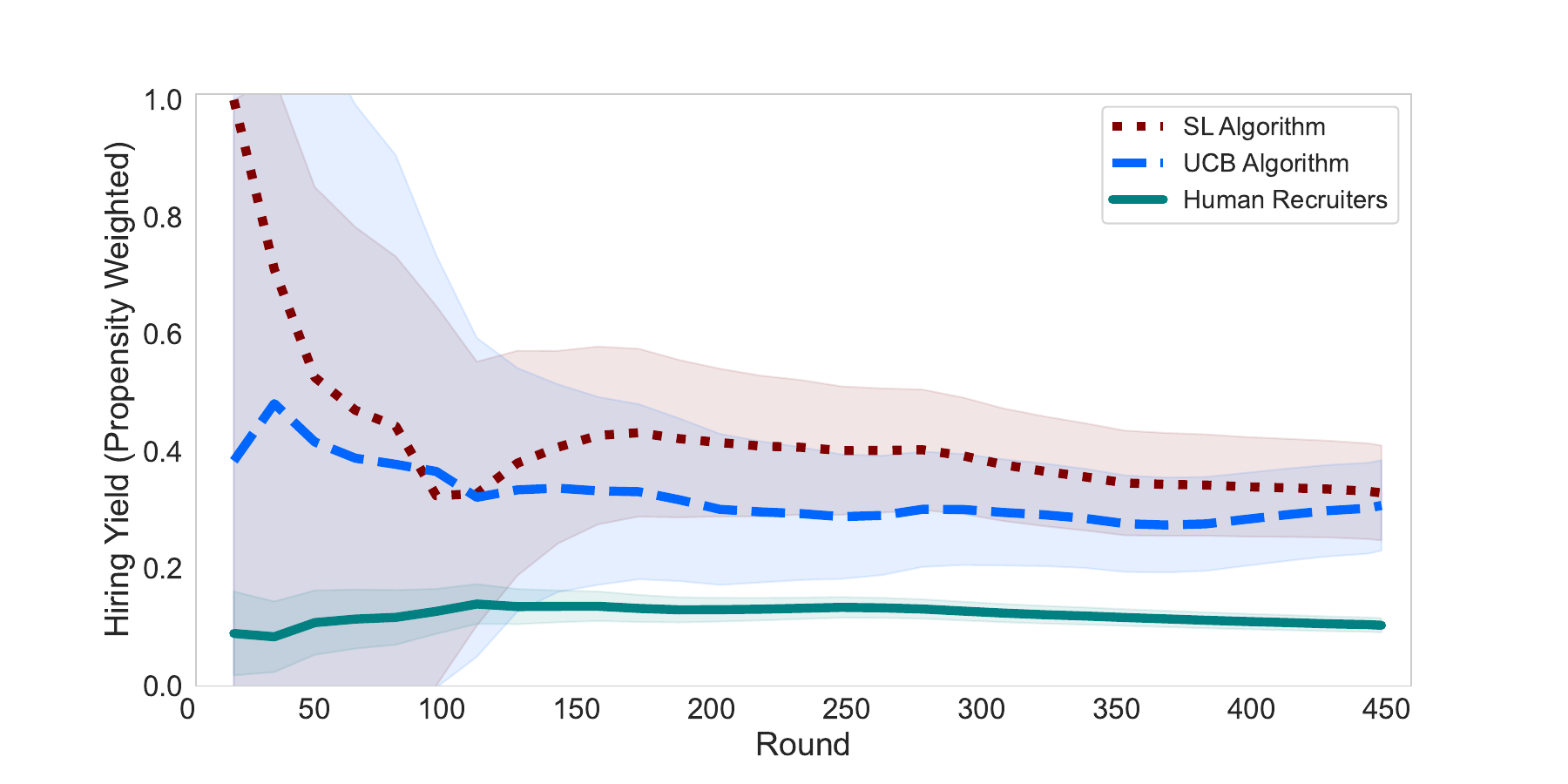} \\ 
    	\textsc{\footnotesize{B. Race/Ethnicity}} \\
				\includegraphics[scale=0.45]{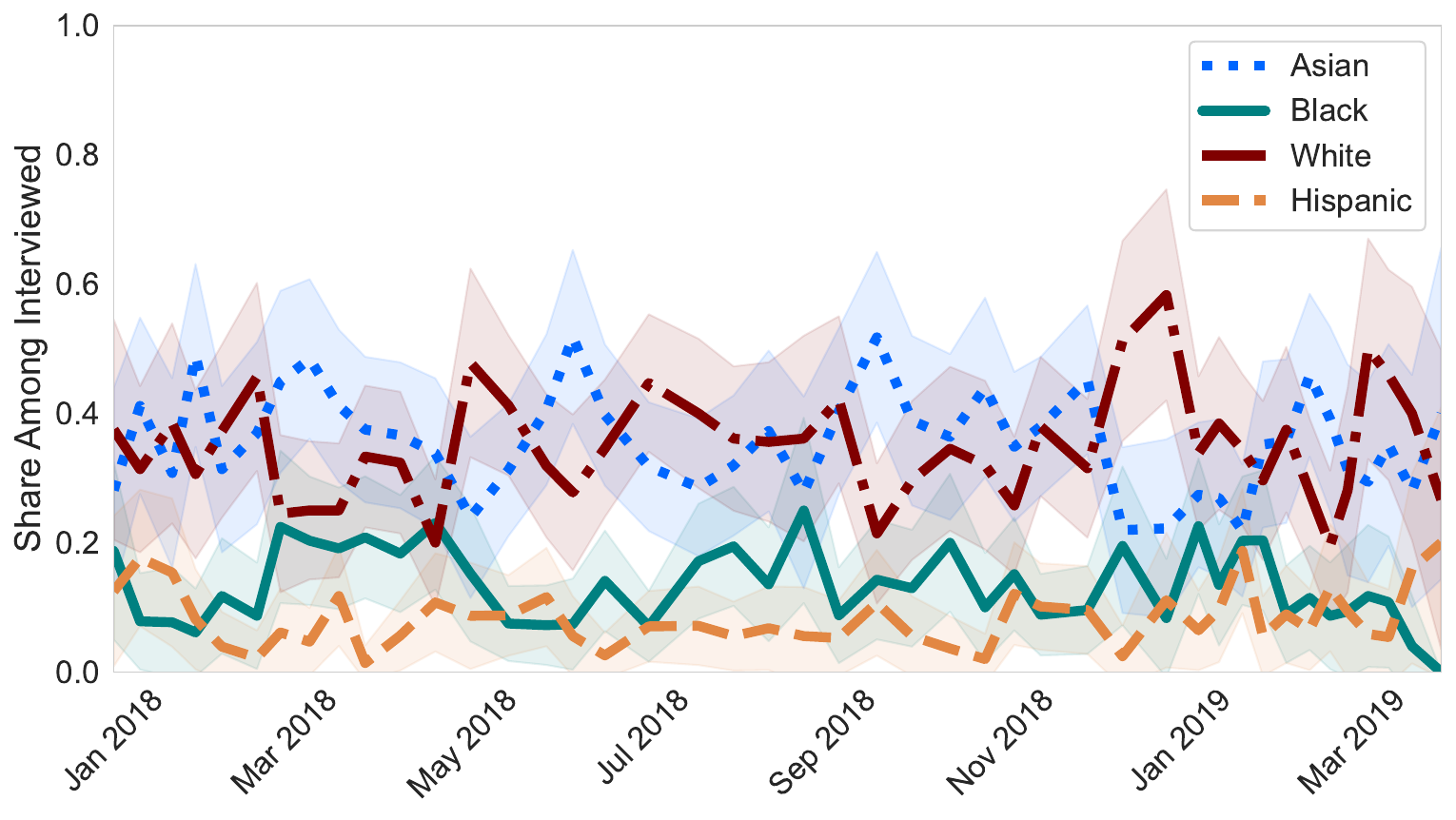} \\
			\end{tabular}
}\\
\label{quality_over_time}
\end{center}
\begin{footnotesize}
\begin{singlespace}
\justifying \noindent \textsc{Notes}: Panel A shows the average hiring yield of each interview screening method calculated cumulatively over time. The ``SL Algorithm'' line plots $E[H|I^{SL}=1]$, the ``UCB Algorithm'' line plots $E[H|I^{UCB}=1]$ and the ``Human Recruiters'' line plots $E[H|I^{H}=1]$. Each estimate of quality is calculated using the inverse propensity weighting described in section \ref{sec:ipw}.  Panel B shows the composition of applicants selected to be interviewed by the UCB model at each point during the analysis period.  We plot a rolling average across a ten-round window and the 95\% confidence intervals around each mean. All data come from the firm's application and hiring records.   
\end{singlespace}
\end{footnotesize}

\end{figure}

Figure \ref{quality_over_time}, shows how the quality (Panel A) and race/ethnicity (Panel B) of selected applicants evolve over our analysis sample.  In Panel A, we compute the expected quality of ML-selected applicants using the inverse propensity weighting approach discussed in Section \ref{sec:ipw}.  Our estimates are in general somewhat noisy and we are unable to observe any statistically significant differences in estimated hiring yield between applicants selected by the SL and UCB models, though both ML models select applicants who are more likely to be hired than those selected by human recruiters.  However, taking the point estimates seriously, the quality of the SL model appears to decline over time, while the quality of UCB choices is more stable.  At the end of our sample, the quality of applicants selected by both models is essentially identical.  

In Panel B, we find no discernible downward trend in the proportion of Black and Hispanic candidates selected by our UCB model over time. This suggests that, in our sample, hiring outcomes for minority applicants are high enough that our models do not update downward upon selecting them.  As discussed in Section \ref{sec:feasible}, one may be concerned that the stability of our demographic results represents a failure to learn due to biases arising from sample selection.  Our tests for selection on unobservables, described in Section \ref{sec:ipwres} suggests that this possibility is not driving our results here.

\subsection{Learning in simulated applicant data} \label{sec:simulations}

We do not see evidence of either the UCB or SL model learning during our analysis sample. This may be because our analysis above is unfortunately limited by sample size and timing: our analysis period spans just under 1.5 years, and we only observe hiring outcomes among candidates interviewed during this period.  Combined, this gives us limited opportunities to observe how our models may evolve over longer periods, or respond to more substantial changes in applicant quality.

To further explore how our UCB and SL models behave, we conduct simulations in which we change the quality of applicants who enter our analysis sample, starting in 2018.  We provide details of how we implement this in Appendix \ref{asec:simulations} but, essentially, we imagine that the quality of one demographic group begins to increase during the analysis period so that, by the end of the period, all applicants from that group having a hiring yield of 1.  In the meantime, we hold the quality of applicants from all other groups constant at their true 2018 mean. Given this stark set up, an efficient model is one that can detect this change in applicant quality and begin interviewing only applicants from that group.  To evaluate this, we consider how each of our ML scoring approaches would evaluate the \textit{same} cohort of candidates at different points in time.  Specifically, we take the actual set of candidates who applied between January 2019 and April 2019 (hereafter, the ``evaluation cohort''), and estimate their ML model scores at different points in 2018.  This allows us to isolate changes in the algorithm's scores that arise from differences in learning and exploration over time, rather than from differences in the applicant pool.  

For intuition, consider the scores of candidates on January 1, 2018, the first day of the analysis period.  In this case, both the SL and UCB algorithms would have the same beliefs about the hiring potential of candidates in the evaluation cohort, because they share the same estimate of $E[Y_{it}|X'_{it}; D_0]$ trained on the initial data $D_0$. The UCB model, however, may have a different score, because it also factors in its exploration bonus.  On December 31, 2018, however, the SL and UCB algorithms would have both different beliefs (based on their potentially different history of selected applicants) and different scores (because the UCB factors in its exploration bonus in addition to expectations of quality).  To better understand how the UCB model differs from the SL, we also consider a third variant, which tracks who the UCB model would have selected based on its estimates of $E[Y_{it}|X'_{it}; D^{UCB}_t]$ alone; this model allows us to track the evolution of the UCB model's beliefs separately from its exploration behavior.  

\begin{figure}[ht!]
	\begin{center}
		\captionsetup{justification=centering}
		\caption{\textsc{Figure \ref{learning_black}: Simulations Increasing the Quality of Black Applicants}}
		\makebox[\linewidth]{
			\begin{tabular}{cc}
				\textsc{\footnotesize{A. Share of Selected Applicants who are Black}} \\
				\includegraphics[scale=0.36]{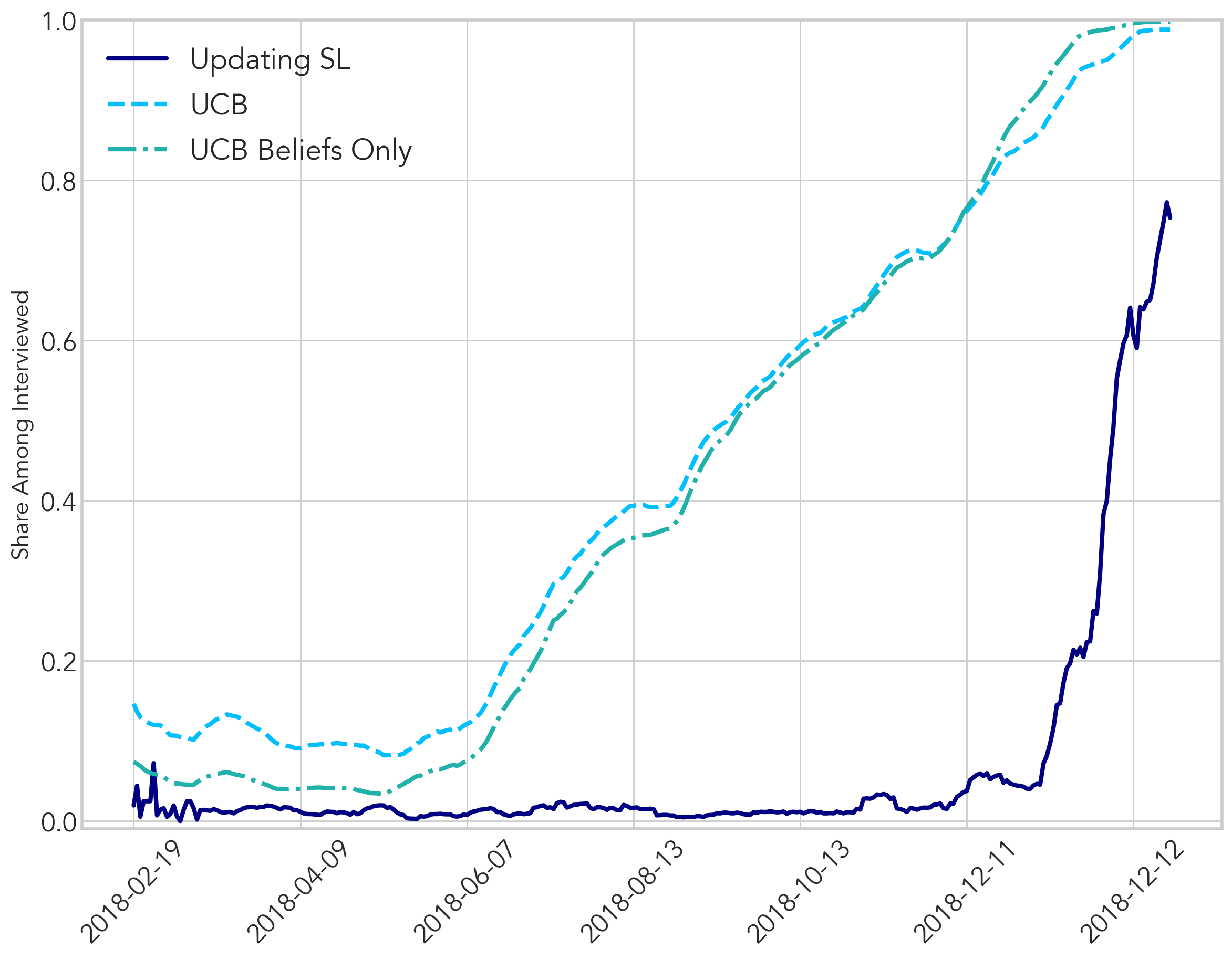}  \\
				\textsc{\footnotesize{B. Average Hiring Yield}}  \\
				\includegraphics[scale=0.36]{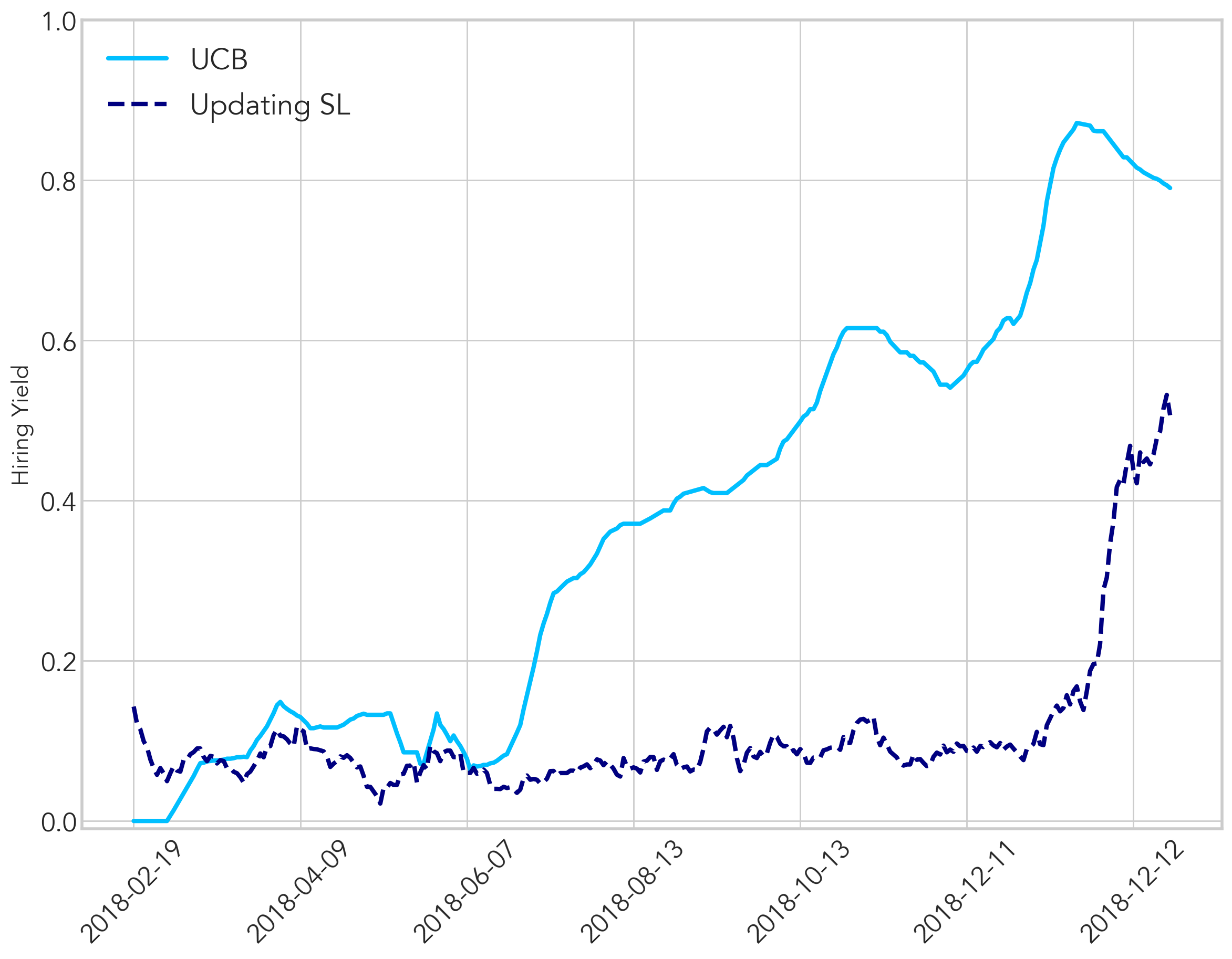}  \\
			\end{tabular}
		}
		\label{learning_black}
	\end{center}
\end{figure}

Figure \ref{learning_black} displays the results of this exercise for the simulation in which we increase the quality of Black applicants.  In Appendix \ref{asec:simulations}, we discuss the results of simulations in which we increase the quality of other demographic groups, as well as the analogous simulations in which we decrease the quality of applicants by demographic group. Panel A focuses on the share of Black applicants who are selected.  We report the results of three different selection criteria.  The blue dashed line reports the selection decisions of the UCB model.  The UCB model rapidly increases the share of Black candidates it selects.  To better understand why this happens, we plot a green dash-dot-dot line, which tracks the UCB model's \textit{beliefs}: that is, the share of Black applicants it would select if its decisions were driven by the $\hat{E}[Y_{it}|X'_{it}; D^{UCB}_{t}]$ component of Equation \eqref{UCBrule} only, leaving out the exploration bonus.  Initially, the blue dashed line is above the green dash-dot-dot line; this means that the UCB model begins by selecting more Black applicants not because it necessarily believes that they have strong hiring potential, but because it is looking to explore.  Over time, the green dash-dot-dot line increases as the models see more successful Black candidates and positively updates it's beliefs.  Eventually, the two lines cross: at this point, the UCB model has strong positive beliefs about the hiring potential of Black applicants, but it holds back from selecting more Black candidates because it would still like to explore the quality of other candidates.  By the end of the simulation period, however, exploration bonuses have declined enough so that the UCB model's decisions are driven by its beliefs, and it selects almost exclusively Black candidates.   

\begin{figure}[ht!]
\begin{center}
\captionsetup{justification=centering}
\caption{\textsc{Figure \ref{pie_race}: Racial Composition}}
\makebox[\linewidth]{
\begin{tabular}{cc}
\textsc{\footnotesize{A. Applicant Pool}} & \textsc{\footnotesize{B.  Actual Interview}} \\
  \includegraphics[scale=0.5]{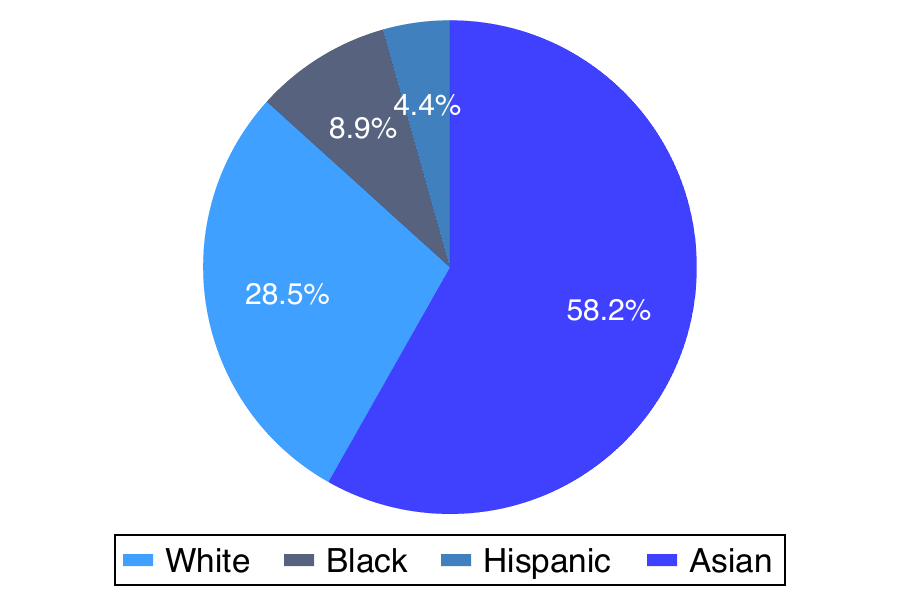} & \includegraphics[scale=0.5]{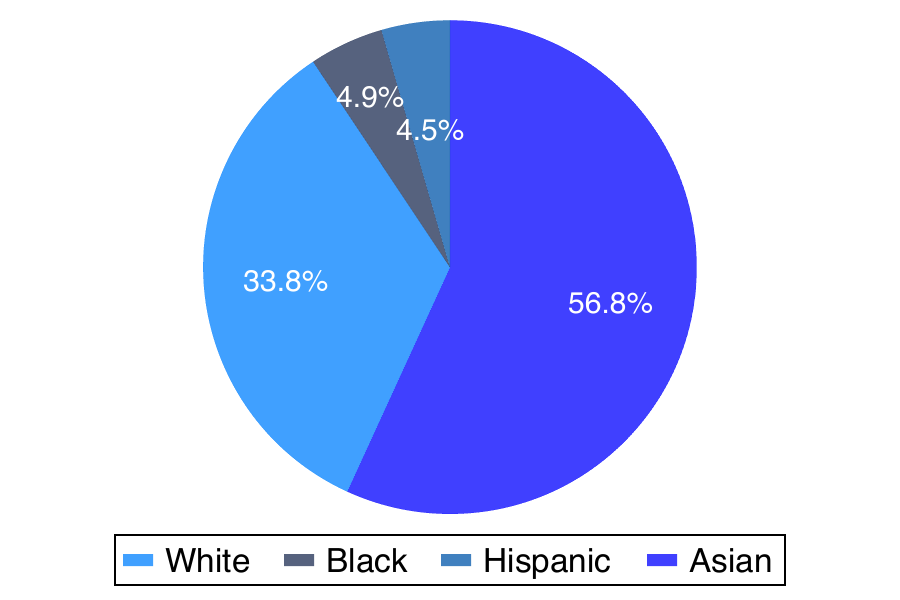}\\
\textsc{\footnotesize{C. SL Model}} & \textsc{\footnotesize{D. UCB Model}} \\
\includegraphics[scale=0.5]{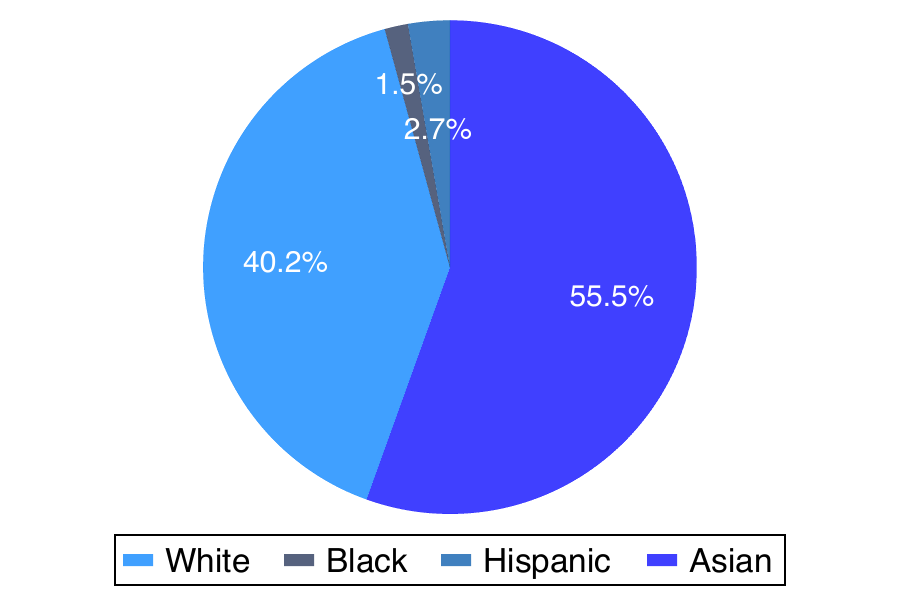} & \includegraphics[scale=0.5]{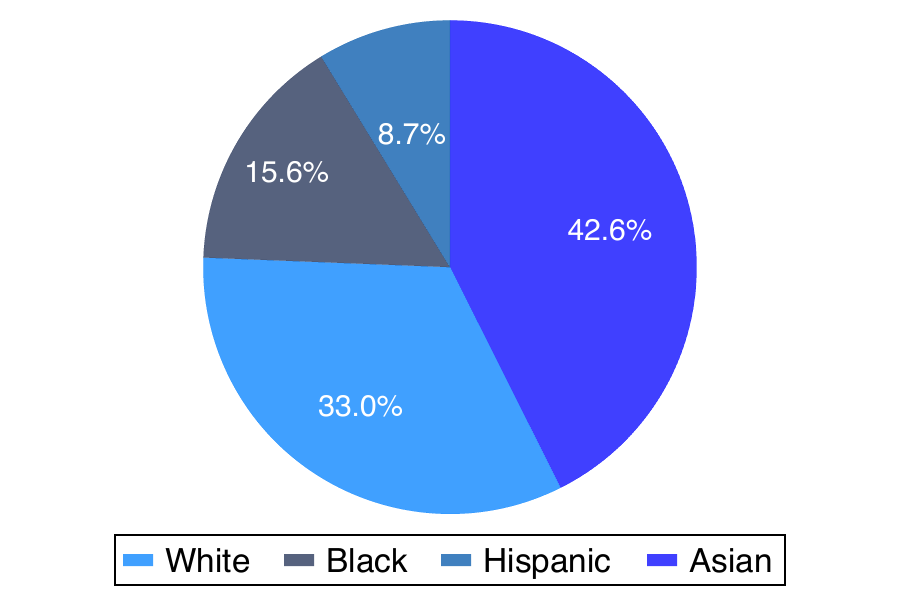}\\
\end{tabular}
}
\label{pie_race}
\end{center}

\begin{footnotesize} 
\begin{singlespace}
\justifying \noindent \textsc{Notes}: Panel A shows the race/ethnicity composition of applicants in our data.  Panel B shows the composition of applicants actually selected for an interview by the firm.  Panel C shows the racial composition of applicants who would be selected if chosen by the supervised learning algorithm described in Equation \eqref{SLrule} predicting hiring potential.  Finally, Panel D shows the composition of applicants who would be selected for an interview by the UCB algorithm described in Equation \eqref{UCBrule} predicting hiring potential.  By construction, all methods are constrained to match the number of applicants interviewed by human recruiters. Applicants' demographic information is collected by our firm during the application process. All data come from the firm's application and hiring records. 
\end{singlespace}
\end{footnotesize}
\end{figure}

The solid blue line shows this same process using the SL model.  While it is eventually able to learn about the simulated increase in the hiring prospects of Black applicants, it does so at a significantly slower rate relative to UCB.  Because supervised learning algorithms focus on maximizing current predicted hiring rates, the SL model does not go out of its way to select Black candidates.  As such, it has a harder time learning that these candidates are now higher quality.  This is unsurprising considering Figure \ref{pie_race}, which shows that SL models are very unlikely to select Black applicants.  

Panel B of Figure \ref{learning_black} plots the analogous change in the quality of selected applicants over time.  While the SL model eventually catches up in terms of quality, we see that the UCB model outperforms earlier because it is able to more quickly identify the group with improved quality.  In Figure \ref{learning_good_quality} of Appendix \ref{asec:simulations}, we show that the gap in performance between the UCB and SL models is highest in simulations where the group whose quality is improving is less likely to be selected at baseline.  This is because the UCB model proactively looks for applicants with rare covariates. 

In Appendix \ref{asec:simulations} we also discuss simulations in which applicant quality decreases. We find that the UCB model will drastically reduce the number of minority applicants it selects once it begins to learn that their quality has fallen.  This differs from a quota-based system that sets minimum levels of representation.

\section{Conclusion}  \label{conclusion}

This paper advances our understanding of how algorithmic design affects access to job opportunities. While previous work has highlighted potential gains from following algorithmic recommendations, we highlight how algorithm design can shape the impact of these decision tools.  We show that exploration-based algorithms can help firms more effectively identify candidates that meet their hiring criteria while simultaneously increasing the representation of minority applicants.  This occurs even though our algorithm is not explicitly charged with increasing diversity, and even when it is blinded to demographic inputs.



Our findings shed light on the relationship between efficiency and equity in hiring.  In our data, supervised learning algorithms increase hiring yield but decrease diversity, relative to the firm's current practices.  A natural interpretation of this finding is that algorithms and human recruiters make different tradeoffs at the Pareto frontier, with humans prioritizing equity over efficiency.  Our UCB results, however, show that such explanations may be misleading. By demonstrating that an algorithmic approach can improve hiring outcomes while expanding representation, we provide evidence that human recruiters operate inside the Pareto frontier: in seeking diversity, they select weaker candidates over stronger ones from the same demographic groups. This leaves room to design more data-driven approaches that better identify strong candidates from under-represented backgrounds.
 

Finally, our findings raise important directions for future research. 
As firms increasingly adopt algorithmic screening tools, it becomes crucial to understand the organizational and general equilibrium effects of such changes in HR practice.  For example, there is considerable debate about the impact of diversity on team performance and how changes in the types of employees may impact organizational dynamics.\footnote{For instance, see \citet{reagans2001} for a discussion of the role of diversity, and, for instance, \citet{atheymentoring} and \citet{fernandez} for a discussion of how changes in firm composition can shift  mentoring, promotion, and future hiring patterns.}  Such changes may also impact the validity of the predictive relation between applicant covariates and outcomes.  In addition, when adopted by a single firm, an exploration-focused algorithm may identify strong candidates who are overlooked by other firms using more traditional screening techniques; yet if all firms adopt similar exploration-based algorithms, the ability to hire such workers may be blunted by supply-side constraints or competition from other firms. These equilibrium effects may reduce the potential benefits of algorithmic selection. While there is limited empirical evidence on the equilibrium effects of algorithms, \citet{raymondjmp} shows that the adoption of algorithmic prediction impacts equilibrium prices and investment in the housing market. Such shifts in the aggregate demand for skills may also have long run impacts on the supply of skills in the applicant pool and on the returns to those skills. Both the magnitude and direction of these potentially conflicting effects deserve future scrutiny.



\clearpage
\begin{spacing}{0.9}
\bibliography{bibliography}
\end{spacing}


\newpage
\clearpage


\clearpage

\clearpage

\begin{singlespace}
	\begin{footnotesize}
		\justifying \noindent \textsc{Notes}: This figure shows our inverse propensity weighting estimates of $E[Y|I^{ML}=1]$ for each algorithmic selection strategy (SL or UCB), alongside actual hiring yields from human recruiter decisions. Our inverse propensity weighting estimation method is described in Section \ref{sec:ipw}. We also plot the 95\% confidence intervals around each estimate of hiring yield. All data come from the firm's application and hiring records.
	\end{footnotesize}
\end{singlespace}


\clearpage
\begin{singlespace}
\begin{footnotesize}

\justifying \noindent \textsc{Notes}: These binned scatterplots show the relationship between the leniency of randomly assigned screeners and the hiring outcomes of the applicants they select to be interviewed.  Panel A plots this relationship, controlling for job level characteristics: type of job, seniority level, work location, and application year.  Panel B plots this relationship after adding additional controls for applicant characteristics: education, work history, and demographics. All data come from the firm's application and hiring records.

\end{footnotesize}
\end{singlespace}

\clearpage
\begin{singlespace}
\begin{footnotesize}

\justifying \noindent \textsc{Notes}: \justifying \noindent \textsc{Notes}: Each panel in this figure shows the results of estimating the characteristics of applicants interviewed on the margin.  In each panel, these characteristics are estimated separately for applicants in the top and bottom half of the UCB algorithm's score.  Panels A, C, and E consider marginal applicants as defined by SL model scores.  Panels B, D, and F consider marginal applicants as defined by UCB model scores.  In Panels A and B, the $y$-axis is the average hiring likelihood of marginally interviewed candidates; Panels C and D focus on the share of selected applicants who are Black or Hispanic; Panels E and F focus on the share of selected applicants who are female.  The confidence intervals shown in each panel are derived from robust standard errors clustered at the recruiter level. All data come from the firm's application and hiring records. 
\end{footnotesize}
\end{singlespace}


\clearpage


\clearpage
\begin{singlespace}
\begin{footnotesize}
\begin{singlespace}
\vspace{-10pt}
\justifying \noindent \textsc{Notes}: Panels A shows the race/ethnicity and gender composition of applicants recommended for interviews by the UCB algorithm when this algorithm explicitly incorporates race/ethnicity and gender in estimation (``UCB'').  Panel B shows the composition of applicants recommended for interviews when the UCB is blinded to race/ethnicity and gender (``Blinded UCB'').  Panel C shows our inverse propensity weighting estimates of $E[Y|I^{ML}=1]$ for the demographically aware UCB, blinded UCB, and actual hiring yields from human selection decisions.  All data come from the firm's application and hiring records.
\end{singlespace}

\end{footnotesize}
\end{singlespace}


\clearpage
\begin{singlespace}
	\begin{footnotesize}
		\begin{singlespace}
			\vspace{-20pt}
			\justifying \noindent \textsc{Notes}:  Panel A shows the race/ethnicity and gender composition of applicants recommended for interviews by the UCB algorithm predicting hiring likelihood.  Panel B considers the alternative of using an SL model that is constrained to select applicants in proportion to their representation in the applicant pool.  Panel C shows our inverse propensity weighting estimates of $E[Y|I^{ML}=1]$ for the UCB versus the SL with quota, alongside actual hiring yields from human selection decisions.  All data come from the firm's application and hiring records. 
		\end{singlespace}
		
	\end{footnotesize}
\end{singlespace}


\clearpage


\clearpage
\begin{singlespace}
	\begin{footnotesize}
		
		\begin{singlespace}
			\justifying \noindent \textsc{Notes}: This figure shows the results of a simulation in which the quality of Black applicants increases over time, as described in Section \ref{sec:simulations} and Appendix \ref{asec:simulations}.  In Panel A, the $y$-axis graphs the share of Black applicants in the ``evaluation cohort'' who are selected to be interviewed by each ML policy.  Panel B plots the overall quality of interview decisions, as measured by hiring yield among interviewed applicants. All data come from the firm's application and hiring records. 
		\end{singlespace}
		
	\end{footnotesize}
\end{singlespace}


\clearpage

\begin{table}[ht!]
\begin{center}
                \caption{\textsc{Table \ref{sumstat_app}: Applicant Summary Statistics}}
                  \vspace{20pt}        
\scalebox{1}{\makebox[\linewidth]{\input{Output/Dec2022/summary.tex}}}
   \label{sumstat_app}
\end{center}

 \begin{singlespace}
 \footnotesize
                \justifying \noindent \textsc{Notes}: This table shows applicants' demographic characteristics, education histories, and work experience. The sample in Column 1 consists of all applicants who applied to a position during our training period (2016 and 2017). Column 2 consists of applicants who applied during the analysis period (2018 to Q1 2019). Column 3 presents summary statistics for the full pooled sample.  All data come from the firm's application and hiring records.
        \end{singlespace}
        \normalsize

\end{table}

\clearpage
\begin{table}[ht!]
        \begin{center}
                \caption{\textsc{Table \ref{disagreement}: Predictive accuracy of Human vs. ML models, among Interviewed Applicants}}
                               \vspace{10pt}
\makebox[\linewidth]{\includegraphics[scale=.9]{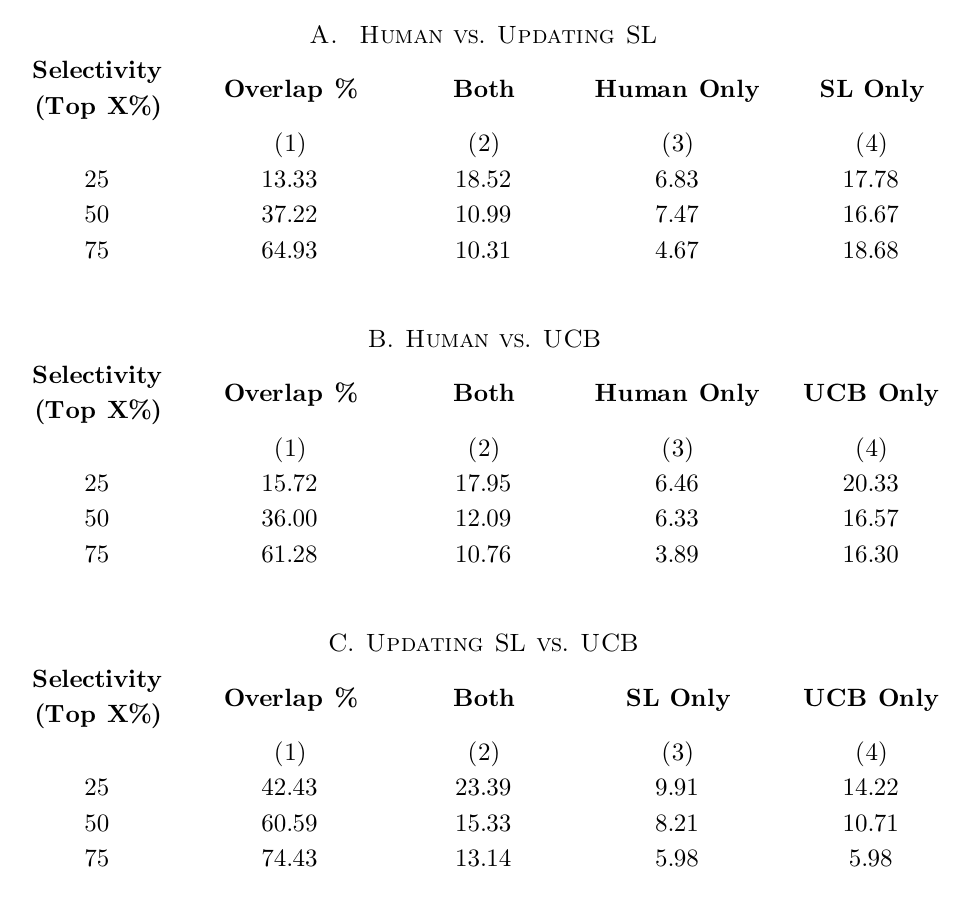}}\\

   \label{disagreement}
        \end{center}
        \begin{singlespace}
        \footnotesize
                \justifying \noindent \textsc{Notes}: This table shows the hiring rates of each algorithm when they make the same recommendation or differing recommendations. The top panel compares the human versus SL algorithm, the middle panel compares the human versus the UCB algorithm, and the lower panel compares the SL versus the UCB algorithm. ``Human'' refers to our model of human recruiter interview propensity introduced in Section \ref{sec:intonly}. Each row of a given panel conditions on selecting either the top 25\%, 50\%, 75\% of applicants according to each of the models.  For the two algorithms being compared in a given panel, Column 1 shows the percent of selected applicants that both algorithms agree on.  Column 2 shows the share of applicants hired when both algorithms recommend an applicant, and Columns 3 and 4 show the share hired when applicants are selected by only one of two algorithms being compared.  All data come from the firm's application and hiring records.                 

        \end{singlespace}
        \normalsize
\end{table}

\clearpage
\begin{table}[ht!]

        \begin{center}
                \caption{\textsc{Table \ref{iv_validity}: Instrument Validity}}

\scalebox{.9}{\makebox[\linewidth]{\input{Output/Dec2022/IVvalidation.tex}}}
         
        \label{iv_validity}
        \end{center}
        \begin{singlespace}
        \footnotesize
                \justifying \noindent \textsc{Notes}: This table shows the results of regressing applicant characteristics on interviewer leniency, defined as the jack-knife mean-interview rate for the recruiter assigned to an applicant, controlling for fixed effects for job family, management level, application year and location of the job opening. The leave-out mean is defined as the share of interviews the assigned recruiter grants, excluding the focal applicant. This leave-out mean is standardized to be mean zero and standard deviation one. The outcome in Column 1 is an indicator variable for being interviewed. The outcomes in Columns (2)--(8) are indicators for baseline characteristics of the applicant. The sample is restricted to recruiters who screened at least 50 applicants. Standard errors are clustered at the recruiter level. All data come from the firm's application and hiring records. 
        \end{singlespace}
        \normalsize
\end{table}

%
%
%
\clearpage
\begin{table}[ht!]

        \begin{center}
                \caption{\textsc{Table \ref{iv_results}:  Impacts of following ML recommendations, IV analysis}}

\textsc{A.  Hire Rates}\\
\vspace{5pt}
\scalebox{1}{\makebox[\linewidth]{\input{Output/Dec2022/tab_IVhired.tex}}}\\ 
\vspace{10pt}
\textsc{B.  Offer Rates}\\
\vspace{5pt}
\scalebox{1}{\makebox[\linewidth]{\input{Output/Dec2022/tab_IVoffered.tex}}}\\
\vspace{10pt}
\textsc{C.  Share Black or Hispanic}\\
\vspace{5pt}
\scalebox{1}{\makebox[\linewidth]{\input{Output/Dec2022/tab_IVbh.tex}}}\\
\vspace{10pt}
\textsc{D.  Share Female}\\
\vspace{5pt}
\scalebox{1}{\makebox[\linewidth]{\input{Output/Dec2022/tab_IVf.tex}}}

        \label{iv_results}
        \end{center}
        \begin{singlespace}
        \footnotesize
                \justifying \noindent \textsc{Notes}:  This table examines the characteristics of marginally interviewed applicants according to our IV strategy described in the text.  Specifically, each number represents the result of the regressions outlined in Equation \eqref{eq:marginal1}.  The reported coefficients are the IV estimates of the coefficient on whether an applicant is interviewed and can be interpreted as the average outcome variable among treatment compliers.  For example, the coefficients in Panel A Columns 1 and 2 represent the estimated average hiring rates of IV compliers with low and high UCB scores, respectively. Applicants receiving high or low scores are those who are above or below the median, respectively. All data come from the firm's application and hiring records. 
        \end{singlespace}
        \normalsize
\end{table}

\clearpage

\begin{table}[ht!]
\begin{center}
       \caption{\textsc{Table \ref{tab_jobperformance}: Correlations between Human Scores and on-the-job Performance}}
                \vspace{15pt}
           \textsc{A.  Human Scores}\\
           \vspace{5pt}
		\scalebox{1}{\makebox[\linewidth]{\input{Output/Dec2022/tab_corr_onthejob_human.tex}}}\\
   		 \vspace{15pt} 
	  \textsc{B.  SL Scores}\\
           \vspace{5pt}
		\scalebox{1}{\makebox[\linewidth]{\input{Output/Dec2022/tab_corr_onthejob_ssup.tex}}}\\
   		 \vspace{15pt}
	\textsc{C. UCB Scores}\\
 	    \vspace{5pt}
 		   \scalebox{1}{\makebox[\linewidth]{\input{Output/Dec2022/tab_corr_onthejob_ucb.tex}}}
        \label{tab_jobperformance}
 \end{center}
        \begin{singlespace}
        \footnotesize
                \justifying \noindent \textsc{Notes}: This table presents the results of regressing measures of on-the-job performance on algorithm scores, for the sample of applicants who are hired and for which we have available performance data. ``Top Rating'' refers to receiving a 3 on a scale of 1-3 in a mid-year evaluation. ``Promoted'' indicates an applicant receives a promotion. Robust standard errors are shown in parentheses. All data come from the firm's application and hiring records.  
        \normalsize
        \end{singlespace}
\end{table}


\begin{appendix}
\renewcommand{\thefigure}{A.\arabic{figure}}
\setcounter{figure}{0}
\renewcommand{\thetable}{A.\arabic{table}}
\setcounter{table}{0}

\newpage
\clearpage
\noindent \textbf{\Large{Appendix Materials -- For Online Publication}}

\newpage
\clearpage
\section{Data and Model Appendix}\label{asec:datamodelsappendix}

\subsection{Sample Construction}\label{asec:sample_details}
We begin with 258,527 job applications submitted to our firm over the period between 2016 and 2019. We exclude all applications associated with college campus recruiting, because these involve on-campus visits and in-person information sessions. We also exclude job postings for temporary hires, duplicate applications, or those for which the job posting was canceled. This leaves us with a data set of 88,666 applications between 2016 and 2019 for non-campus, permanent positions. 

We merge the application data with a set of internal company data sets that allow us to track who receives interviews, interview performance, job offers, hires and on-the-job performance data. To do this, we use a system-generated candidate id in the job application dataset that we can match to interviews, offers, hires and on-the-job information. All data on job applications, interviews, and hires come from the Fortune 500 firm's human resources software systems. Our data on job applications include self-reported demographic information, work history and education parsed from submitted resumes.    

\subsection{Construction of Key Variables}\label{asec:key_vars}
We have information on the applicants' educational backgrounds, their work experience, referral status, basic demographics, as well as the type of position to which they applied. All information is self-reported in a set of standardized application questions or is provided in the resume. In this section, we describe the construction of key variables using the application information.  

\subsubsection{Candidate Demographics}\label{asec:data_demodetails}
Like many other large employers, our firm collects self-reported race, ethnicity, disability, and gender information in the job application. From the set of questions and applicant responses, we code this information into the race/ethnicity, gender, and disability categories that we use in our analysis. 

Applicants report their gender as male, female, non-binary or ``I do not wish to disclose.'' Less than 3\% of our applicants choose not to disclose their gender or identify as non-binary. Less than 1\% of our sample are missing self-reported gender. We do not drop the small number of applicants who choose not to disclose their gender. For our analysis, we have one categorical variable indicating if a candidate is female, which is an indicator function for self-reported gender as female. We also include dummy variables to indicate if no gender information was disclosed or the applicant identifies as non-binary. 


Applicants choose from a set of mutually exclusive categories when reporting race/ethnicity. Applicants who choose ``Black or African American'' are coded as ``Black''. Applicants who report ``White (United States of America)'', ``White - British (United Kingdom)'', ``Not a visible minority (Canada)'', or ``White'' are coded as ``White''. Applicants who report ``Asian'' or ``Asian - Other (Philippines)'' are coded as ``Asian''. Applicants who report ``Hispanic or Latino'' are coded as ``Hispanic''. Applicants are not asked to self-report as Hispanic or Latino separately from race. However, applicants can choose to identify as multiple races or ``Other Ethnicity''. Individuals who report ``Two or More Races'', ``American Indian or Alaska Native'', ``Native Hawaiian or Other Pacific Islander'' are labeled as ``Other Ethnicity.'' 2.3\% of our sample fall into the ``Other Ethnicity'' category. Individuals who choose not to disclose are coded as ``I do not wish to disclose.'' Of our non-campus sample, 7\% percent choose not to disclose race or are missing race information. As a result, we have a series of indicator variables including if an applicant is ``Black'', ''Hispanic'', ''White'', ''Asian'', ``Other'', ''Non-disclosed'', or ''Missing''. Unfortunately, because race and ethnicity are reported in a single question, we cannot separate race and ethnicity categories in more granular detail.
We also code if applicants self-report as a protected veteran. 

\subsubsection{Educational Background}\label{asec:data_educdetails}
Educational information is extracted from the application resumes. 
This information includes the school the applicant attended, degrees achieved, and major (if applicable). The institution names, degree names, and majors are standardized and cleaned. 

We rely on the College Board's list of college majors, available \href{https://satsuite.collegeboard.org/media/pdf/college-majors-academic-area-study-sat-sd.pdf}{here} to create our standardized list of majors. When standardizing, we clean up spelling errors and inconsistent abbreviations. We then match each major to our standardized list of majors. For instance, an applicant with the major ``Economics'' would be matched to ``Finance/Economics,'' and ``Software Engineering'' would be matched to ``Computer Science.'' Across our sample, the most popular majors out of a list of 43 possible majors include: ``Business Administration/Management'', ``Finance/Economics'', ``Computer Science'',  ``Statistics'', ``Accounting'', ``Mathematics'', ``Management Information Systems'', and ``Applied Mathematics.''  We also create an indicator variable if the applicant majored in a quantitative field (defined as having a degree in a science/social science field).

We also clean and standardize institution names, and add geographic and institutional rank information. We standardize institution names so that, for instance, ``UMiami'' becomes ``University of Miami.'' Next, we use college rankings from US News and World Reports to code several additional variables: ``US Top 50'', ``US Top 25'', ``International Elite'', and if the college is a community college. ``US Top 25'' corresponds to US schools that are ranked in the top 25 according to US News and World Reports. ``US Top 50'' indicates if schools are in the top 50. The ``International Elite'' variable indicates top 50 ranked international schools in their respective country according to US News and World Reports.  Examples of popular ``International Elite'' schools in our sample include ``Weizmann Institute of Science'', ``University of Toronto'', ``Zhejiang University'', ``University of Oxford'', ``University of Cambridge'', ``Tsinghua University'', and the ``Indian Institute of Technology.'' 

We also hand-match each school to the country in which the school is located. From this, we hand-code the geographic region of the school (China, India, Europe, Latin America, Middle East/Africa, other Asian countries, and U.S.-based). We retain these seven variables that indicate the geographic region of the educational background. We also code a candidate as an international candidate if they attended any university outside the US (in India, Other Asia, the Middle East/Africa, Europe, China, or Latin America). 

We also include the number of degrees and calculate the highest degree of educational attainment. 68\% of our sample have an advanced degree. We construct a non-traditional education background to indicate whether an applicant attended a community college or received an associate's degree.

\subsubsection{Work Experience}\label{asec:data_workdetails}
From the work history and job titles provided by applicants, we code additional variables to determine if the applicant has work experience at another Fortune 500 firm, held a campus job, worked in finance, or held a quantitative role (such as a data scientist, programmer, software developer, or engineer). We also construct variables to indicate service sector experience, or campus jobs. 

\subsubsection{Job Category}\label{asec:data_jobdetails}
Finally, we create variables that describe the position for which the candidate has applied. Candidates apply for jobs through our firm's career posting portal for jobs in four job families and with various levels of seniority. Job seniority ranges from "Associate" (most junior) to "Executive VP" (most senior). Jobs fall into one of four families: Financial Analysis, Data Science/Quant, Data Analysis, and Business Analysis. 

Data Science/Quant roles target students from quantitative fields to build machine learning models and software tools that the firm uses in its day-to-day business. Such positions typically require experience in open source programming languages for large-scale data analysis, machine learning, and relational databases. Business analysis roles involve product management and strategy skills. These positions involve collaboration with data scientists, project managers, and quantitative analysts to manage the building and validation of statistical and machine learning models, and to communicate technical subject matter in a more general way. Financial Analysis roles are accounting or finance related work and require experience with budgeting, forecasting, variance analysis, and financial reporting.  Data analysis roles focus on analyzing data to tackle a variety of business problems. The required background often includes experience with statistical or econometric models, manipulating and performing analysis with large data sets, and data governance and predictive analytics. We also track if an applicant has received a referral from an existing employee for a given role. 

\subsubsection{Employee Performance Data}\label{asec:data_onthejobdetails}
We have a limited data set of on-the-job performance data for hired employees. Employees receive performance ratings during mid-year reviews. Mid-year reviews are intended to provide employees with useful feedback on their progress and performance. Ratings range from 1 to 3, where 1 indicates ``below average'' performance, 2 indicates ``average'' and 3 for employees who are ``above average.'' The mean performance rating for the entire sample is 2.1. We also have promotion dates for employees who receive promotions. Due to data limitations in our data provider's personnel management software systems, we only have systematic performance ratings or promotion data for a subset of our sample. 



\subsection{Models and Training}\label{asec:m_details}

\subsubsection{Input Features}\label{asec:m_featuresdetails}
To transform raw information into usable inputs for a machine learning model, we create a series of categorical and numerical variables from our constructed variables on our applicants. These serve as ``features'' for each applicant. We standardize all non-indicator features to ensure they fall within the same value range, and we use one-hot encoding to process categorical features. This means we create a new binary feature for each unique category value: a category is marked as 1 when the category value is present; otherwise, it's 0. 

Because our interest lies in screening applicants for interviews, we only use the information available as of the application date as predictive features. Our final model includes 106 input features. Table \ref{sumstat_app} shows the summary statistics for both the training and the test samples.

We also note that an approach commonly used by commercial hiring algorithms is to set $H_{it}=0$ for all applicants who are not interviewed.  We choose not to follow this approach, as it runs counter to our view that $H_{it}$ should be thought of as a potential outcome. Furthermore, \citet{rambachan2019bias} shows that this approach often leads to algorithms that are more biased against racial minorities.  
%

\subsubsection{Machine Learning Training Procedure}\label{asec:m_trainingdetails}
We divide our data into two sets for the construction of our machine learning algorithms--the ``training'' set and the ``test'' set. The first set comprises 48,719 applicants who apply in 2016 and 2017, of which 2,617 received an interview. The second set consists of the 39,947 applications received in 2018-2019, with 2,275 applicants interviewed. We create our initial ``training'' dataset using the 2016-2017 period rather than taking a random sample. This approach more closely approximates the actual application of machine learning in hiring, where firms tend to use historical data to train a model that is then applied prospectively. All our results are reported on the 2018-2019 ``test'' sample.

\subsubsection{Hired/Offered SL Model}\label{asec:m_sldetails}
Our aim is to formulate an estimate of $\hat{s}(X)$ based on the observed covariates, $X$, available at the candidate's application (see \ref{sec:framework}). For our supervised learning model, we employ L1-regularized logistic regression (LASSO), configured with four-fold cross validation for tuning the LASSO penalty term \citep{tibshirani}. The model is trained on job applicants from the first two years of our sample period, from 2016 to 2017. 

L1 regularization, also known as Lasso regularization, is a method used in the Lasso model to help prevent overfitting, which occurs when a model learns too much detail and noise and performs poorly on new, unseen data.  Regularization in machine learning involves adding an extra penalty to the different parameters of the machine learning model to reduce its freedom and therefore avoid overfitting. In the case of L1 regularization, this penalty involves adding the absolute value of each parameter's magnitude to the loss function \citep{tibshirani}.  

To fine-tune the Lasso's optimal value of lambda, we utilize the \texttt{LogisticRegressionCV} class from \texttt{sklearn}, a commonly-used machine learning library in Python \citep{pedregosa2011scikit}. We use ``L1'' regularization and maximize accuracy during cross-validation by performing a grid search over penalty terms between $.0001$ and $10000$.

For each applicant, we use a complete set of constructed variables concerning education, work history, job category, and race/gender (in some specifications). However, for models tracking hires and job offers, we train only on the subset of applicants who have received interviews. Thus, we know whether applicants are offered a job or have been hired. Yet, it should be mentioned that we only use the information available at the time of application for our model and do not include any interview performance data. For our race/ethnicity-blind models, we exclude all features related to race/ethnicity and gender.

Following best practices as described in \citet{Kaebling2019}, we randomly sub-sample our training data to create a balanced sample, half of whom are hired/offered/interviewed and half of whom are not. Because our outcome variables are rare (e.g. only 9\% of those interviewed are hired), our model could achieve high accuracy with a model that always predicted an outcome of never-hired, with coefficients of zero on all features. To avoid this, we create balanced samples (50-50 split of the outcome), so that our accuracy metrics reflect useful information. To create our balanced sample, we randomly sample a subset of our dataset such that the mean share of the subset that is hired/offered/interviewed is $0.5$. We use the \texttt{RandomUnderSampler} class from the \texttt{imblanced-learn} python package developed in \citet{imblearn}. We train all models and report all accuracy metrics based on balanced samples. We use balanced sampling, rather than a reweighted loss function, so that we follow the same sample balancing procedures for both our SL and UCB models. We detail the training, model fitting and updating procedure for our UCB and SL models in Algorithm \ref{alg:update}.

Our results are also robust to different model choices. For instance, using an ensemble that combines a lasso model and random forest approach, delivers slightly higher predictive validity (0.67 vs. 0.64 AUC). In our paper, we choose to stick to the simple logit model for transparency and to ensure consistency with our UCB model, which is based on \citet{li2017}'s implementation that uses a logit model to predict expected quality.

We employ a standard supervised learning approach to build two models. We first to predict an applicant's likelihood of being hired, given that they were interviewed, and a second model that predicts receiving an offer, also conditional on being interviewed. Our outcome variables are ``hired if interviewed'' and ``offered if interviewed.'' For consistency with our training procedure, we evaluate out-of-sample performance on randomly-selected balanced samples from our 2018-2019 ``testing'' period, having trained our models using the 2016-2017 sample. 

\begin{figure}[h!]
	\begin{center}
		\captionsetup{justification=centering}
		\caption{\textsc{Figure \ref{roc}: Model Performance: predicting hiring, conditional on receiving an interview}}
		\begin{tabular}{c}
			\includegraphics[scale=0.45]{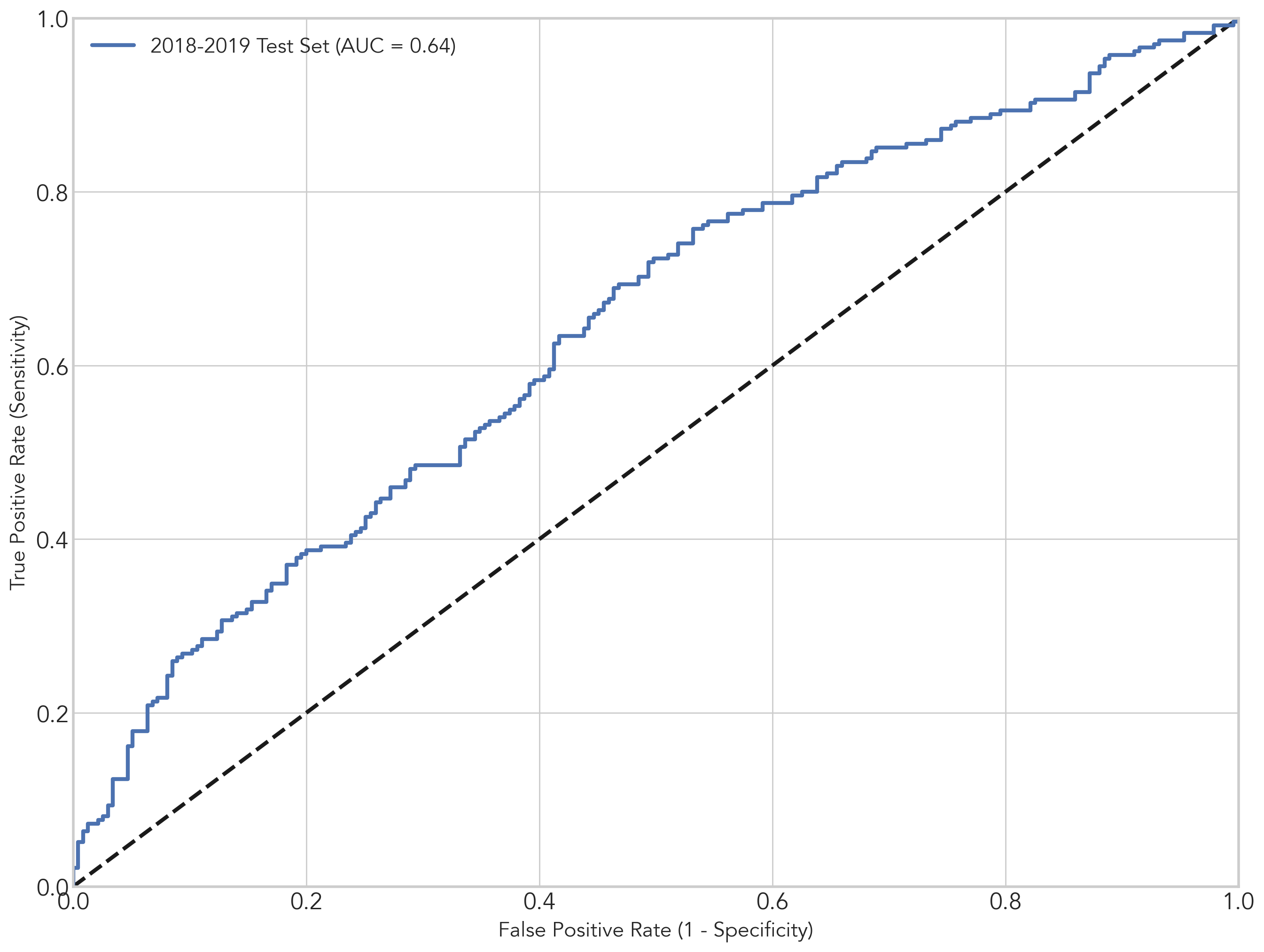}\\
		\end{tabular}
		\label{roc}
	\end{center}
	\begin{singlespace}
		\begin{footnotesize}
			\footnotesize
			\justifying \noindent \textsc{Notes}: This figure shows the Receiver-Operating Characteristic (ROC) curve for the baseline supervised learning model, which predicts hiring potential.  The ROC curve plots the false positive rate on the $x$-axis and the true positive rate on the $y$-axis. For reference, the 45 degree line is shown with a black dash in each plot. All data come from the firm's application and hiring records. 
			
		\end{footnotesize}
	\end{singlespace}
	
\end{figure}

Appendix Figure \ref{roc} plots the Receiver Operating Characteristic (ROC) curve along with the associated AUC or Area Under the Curve.  The AUC is a standard measure of predictive performance that quantifies the tradeoff between a model's true positive rate and its false positive rate.  Formally, the AUC is defined as $\Pr(s(X'_{it}) > s(X'_{jt}) | Y_{it}=1, H_{jt}=0)$.  If the AUC equals 0.5, the model's ability to distinguish between the two classes (hired or not hired) is equivalent to random chance. If it equals 1.0, the model achieves perfect classification. Hence, our hiring model ranks a randomly selected ``hired'' applicant above a randomly selected ``not-hired'' applicant 64\% of the time. We also plot the associated confusion matrix in Appendix Figure \ref{cm}, which further elaborates on the model's classification performance. The confusion matrix displays the percentage of correctly classified candidates, as well as the number of type I and II errors. We construct this model with the constraint that our model ``selects'' the same number of applicants as the actual number of interviewees. Our model correctly classifies 67\% of ``not-hired'' applicants as ``not hired'' and 49\% of ``hired'' applicants as ``hired.'' 

The Receiver Operating Characteristic (ROC) curve for our offer model is shown in Appendix Figure \ref{roc_offer}, and the confusion matrix is presented in Appendix Figure \ref{cm_offer}. Our offer model is capable of ranking a randomly selected ``offered'' applicant higher than a randomly selected ``not-offered'' applicant 68\% of the time. Furthermore, we correctly classify 69\% of ``not-offered'' applicants as ``not-offered'' and 62\% of applicants who receive an ``offer'' as ``offered.''

\subsubsection{Interviewed SL Model}\label{asec:m_humandetails}
\begin{figure}[ht!]
\begin{center}
\captionsetup{justification=centering}
\caption{\textsc{Figure \ref{roc_human}: Model Performance: Predicting interview selection}}
\begin{tabular}{c}
\includegraphics[scale=0.38]{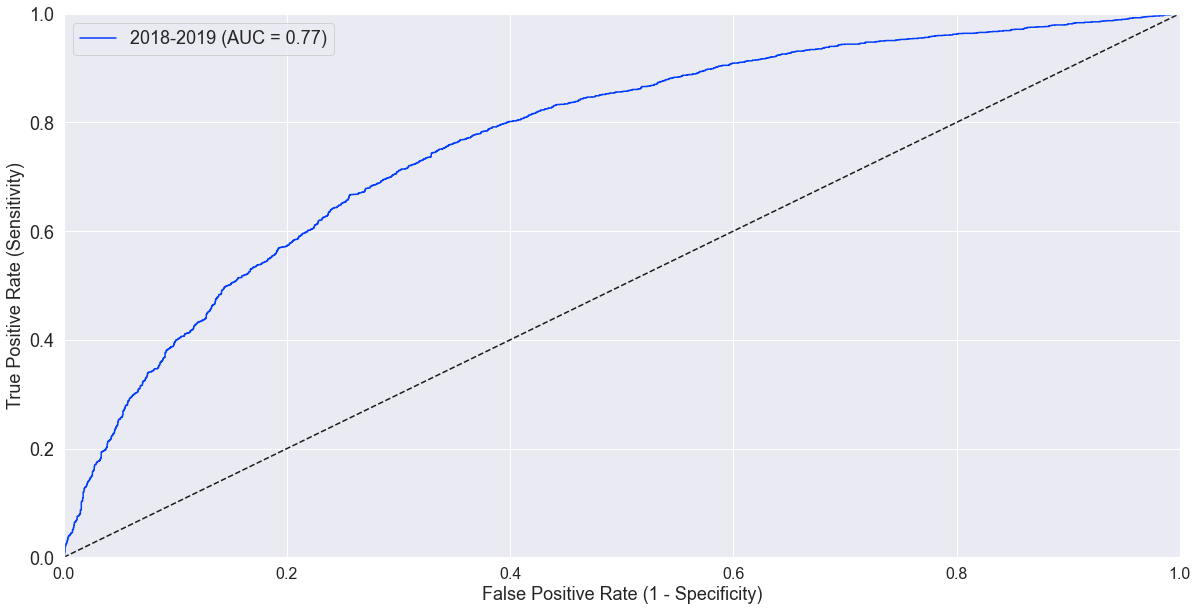} 
\end{tabular}
 \label{roc_human}
\end{center}
\begin{singlespace}
\begin{footnotesize}
\footnotesize
\justifying \noindent \textsc{Notes}: This figure shows Receiver-Operating Characteristic (ROC) curve for the human decision making model, which is trained to predict an applicant's likelihood of being selected for an interview. The ROC curve plots the false positive rate on the $x$-axis and the true positive rate on the $y$-axis.  For reference, the 45 degree line is shown with a black dash in each plot.  All data come from the firm's application and hiring records. 

\end{footnotesize}
\end{singlespace}

\end{figure}

\begin{figure}[ht!]
	\begin{center}
		\captionsetup{justification=centering}
		\caption{\textsc{Figure \ref{cm_human}: Confusion Matrix Model Performance: Predicting interview selection}}
		\begin{tabular}{c}
			\includegraphics[scale=0.5]{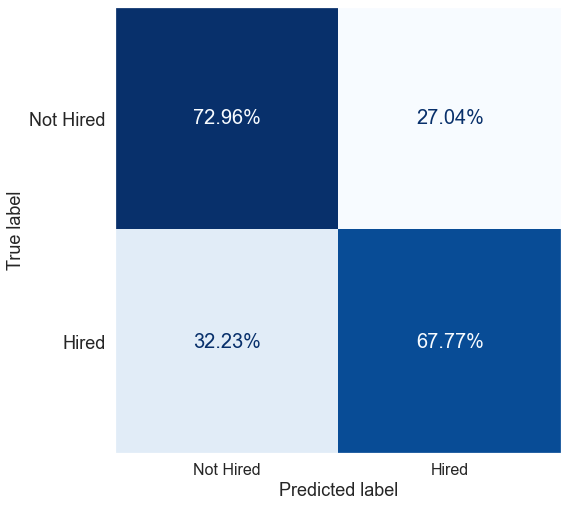} 
		\end{tabular}
		\label{cm_human}
	\end{center}
	\begin{singlespace}
		\begin{footnotesize}
			\footnotesize
			\justifying \noindent \textsc{Notes}: This figure shows a confusion matrix for the human decision making model, which is trained to predict an applicant's likelihood of being selected for an interview. The confusion plots the predicted label on the $x$-axis and the true label rate on the $y$-axis.  Correctly classified applicants are in the bottom right cell, ``true positives" and top left, ``true negatives".  All data come from the firm's application and hiring records. 
			
		\end{footnotesize}
	\end{singlespace}
	
\end{figure}

To compare our model's preferences with the human recruiters, we train an additional model to predict an applicant's likelihood of being selected for an interview by a human recruiter.  This is necessary because all applicants in this interviewed sample are---by definition---selected by human recruiters, so we need to generate additional variation in human preferences.  Specifically, we generate a model of $E[I|X]$ where $I \in \{0,1\}$ are realized human interview outcomes, using the same L1-regularized logistic model as our updating supervised learning model and cross validation on the training sample to tune the lasso penalty term. By choosing the same model structure for both interview, offer, and hire, we avoid differences in predictive accuracy driven by the choice of machine learning algorithm. However, unlike our updating model, we do not dynamically update human preferences---we train the model using all applicants from the 2016-2017 test period and then generate predicted interview propensities for all applicants in the 2018-2019 period.  Because we are trying to predict interview outcomes as opposed to hiring outcomes conditional on interview, our training sample consists of all applicants, rather than only those who are interviewed. This model allows us to order interviewed applicants in terms of their ``human score,'' $s^H$, in addition to their algorithmic scores, $s^{SL}$ and $s^{UCB}$.\footnote{Our IV results do not require us to model human interview practices.}  Appendix Figure \ref{roc_human} plots the ROC associated with this model.  Our model ranks a randomly chosen interviewed applicant ahead of a randomly chosen applicant who is not interviewed 76\% of the time. Appendix Figure \ref{cm_human} shows the interviewed model's confusion matrix. We correctly classify 73\% of applicants not selected for an interview as ``not-interviewed'' and 68\% of applicants who receive an interview as ``interviewed.''

\subsubsection{UCB Model}\label{asec:m_ucbdetails}
For our upper confidence bound (UCB) contextual model, we use an Upper Confidence Bound Generalized Linear Model (UCB-GLM) described in \citet{li2017}. This model is particularly well suited for a setting with a binary reward (e.g. hired/not hired or offered/not offered). This model also uses a regularized logistic regression to predict quality ($\hat{E}[H_{it}|X'_{it}; D^{UCB}_{t}]$) on the 2016-2017 set of interviewed applicants \citep{cortes_adapting_2019}. The UCB has access to the same training data as the SL model and the same functional form, so at the beginning of our analysis period, it has the same ``beliefs'' about applicant quality. However, when evaluating applicant quality during the 2018-2019 period, the UCB model calculates applicant quality as described in Equation \ref{UCBrule}, using the upper bound of the 95th percent confidence interval.\footnote{Cross-validation metrics in bandits can be overly influenced by past decisions (arm pulls) within the data. The model might ``overfit'' to these past decisions, leading to suboptimal choices in the future when encountering new situations. As a result, we do not use cross validation to select the 95th percentile confidence interval parameter.}

We use the same transformed and standardized data set to construct our UCB model. Like our SL model, this model also dynamically updates as it makes candidate selections. For additional guidance on using GLM-UCB models see \citet{filippi2010} and \citet{li_contextual-bandit_2010}. 

To implement the linear UCB contextual bandit, we use the python package \texttt{contextualbandits}. This package provides an implementation of LinUCB algorithm described in \citet{li_contextual-bandit_2010} in the \texttt{LogisticUCB} class \citep{cortes_adapting_2019}.\footnote{We use the package version 0.2.5.} Following the choices in \citet{li_contextual-bandit_2010} and for simplicity, we use the 95th percentile upper confidence bound in our contextual bandit simulations.  During our analysis, we use the ``Cholesky decomposition'' method to fit the model, which is more appropriate for calculating the precision matrix on larger data. Like our SL models, we train our UCB model on balanced samples. Our UCB model is also re-trained iteratively as the model selects new candidates to hire. When implementing the race/ethnicity-blind version of the model, we exclude features associated with race/ethnicity and gender. We detail the training, model fitting, and updating procedure for our UCB and SL models in Algorithm \ref{alg:update}. In the algorithm, we refer to the scoring function implementing our UCB approach as $s^{UCB}$ and the supervised learning LASSO model as $s^{SL}$.

\begin{algorithm}
\caption{UCB ($s^{UCB}$) and SL ($s^{SL}$) Models Training and Selection Procedure}\label{alg:update}
\begin{algorithmic}
\For{the $2016-2017$ interviewed training sample:}
\State $s^{UCB} \gets \hat{E}[Y|X';D^{train}]$
\State $D^{UCB}_{0} \gets D^{train}$
\State $s^{SL}\gets  \hat{E}[Y|X'; D^{train}]$
\State $D^{SL}_{0} \gets D^{train}$
\EndFor
\For{\textbf{each} period $t = 1, \dots, T$}
\For{\textbf{each} applicant $i \in t$}
\State $s_{it}^{UCB} \gets s^{UCB}(X'_{it}; D^{UCB}_{t}) $
\State $s_{it}^{SL} \gets s^{SL}(X'_{it}; D^{SL}_{t}) $
\EndFor
\State select $I^{UCB}_t$, $I^{SL}_t$
\State $D^{SL}_{t} \gets D^{SL}_{t-1} \cup (I^{SL}_{t} \cap I_{t})$
\State $D^{UCB}_{t} \gets D^{UCB}_{t-1} \cup (I^{UCB}_{t} \cap I_{t})$
\State refit $s^{SL} $  with $D^{SL}_{t}$
\State refit $s^{UCB}$ with $D^{UCB}_{t}$
\EndFor
\end{algorithmic}
\end{algorithm}

\newpage
\clearpage
\subsection{Data and Model Appendix: Figures and Tables}

\newpage
\clearpage

\newpage
\clearpage
\begin{figure}[ht!]
	\begin{center}
		\captionsetup{justification=centering}
		\caption{\textsc{Figure \ref{cm}: Confusion Matrix Model Performance: Predicting hiring, conditional on receiving an interview}}
		\begin{tabular}{c}
			\includegraphics[scale=0.8]{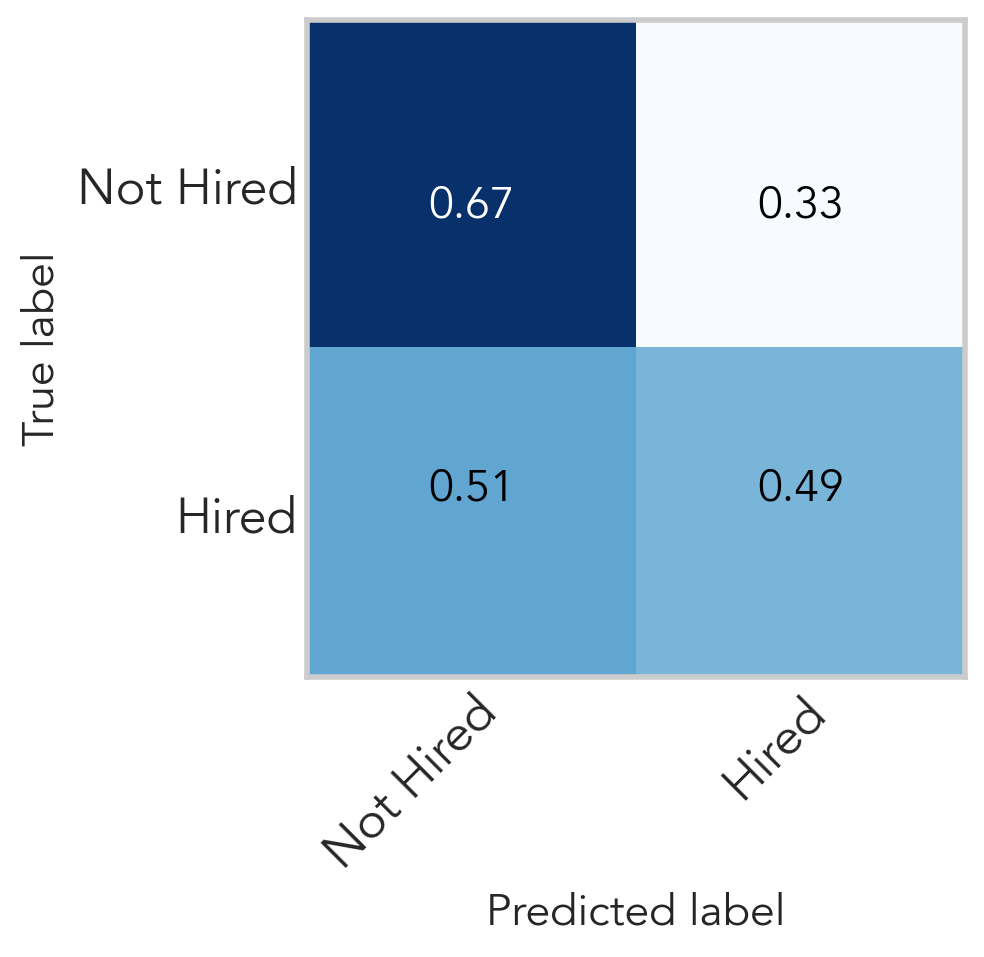} 
		\end{tabular}
		\label{cm}
	\end{center}
	\begin{singlespace}
		\begin{footnotesize}
			\footnotesize
			\justifying \noindent \textsc{Notes}: This figure shows a confusion matrix for the baseline supervised learning model, which predicts hiring potential. The confusion plots the predicted label on the $x$-axis and the true label rate on the $y$-axis.  Correctly classified applicants are in the bottom right cell, ``true positives" and top left, ``true negatives". Examples that are incorrectly classified are in the top right cell (``false positives") and the bottom left (``false negatives"). All data come from the firm's application and hiring records. 
			
		\end{footnotesize}
	\end{singlespace}
	
\end{figure}

\newpage
\clearpage

\newpage
\clearpage


\clearpage
\begin{figure}[h!]
	\begin{center}
		\captionsetup{justification=centering}
		\caption{\textsc{Figure \ref{roc_offer}: Model Performance: predicting offer, conditional on receiving an interview}}
		\begin{tabular}{c}
			\includegraphics[scale=0.45]{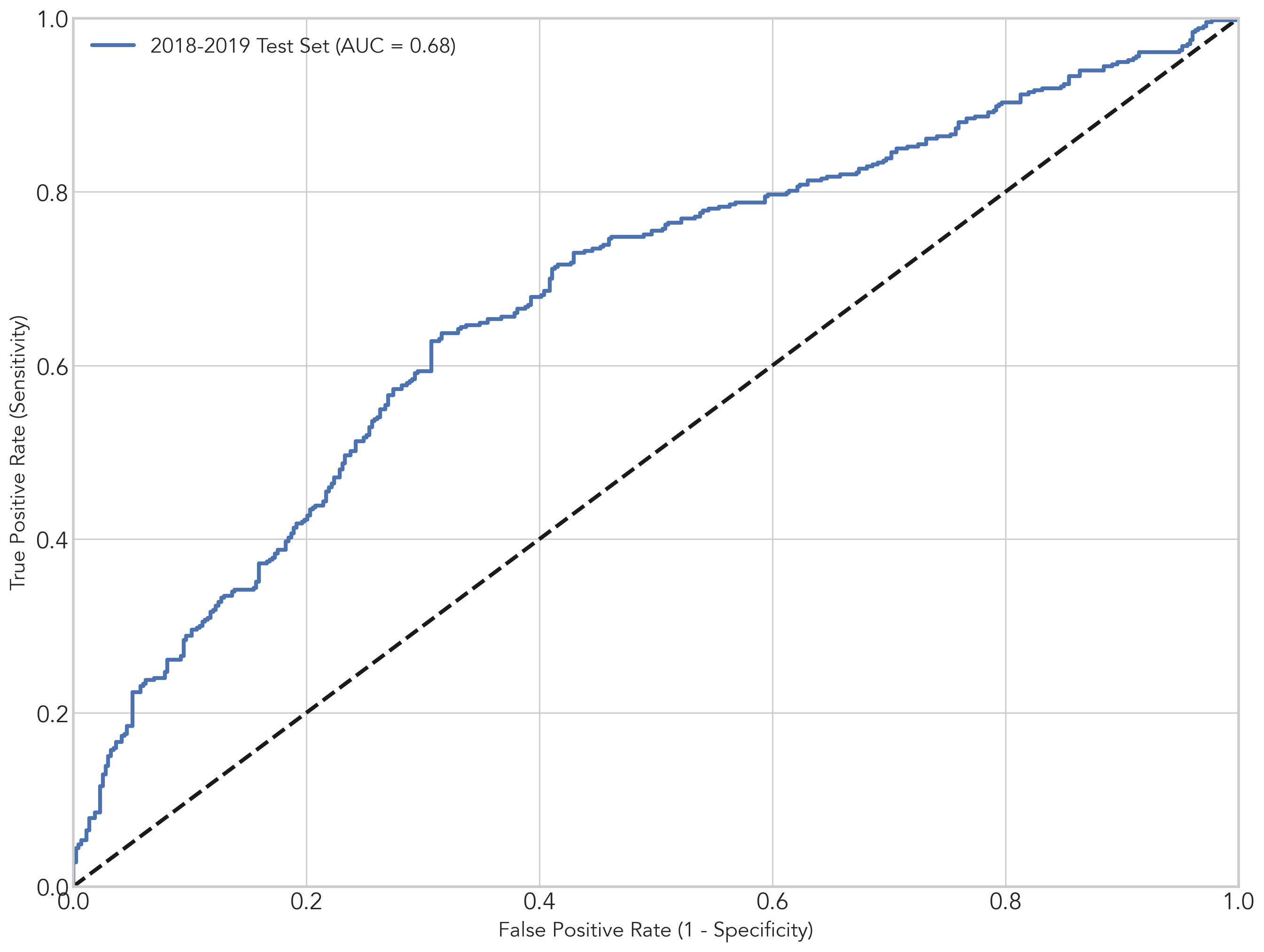}\\
		\end{tabular}
		\label{roc_offer}
	\end{center}
	\begin{singlespace}
		\begin{footnotesize}
			\footnotesize
			\justifying \noindent \textsc{Notes}: This figure shows the Receiver-Operating Characteristic (ROC) curve for the baseline supervised learning model, which predicts offer potential.  The ROC curve plots the false positive rate on the $x$-axis and the true positive rate on the $y$-axis. For reference, the 45 degree line is shown with a black dash in each plot. All data come from the firm's application and hiring records. 
			
		\end{footnotesize}
	\end{singlespace}
	
\end{figure}

\newpage
\clearpage
\begin{figure}[ht!]
	\begin{center}
		\captionsetup{justification=centering}
		\caption{\textsc{Figure \ref{cm_offer}: Confusion Matrix Model Performance: Predicting Offer}}
		\begin{tabular}{c}
			\includegraphics[scale=0.8]{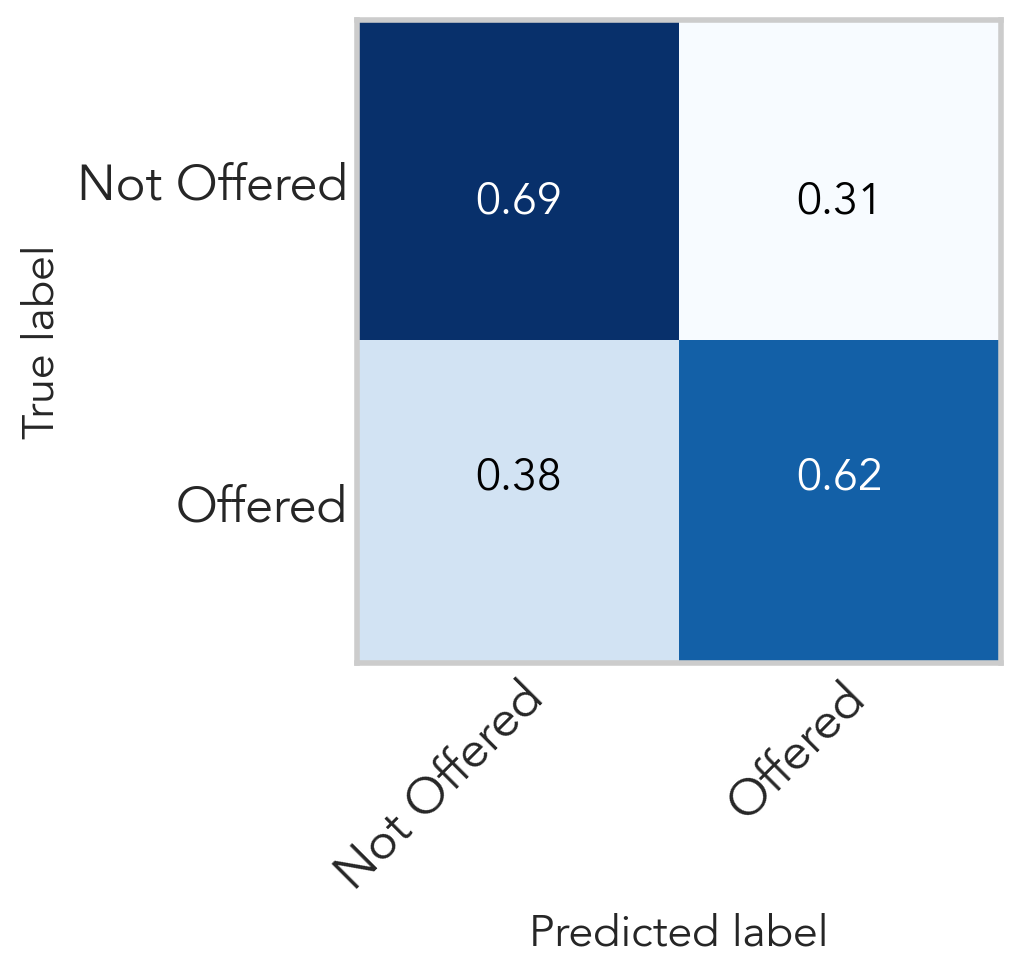} 
		\end{tabular}
		\label{cm_offer}
	\end{center}
	\begin{singlespace}
		\begin{footnotesize}
			\footnotesize
			\justifying \noindent \textsc{Notes}: This figure shows a confusion matrix for the supervised learning offer model, which is trained to predict an applicant's likelihood of receiving an offer. The confusion plots the predicted label on the $x$-axis and the true label rate on the $y$-axis.  Correctly classified applicants are in the bottom right cell, ``true positives" and top left, ``true negatives". All data come from the firm's application and hiring records. 
			
		\end{footnotesize}
	\end{singlespace}
	
\end{figure}

\newpage
\clearpage
\begin{table}[ht!]
	\begin{center}
		\caption{\textsc{Table \ref{atab:features}: Applicant Features and Summary Statistics}}
		\vspace{20pt}        
		\scalebox{1}{\makebox[\linewidth]
  {\input{Output/Dec2022/summary_features}}}
		\label{atab:features}
	\end{center}
	
	\begin{singlespace}
		\footnotesize
		\justifying \noindent \textsc{Notes}: This table shows more information on applicants' characteristics, education histories, and work experience.  The sample in Column 1 consists of all applicants who applied to a position during our training period (2016 and 2017). Column 2 consists of applicants who applied during the analysis period (2018 to Q1 2019). Column 3 presents summary statistics for the full pooled sample.  All data come from the firm's application and hiring records.
		
	\end{singlespace}
	\normalsize
	
\end{table}

\newpage
\clearpage
\section{Diversity of selected applicants}\label{asec:diversity}

\newpage
\clearpage
\begin{figure}[ht!]
\begin{center}
\captionsetup{justification=centering}
\caption{\textsc{Figure \ref{pie_gender}: Gender Composition}}
\makebox[\linewidth]{
\begin{tabular}{cc}
\textsc{\footnotesize{A. Applicant Pool}} & \textsc{\footnotesize{B.  Human}} \\
  \includegraphics[scale=0.5]{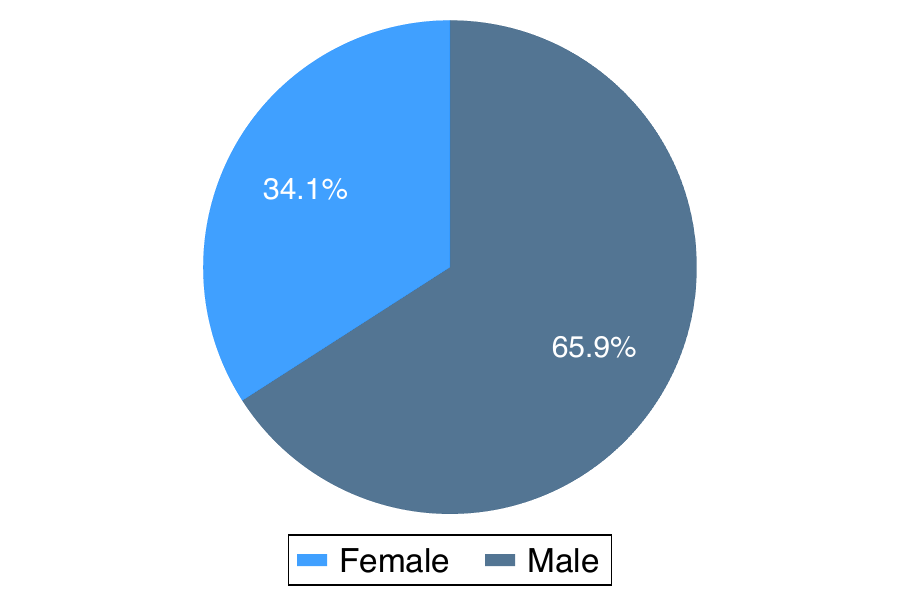} & \includegraphics[scale=0.5]{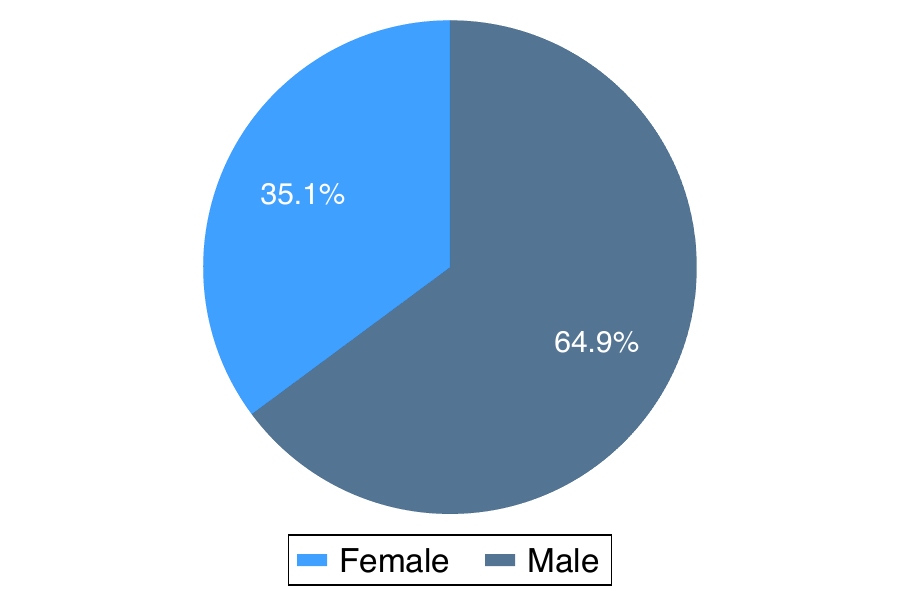}\\
\textsc{\footnotesize{C. SL Model}} & \textsc{\footnotesize{D. UCB Model}} \\
\includegraphics[scale=0.5]{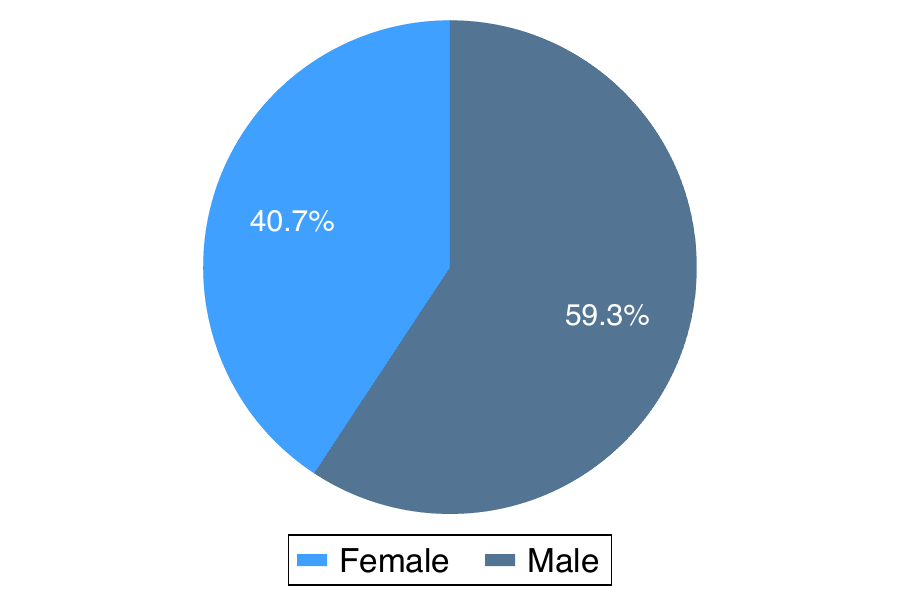} & \includegraphics[scale=0.5]{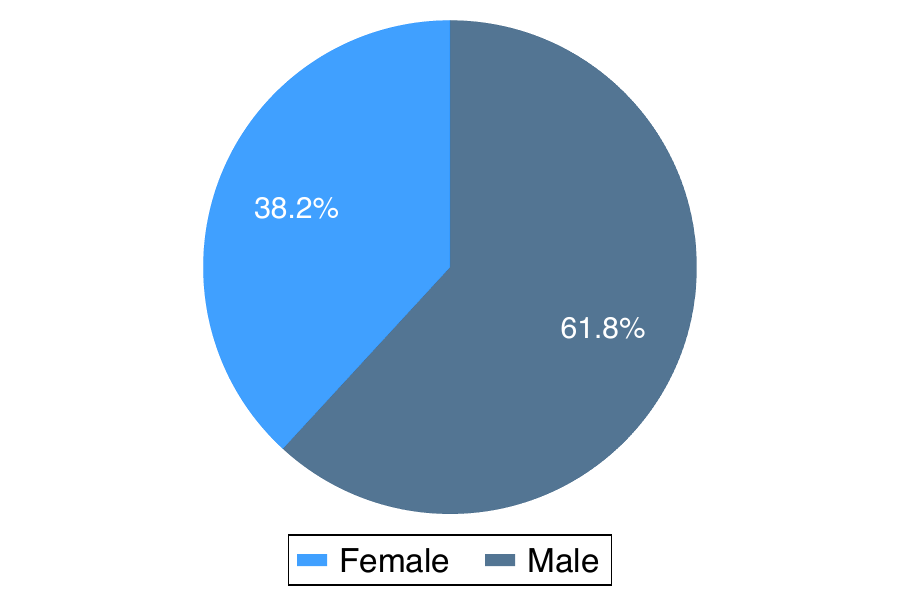}\\
\end{tabular}
}
\label{pie_gender}
\end{center}
\end{figure}
\begin{singlespace}
\begin{footnotesize}

\justifying \noindent \textsc{Notes}: Panel A shows the gender composition of applicants in our data.  Panel B shows the composition of applicants actually selected for an interview by the firm.  Panel C shows the composition of applicants who would be selected if chosen by the supervised learning algorithm described in Equation \eqref{SLrule} that predicts hiring likelihood.  Finally, Panel D shows the composition of applicants who would be selected for an interview by the UCB algorithm described in Equation \eqref{UCBrule} that predicts hiring likelihood.  All data come from the firm's application and hiring records. 

\end{footnotesize}
\end{singlespace}

\clearpage
\begin{figure}[ht!]
	\begin{center}
		\captionsetup{justification=centering}
		\caption{\textsc{Figure \ref{ucb_bonuses_static}: UCB Bonuses}}
		\makebox[\linewidth]{
			\begin{tabular}{c}
				\includegraphics[scale=0.7]{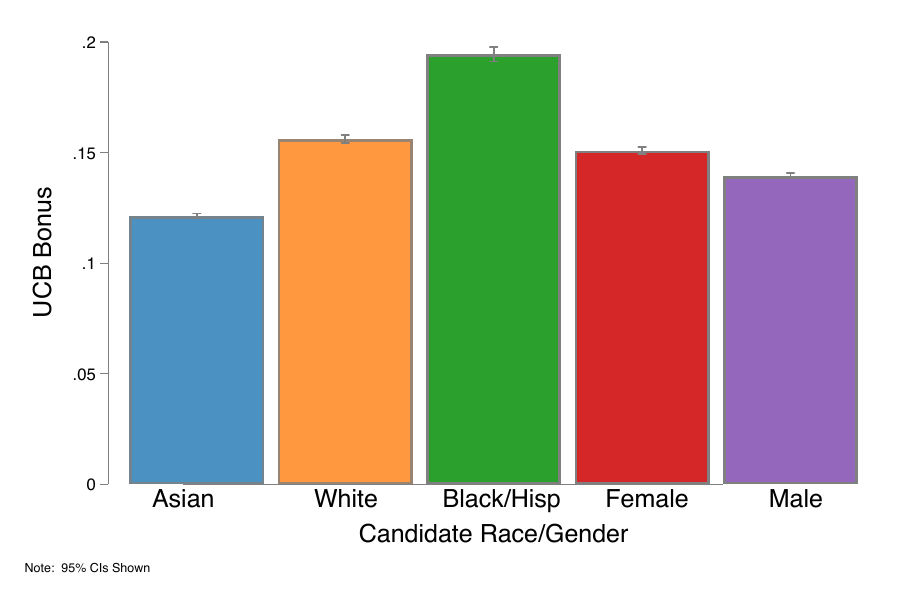} \\
			\end{tabular}
		}
		\label{ucb_bonuses_static}
	\end{center}
	
	\begin{footnotesize} 
		\begin{singlespace}
			\justifying \noindent \textsc{Notes}: This figure shows UCB hiring model exploration bonuses averaged over the testing period (2018-2019), by demographics group. All data come from the firm's application and hiring records.
    \end{singlespace}
	\end{footnotesize}
\end{figure}

	


\clearpage
\begin{figure}[ht!]
\begin{center}
\captionsetup{justification=centering}
\caption{\textsc{Figure \ref{pie_diversity_interviewed}: Demographic Diversity: Selecting Top 50\% Among Interviewed}}
\makebox[\linewidth]{
\begin{tabular}{cc}
\textsc{\footnotesize{A. SL Hired Model, Race/Ethnicity}} & \textsc{\footnotesize{B. UCB Hired Model, Race/Ethnicity}} \\
\includegraphics[scale=0.4]{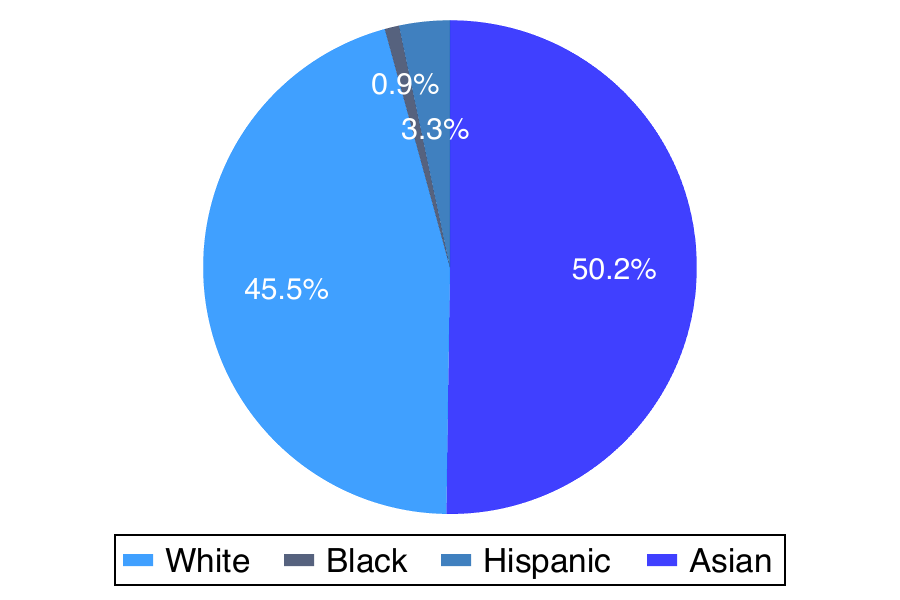} & \includegraphics[scale=0.4]{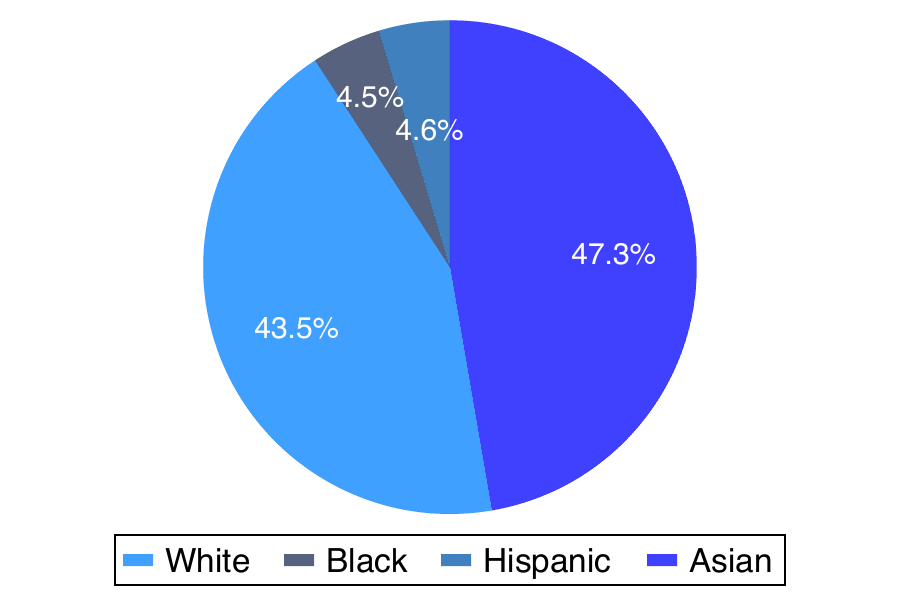}\\
\textsc{\footnotesize{C. SL Offered Model, Race/Ethnicity}} & \textsc{\footnotesize{D. UCB Offered Model, Race/Ethnicity}} \\
\includegraphics[scale=0.4]{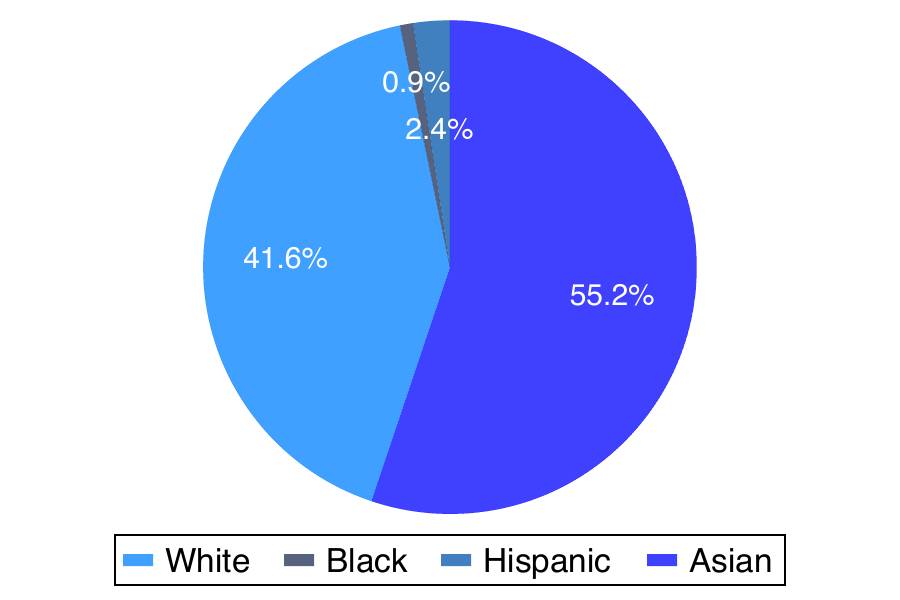} & \includegraphics[scale=0.4]{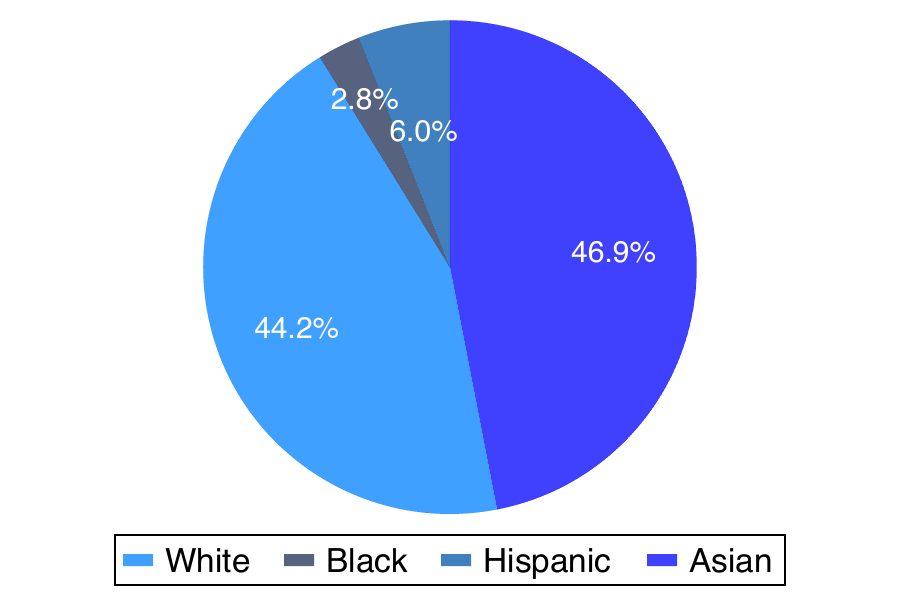}\\

\textsc{\footnotesize{E. SL Hired Model, Gender}} & \textsc{\footnotesize{F. UCB Hired Model, Gender}} \\
\includegraphics[scale=0.4]{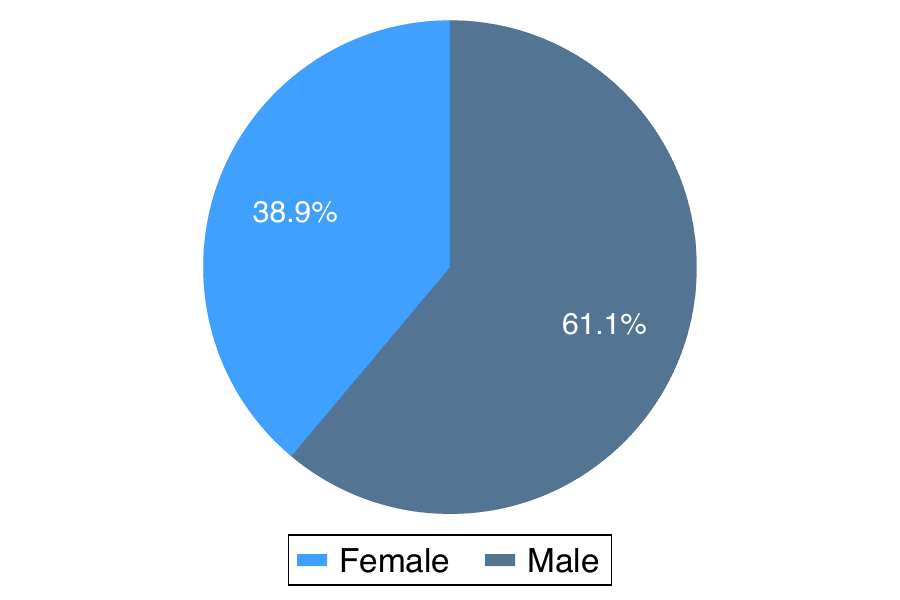} & \includegraphics[scale=0.4]{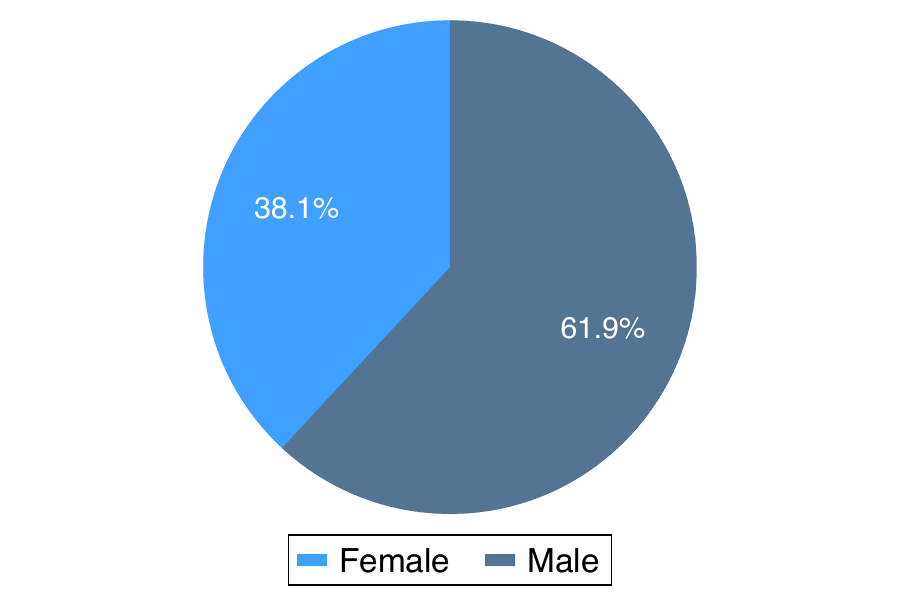}\\
\textsc{\footnotesize{G. SL Offered Model, Race/Ethnicity}} & \textsc{\footnotesize{H. UCB Offered Model, Gender}} \\
\includegraphics[scale=0.4]{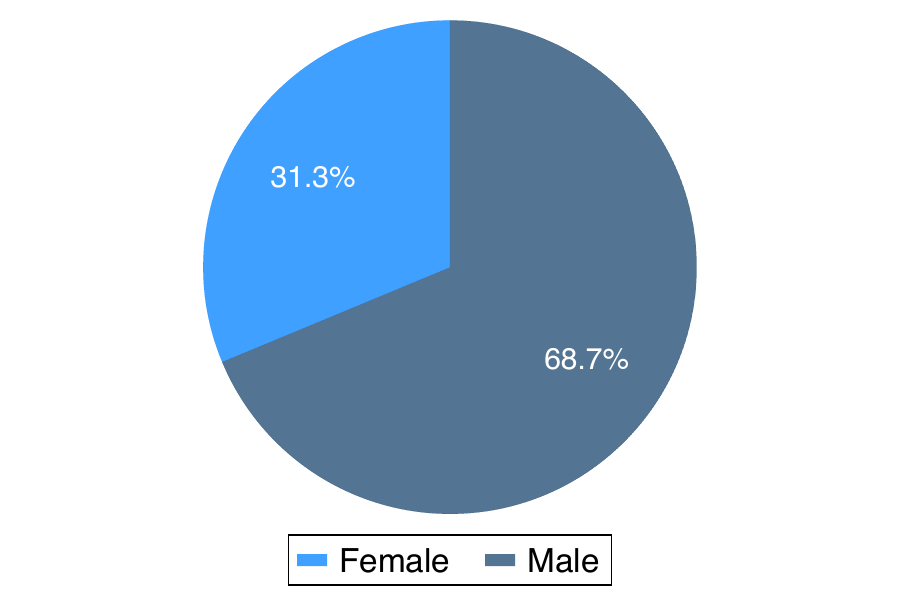} & \includegraphics[scale=0.4]{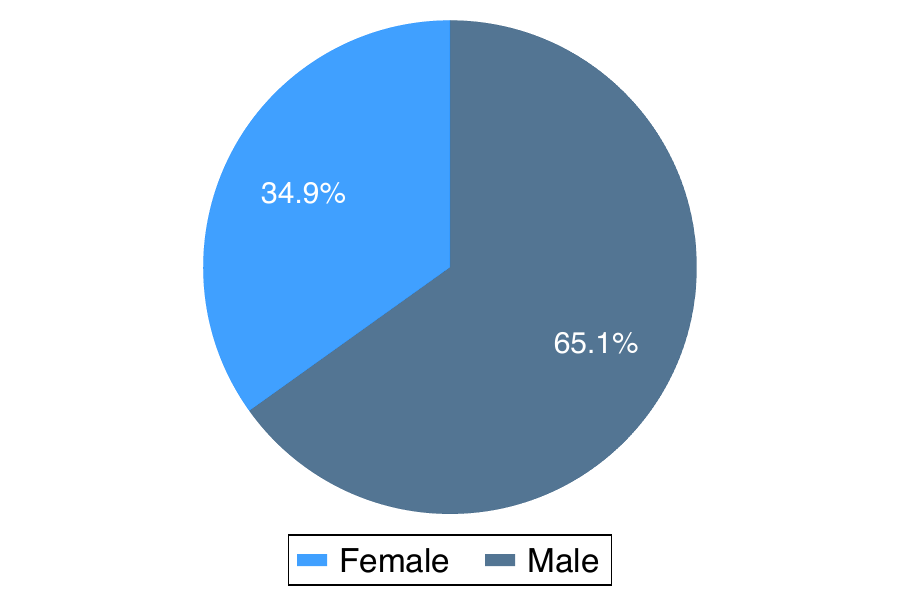}\\

\end{tabular}
}
\label{pie_diversity_interviewed}
\end{center}
\end{figure}
\begin{singlespace}
\begin{footnotesize}

\justifying \noindent \textsc{Notes}: These panels consider the demographic diversity of candidates, selecting amongst the interviewed candidates.  Here, we consider the scenario in which we select the top half of candidates as ranked by each ML score. Results are similar if we use other selection rules.  All data come from the firm's application and hiring records.

\end{footnotesize}
\end{singlespace}

\newpage
\clearpage
\section{Quality of selected applicants, non IV}\label{asec:quality}

\subsection{Interviewed Subsample Results}

\begin{table}[ht!]

        \begin{center}
                \caption{\textsc{Table \ref{allalgoscorr}: Correlations between algorithm scores and hiring/offer likelihood}}

\scalebox{1}{\makebox[\linewidth]{\input{Output/Dec2022/tab_corr_H1_scores.tex}}}
         
        \label{allalgoscorr}
        \end{center}
        \begin{singlespace}
        \footnotesize
                \justifying \noindent \textsc{Notes}: This table presents the results of regressing an indicator for being hired on the algorithm scores in columns 1-3 and receiving an offer in columns 4-6. Control variables include fixed effects for job family, application year-month, and seniority level. ``Human'' is scores from a model that predicts interview likelihood. ``SL Hired'' refers to the supervised learning model that predicts hiring likelihood. ``SL Offered'' refers to the supervised learning model that predicts offer. ``UCB Hired'' and ``UCB Offered'' denote the UCB models for the hiring and offer outcome respectively. All data come from the firm's application and hiring records.  Robust standard errors are shown in parentheses. Sample includes interviewed applicants in our analysis period. All data come from the firm's application and hiring records. 
        \end{singlespace}
        \normalsize
\end{table}

\newpage
\clearpage
\subsection{Inverse Propensity Weighting} \label{asec:ipw}

\subsubsection{Derivation for Equation \eqref{decompeqn}}
Equation \eqref{decompeqn} can be derived as follows:

\begin{eqnarray}
	E[Y|I^{ML}=1] &=& \sum_{X} p(X|I^{ML}=1)E[Y|I^{ML}=1, X] \nonumber \\
	&=&	\sum_{X} \frac{p(I^{ML}=1|X)p(X)}{p(I^{ML}=1)}E[Y|I^{ML}=1, X]  \nonumber\\
	&=&	\frac{1}{p(I^{ML}=1)}\sum_{X} p(I^{ML}=1|X)p(X)E[Y|I^{ML}=1, X] \nonumber \\
	&=&	\frac{1}{p(I^{ML}=1)}\sum_{X} p(I^{ML}=1|X)p(X)E[Y|I^{ML}=1, X]\frac{p(X|I=1)p(I=1)}{p(I=1|X)p(X)}  \nonumber\\
	   & & \left(\text{Multiply by }1 = \frac{p(X|I=1)p(I=1)}{p(I=1|X)p(X)}\right) \nonumber\\
	&=&	\frac{p(I=1)}{p(I^{ML}=1)}\sum_{X} E[Y|I^{ML}=1, X]\frac{p(I^{ML}=1|X)p(X|I=1)}{p(I=1|X)}  \nonumber\\
	&=&	\frac{p(I=1)}{p(I^{ML}=1)}\sum_{X} E[Y|I=1, X]\frac{p(I^{ML}=1|X)p(X|I=1)}{p(I=1|X)} \nonumber \\
	&& \text{(Assuming selection on observables)} \nonumber\\
	&=&	\frac{p(I=1)}{p(I^{ML}=1)}\sum_{X} p(X|I=1)E[Y|I=1, X]\frac{p(I^{ML}=1|X)}{p(I=1|X)}\\
	&=&	\frac{p(I=1)}{p(I^{ML}=1)}E\left[\frac{p(I^{ML}=1 | X)}{p(I=1|X)}Y|I=1\right] 
\end{eqnarray} 

\newpage
\clearpage
\subsubsection{Common Support}


\begin{figure}[ht!]
	\begin{center}
		\captionsetup{justification=centering}
		\caption{\textsc{Figure \ref{common_support}: Distribution of Human Selection Propensity, among ML-selected Applicants}}
		\makebox[\linewidth]{
			\begin{tabular}{c}
				\textsc{\footnotesize{A. SL Selected Candidates}}  \\
				\includegraphics[scale=0.3]{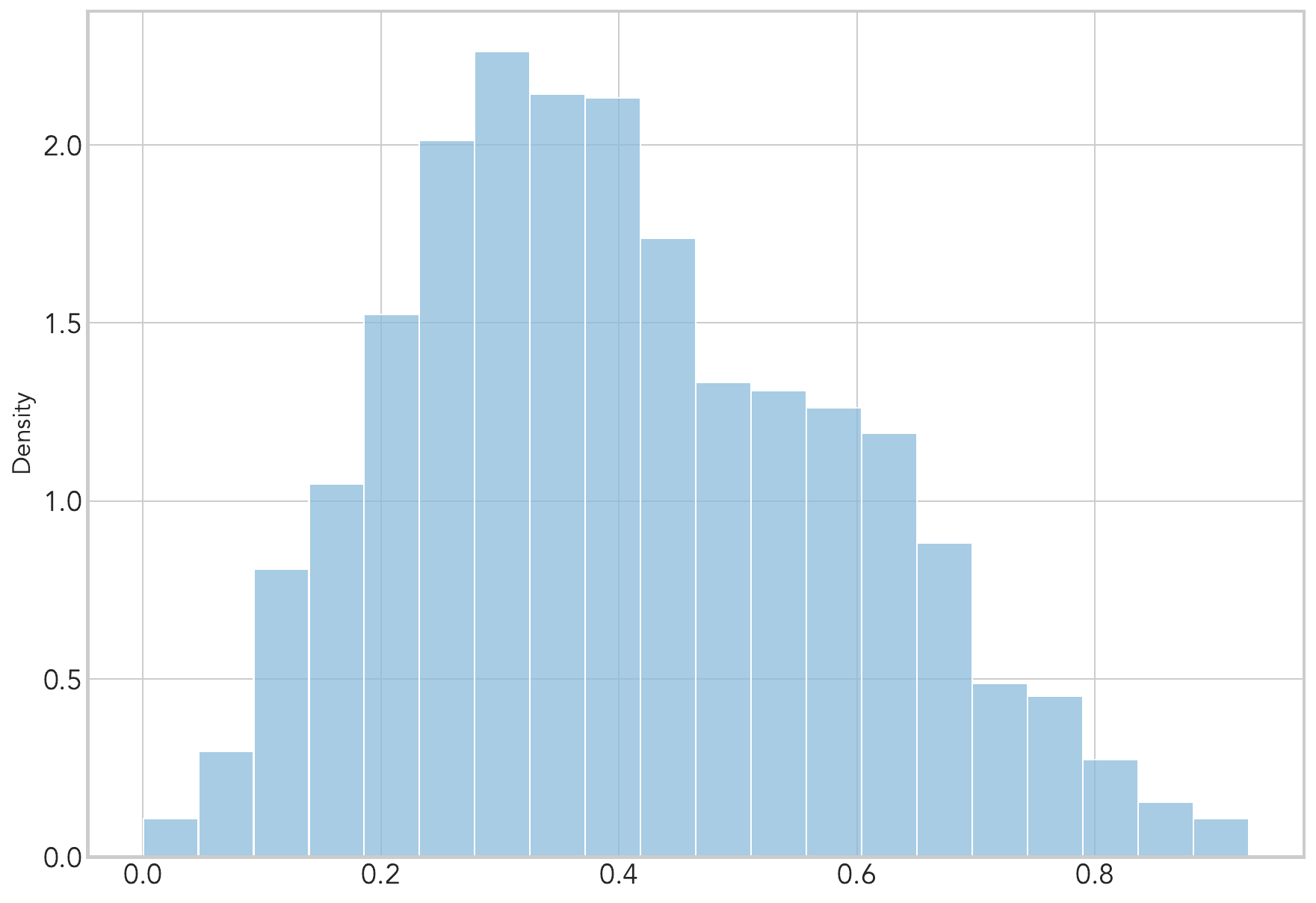} \\ 
				\textsc{\footnotesize{B. UCB Selected Candidates}}  \\ 
				\includegraphics[scale=0.3]{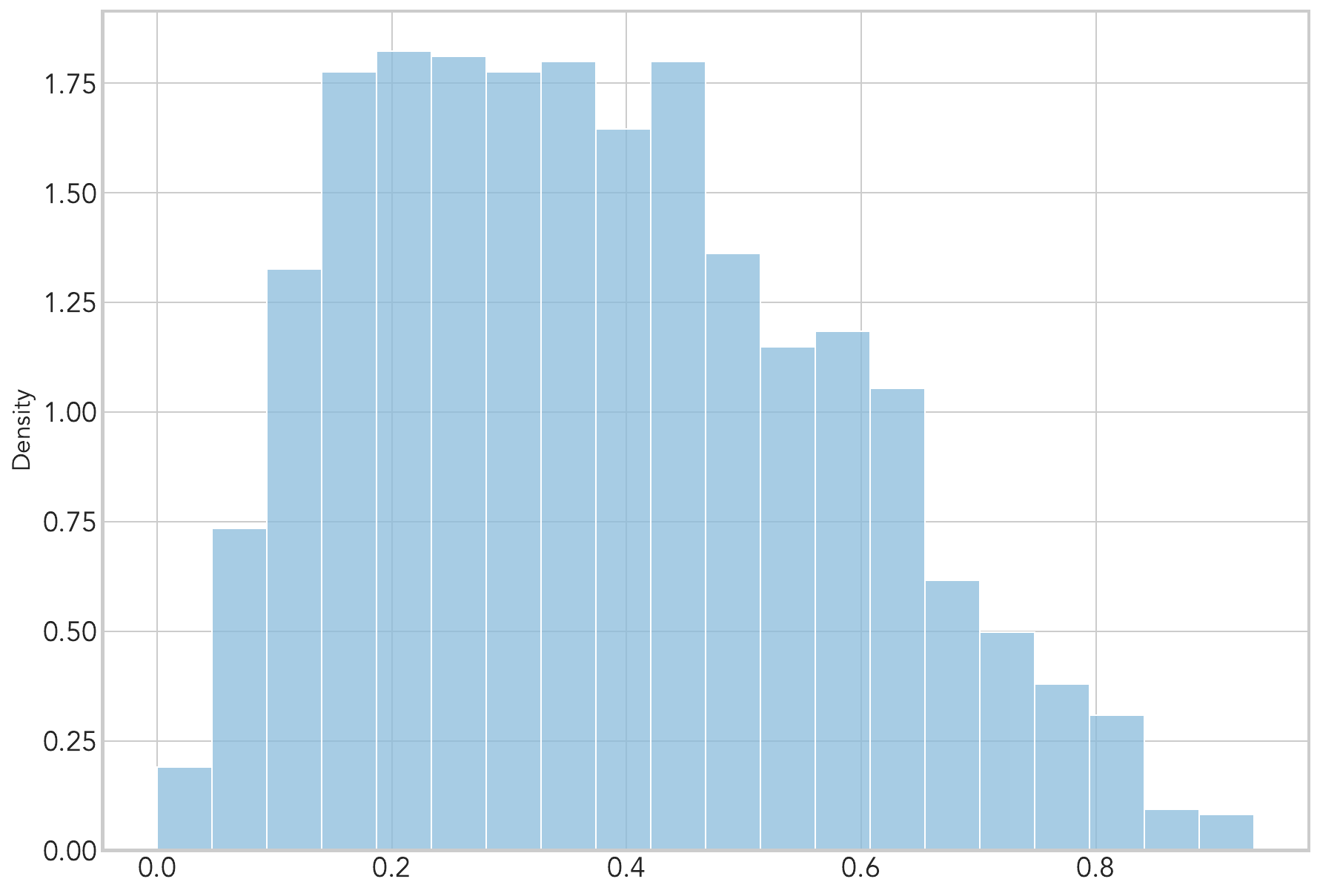} \\
			\end{tabular}
		}
		\label{common_support}
	\end{center}
\end{figure}
\begin{footnotesize}
	\begin{singlespace}		
		\justifying \noindent \textsc{Notes}: 
		This figure shows the distribution of propensity scores for human recruiter selection into the interview set, $p(I=1|X)$. Panel A plots the distribution for applicants selected for an interview by the SL model that predicts hiring likelihood, $I^{SL}$. Panel B plots the distribution for applicants selected for an interview by the UCB model that predicts hiring likelihood, $I^{UCB}$. All data come from the firm's application and hiring records.
	\end{singlespace}	
\end{footnotesize}

\begin{figure}[ht!]
	\begin{center}
		\captionsetup{justification=centering}
		\caption{\textsc{Figure \ref{common_support_offer}: Distribution of Human Selection Propensity, among ML-selected Applicants, Offer Model}}
		\makebox[\linewidth]{
			\begin{tabular}{c}
				\textsc{\footnotesize{A. SL Selected Candidates}}  \\
				\includegraphics[scale=0.3]{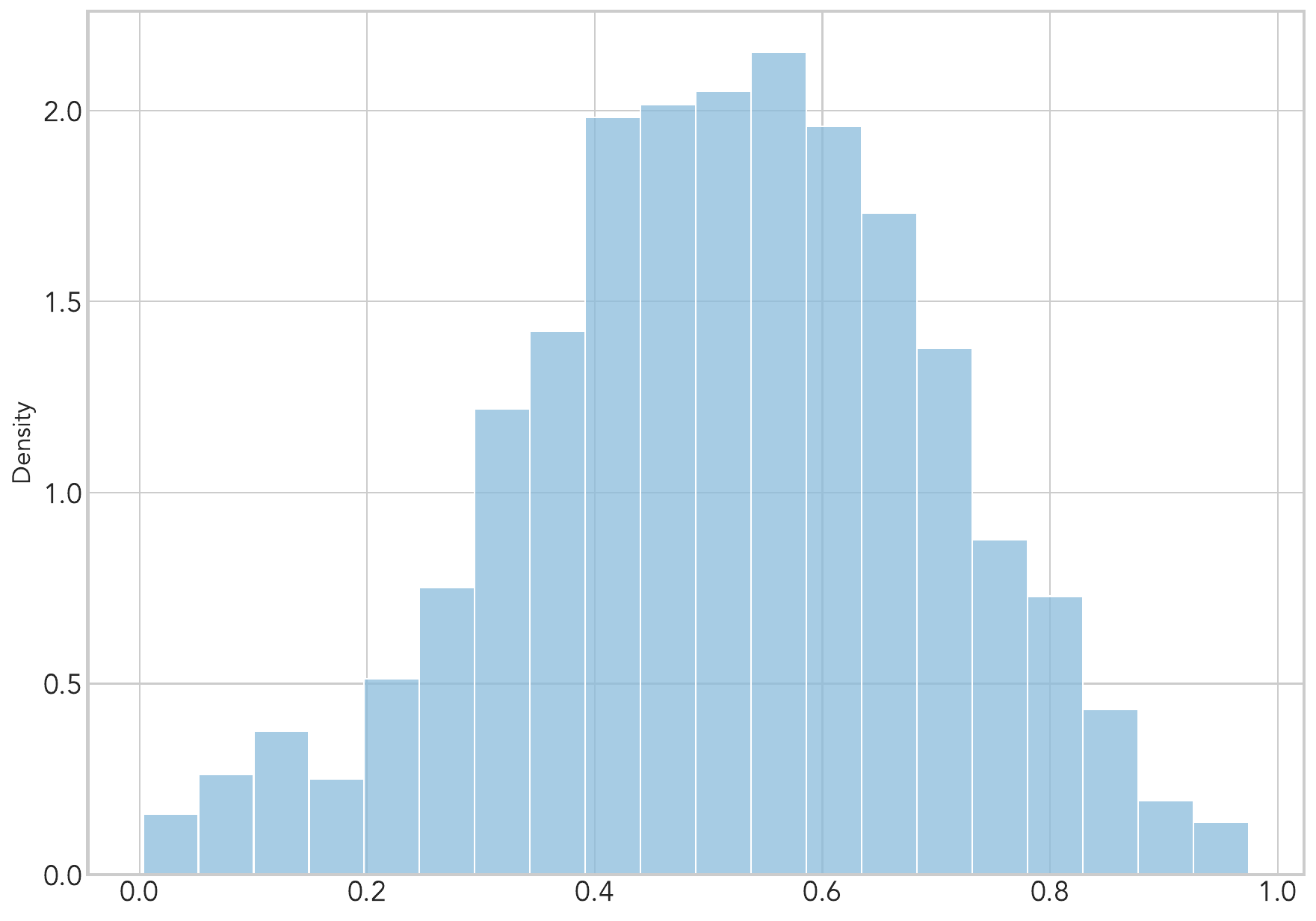} \\ 
				\textsc{\footnotesize{B. UCB Selected Candidates}}  \\ 
				\includegraphics[scale=0.3]{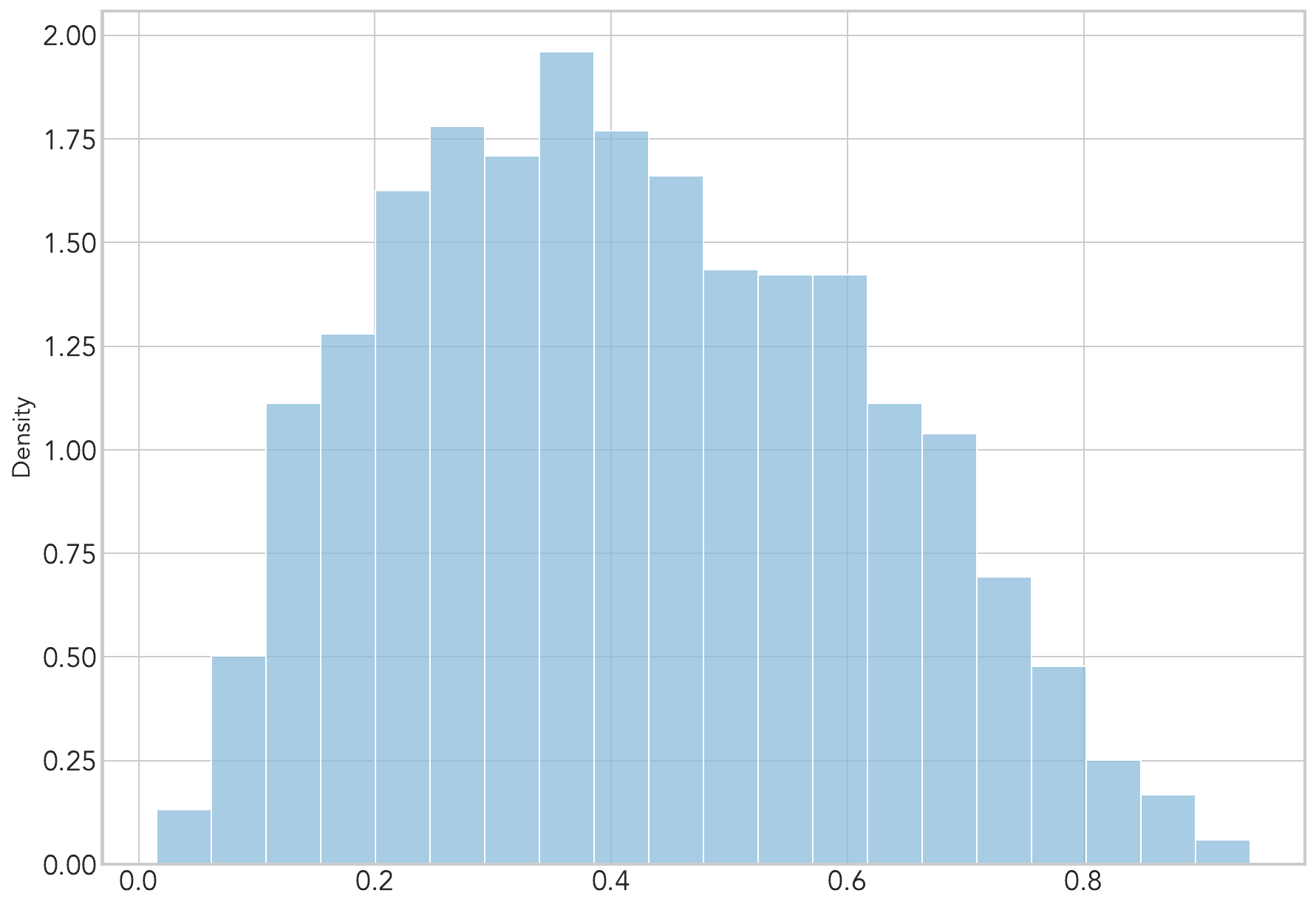} \\
			\end{tabular}
		}
		\label{common_support_offer}
	\end{center}
\begin{footnotesize}
	\begin{singlespace}		
		\justifying \noindent \textsc{Notes}: 
		This figure shows the distribution of propensity scores for human recruiter selection into the interview set, $p(I=1|X)$. Panel A plots the distribution for applicants selected for an interview by the SL model that predicts offer likelihood, $I^{SL}$. Panel B plots the distribution for applicants selected for an interview by the UCB model that predicts offer likelihood, $I^{UCB}$. All data come from the firm's application and hiring records.
	\end{singlespace}	
\end{footnotesize}
\end{figure}


\newpage
\clearpage
\section{IV Analysis} \label{asec:ivvalidity}

\subsection{IV details}

\begin{figure}[ht!]
\begin{center}
\captionsetup{justification=centering}
\caption{\textsc{Figure \ref{hist_iv}: Distribution of Interview Rates}}
\makebox[\linewidth]{\includegraphics[scale=1]{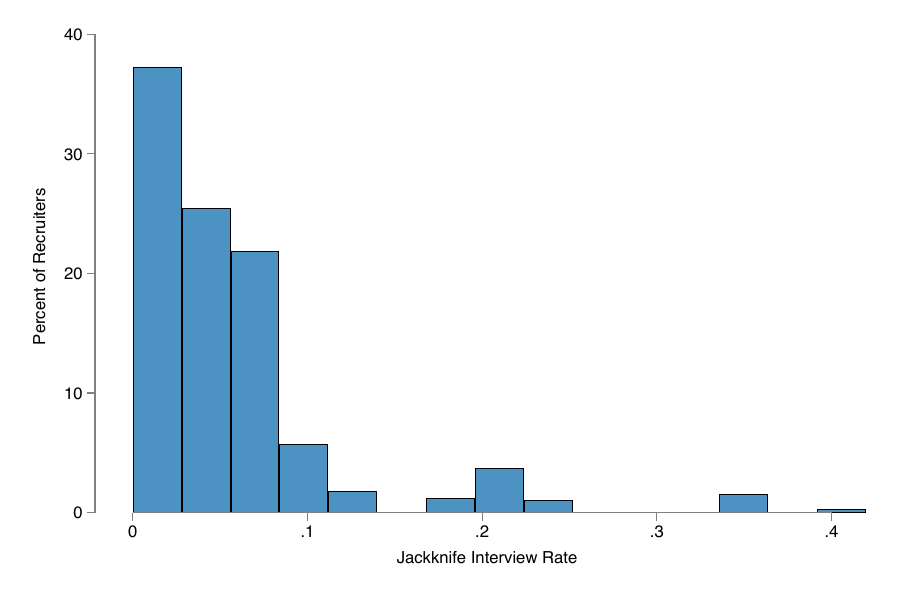} }\\
\label{hist_iv}
\end{center}
\begin{footnotesize}
\begin{singlespace}
\justifying \noindent \textsc{Notes}: This histogram shows the distribution of jack-knife interview rates for the 54 recruiters in our data who evaluate more than 50 applicants. This leave-out mean is defined as the share of interviews the assigned recruiter grants, excluding the focal applicant.  All data come from the firm's application and hiring records.
\end{singlespace}
\end{footnotesize}

\end{figure}

\newpage
\clearpage
\begin{figure}[ht!]
\begin{center}
\captionsetup{justification=centering}
\caption{\textsc{Figure \ref{I_ML_byleniency}}: Relationship between ML scores and interview likelihood by recruiter assignment}
\makebox[\linewidth]{
\begin{tabular}{c}
\textsc{\footnotesize{A. UCB Scores}} \\
\includegraphics[scale=0.8]{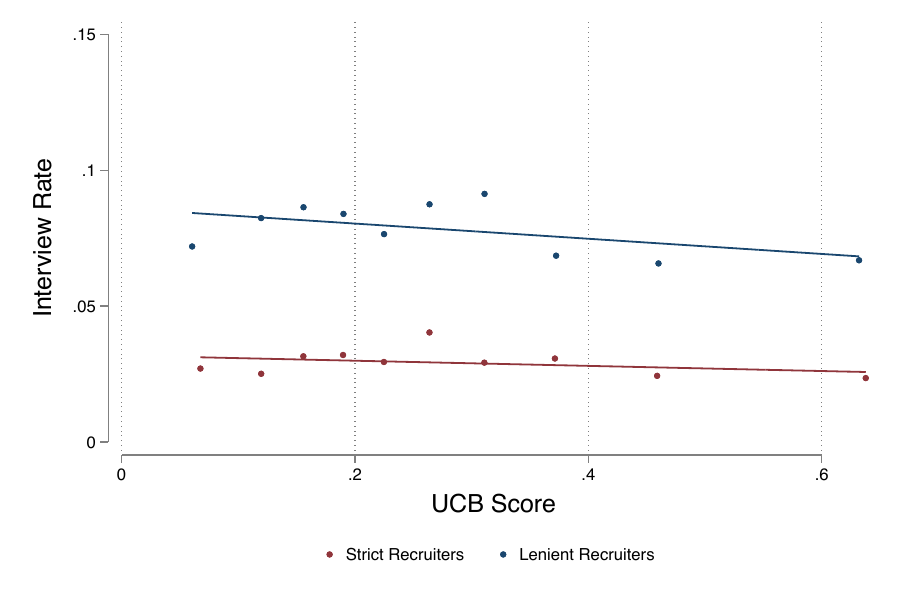}\\
\textsc{\footnotesize{B. SL Scores}} \\
\includegraphics[scale=0.8] {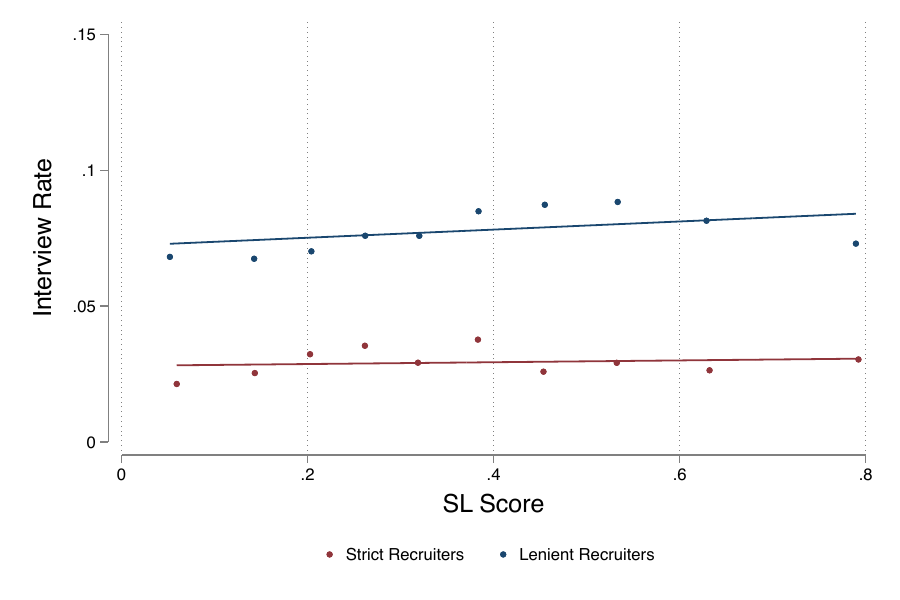}\\
\end{tabular}
}
\label{I_ML_byleniency}

\end{center}
\end{figure}
\begin{singlespace}
\begin{footnotesize}
\justifying \noindent \textsc{Notes}: These figures plot binned scatterplots of the relationship between ML scores on the x-axis and interview likelihood. ``Lenient'' and  ``Strict'' recruiters refer to those with an above and below median jack-knifed interview rate, respectively. This leave-out mean is defined as the share of interviews the assigned recruiter grants, excluding the focal applicant.  Panel A shows the relationship between interview rate and UCB score of hiring likelihood, plotted separately for applicants assigned to strict and lenient recruiters. Panel B plots the same analysis by our SL model score. We restrict our analysis to recruiters who evaluate more than 50 applicants. We control for work location and application year.  All data come from the firm's application and hiring records. 
\end{footnotesize}
\end{singlespace}

\newpage
\clearpage
\subsection{Monotonicity}


We are concerned about violations of monotonicity, that is, when lenient screeners have systematically different preference ordering of applicants, relative to strict screeners. 
 We examine this in several ways.  
 
First, following the literature, we examine whether the positive correlation between our leniency instrument and the likelihood of the interview differs between subgroups of our sample (following, for example, \citet{frandsenjudgeiv, leslieandpope, dobbieetal2}). Appendix Table \ref{ivtests1} shows that our leniency instrument is positively correlated with an applicant's interview likelihood and that positive correlation is similar across demographic and educational groups.  

Second, we examine the preferences of lenient and strict screeners directly. Using our training period sample (2016-2017), we build two models predicting an applicant's likelihood of being interviewed: one trained on data from lenient screeners and one model is trained from strict screeners. In Appendix Table \ref{ivtests2}, we show that the within-individual correlation between these two selection propensities in our analysis data (2018-2019) is high: applicants favored by strict reviewers are likely to be favored by lenient reviewers as well. 

Together, these results provide evidence that screeners have similar preferences over individuals in terms of their observed covariates.  However, it is still possible that strict recruiters screen applicants differently on the basis of their unobserveables.  To examine this, we train a model to predict hiring rates (conditional on interview) on a subsample of applicants who are selected to be interviewed by lenient recruiters (those with above median interview rates).  We then apply this model to hold out samples of applicants selected by strict recruiters as well as those selected by lenient recruiters. If strict and lenient recruiters are using unobservables in a similar way, then we would expect the relationship between actual hiring outcomes and model predictions to be similar among applicants selected by either strict or lenient recruiters \citep{mullainathan2018}.   

In Appendix Figure \ref{lenient_recruiter_model}, we plot the relationship between predicted hiring likelihood based on a model trained on applicants interviewed by lenient recruiters only.   If lenient recruiters were better at using unobservables, then a model of hiring yield trained on their decisions should overestimate the actual hiring yields of applicants selected by more strict reviewers.  Instead, we see that a model trained on the lenient subsample is well-calibrated to applicants selected by more strict screeners and mirrors the relationship in the full sample.

\newpage
\clearpage
\begin{table}[ht!]
	\begin{center}
		\caption{\textsc{Table \ref{ivtests1}: Average Monotonicity Test}}
		\vspace{20pt}        
		\scalebox{.9}{\makebox[\linewidth]{\input{Output/InstrumentTestsNov2021/IV_monotonicity_race_gender.tex}}}
		\label{ivtests1}
	\end{center}
	
	\begin{singlespace}
		\footnotesize
		\justifying \noindent \textsc{Notes}: This table presents results of regressing an applicant's interview status on their jack-knifed screener leniency instrument, by subgroups defined by race/ethnicity, education, referral and gender characteristics.  Each column presents the coefficient on the interviewer leniency variable for that subgroup regression.  All specifications include controls for job type, job seniority level, work location and application date. Standard errors are clustered at the screener level.  We find that being assigned to a lenient screener is, on average, positively related to the propensity to get an interview and the relationship is similar across demographic groups.  
	\end{singlespace}
	\normalsize
	
\end{table}

\newpage
\clearpage
\begin{table}[ht!]
	\begin{center}
		\caption{\textsc{Table \ref{ivtests2}: Correlation of Preferences of Lenient and Strict Screeners}}
		\vspace{20pt}        
		\scalebox{1}{\makebox[\linewidth]{\input{Output/InstrumentTestsNov2021/leniency_correlations_randomsplit.tex}}}
		\label{ivtests2}
	\end{center}
	
	\begin{singlespace}
		\footnotesize
		\justifying \noindent \textsc{Notes}: This table presents the results of another test of monotonicity.  Here, we predict two propensity scores for every applicant: that applicant's likelihood of being selected by a strict screener and that applicant's likelihood of being selected by a lenient screener.  We then examine the correlation between these two scores across subgroups in order to ask whether the preferences of lenient and strict screeners are correlated.  To do this, we randomly split our sample into a train and test set for applicants assigned to lenient or strict screeners (above and below the median jack-knifed leniency). We train a regularized logit model that predicts interview propensity for the set of strict screeners and a second model on lenient screeners. During our testing period, we generate an out-of-sample predicted probability of interview for all applicants using the strict interviewer model and the lenient screener model.  We find that the correlation between the predicted probability of interview under the lenient screener and the strict screeners is positive across race/ethnicity, gender, and education groups.
	\end{singlespace}
	\normalsize
	
\end{table}

\newpage
\clearpage
\begin{figure}[ht!]
\begin{center}
\captionsetup{justification=centering}
\caption{\textsc{Figure \ref{lenient_recruiter_model}: Calibration of Hiring Model Trained on Lenient Screeners}}
\makebox[\linewidth]{
			\begin{tabular}{c}
				\includegraphics[scale=0.7]{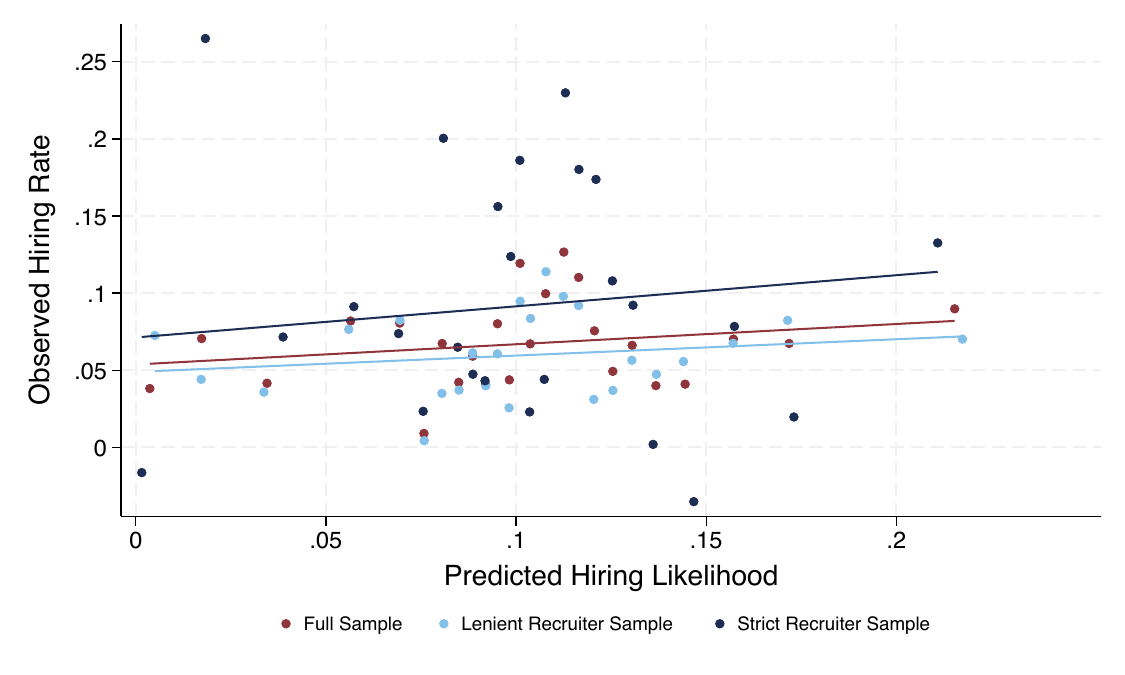} \\ 
			\end{tabular}
}\\
\label{lenient_recruiter_model}
\end{center}
\begin{footnotesize}
\begin{singlespace}
\justifying \noindent \textsc{Notes}: This figure tests if recruiters of different leniency are better at using unobservables in interview screening decisions than the less lenient recruiters. We train an algorithm using just the set of candidates interviewed by the most lenient recruiters, and use that algorithm to generate predicted hiring yield. We compare predicted hiring yield to observed hiring outcomes for three subsamples. The ``Full Sample'' line plots the lenient interviewer model predictions on the full test sample. The ``Lenient Recruiter Sample'' line plots the results for test sample applicants interviewed by lenient recruiters. The ``Strict Recruiter Sample'' line plots the results for the test sample interviewed by strict recruiters. If the lenient recruiters are better able to identify candidates with high-potential unobservables, the imputed hiring outcomes would be lower than the actual hiring outcomes within the set of people interviewed by strict recruiters. Yet, what we see is that the imputed and actual values are well calibrated across the strict recruiter sub-sample, lenient recruiter sub-sample, and the full sample. 
\end{singlespace}
\end{footnotesize}
\end{figure}

\newpage
\clearpage
\subsection{Compliers Analysis} \label{asec:compliers} 

To compute the characteristics of ``marginally'' interviewed candidates, we compute the average characteristics of instrument compliers---those who are marginal in the sense that they would be interviewed if assigned to a lenient recruiter, but not interviewed if they were assigned to a strict recruiter.  To obtain an estimate of these characteristics, we estimate regressions following \citet{benson2019} and \citet{abadie03}:
\begin{eqnarray}
Y_{it} \times I_{it} &=& \alpha_{0} + \alpha_{1} I_{it} + X'_{it}\alpha + \varepsilon_{it} \text{ if } s^{ML}(X'_{it}) \geq \tau  \label{eq:marginal1} \\
Y_{it} \times I_{it} &=& \beta_{0} + \beta_{1} I_{it} + X'_{it}\beta + \varepsilon_{it} \text{ if } s^{ML}(X'_{it}) < \tau  \label{eq:marginal2}
\end{eqnarray} 

In Equation \eqref{eq:marginal1}, $Y_{it} \times I_{it}$ is equal to applicant $i$'s hiring outcome at time $t$ if she is interviewed or to zero if she is not.  This regression is structured so that the OLS coefficient $\hat{\alpha}^{OLS}_{1}$ estimates the average hiring potential among all interviewed applicants with high ML scores.  The IV estimate $\hat{\alpha}^{IV}_{1}$, in contrast, is an estimate of the hiring potential among the high ML-score compliers: $E[Y_{it} | I_{it}^{Z_{it}=1} > I_{it}^{Z_{it}=0}, s^{ML} \geq \tau]$.  Similarly, $\hat{\beta}^{IV}_{1}$ in Equation \eqref{eq:marginal2} is the analogous estimate for low ML-score compliers: $E[Y_{it} | I_{it}^{Z_{it}=1} > I_{it}^{Z_{it}=0}, s^{ML} < \tau]$.  This logic is analogous to the idea that IV estimates identify a LATE amongst compliers.\footnote{In standard potential outcomes notation, the LATE effect is $E[Y^1-Y^0 | I^{Z=1} > I^{Z=0}]$.  In our case, we are only interested in the average potential outcome of compliers: $E[Y^1 | I^{Z=1} > I^{Z=0}]$.  Here, $Y^1$ is equivalent to a worker's hiring outcome if she is interviewed---this is what we have been calling quality, $H$. For a formal proof, see \citet{benson2019}.}
We present the results of this analysis in Section \ref{sec:ivres}.


\newpage
\clearpage
\section{Offer Model} \label{asec:offer}


\begin{figure}[ht!]
\begin{center}
\captionsetup{justification=centering}
\caption{\textsc{Figure \ref{corr_quality_offer}: Correlations between algorithm scores and offer likelihood}}
\makebox[\linewidth]{
\begin{tabular}{c}
\textsc{\footnotesize{A. Human}}  \\
\includegraphics[scale=0.55]{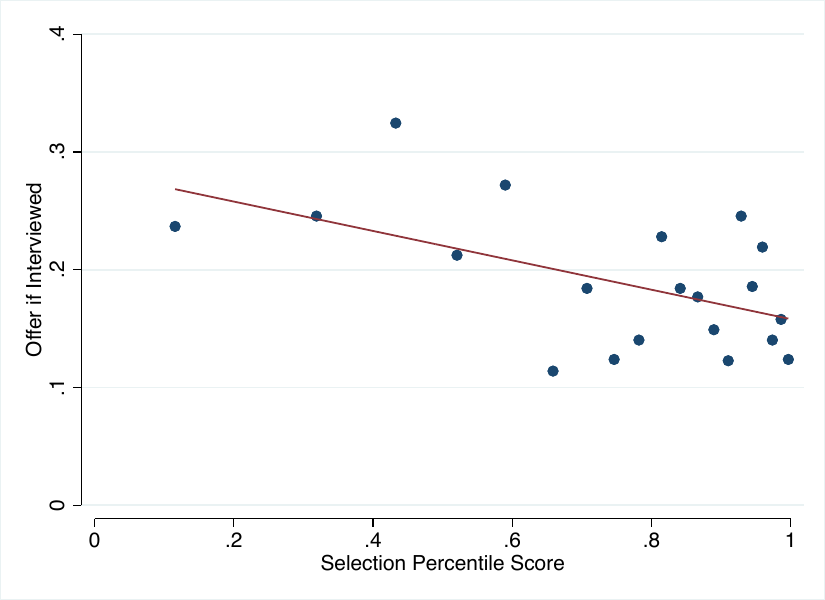} \\
\textsc{\footnotesize{B. SL Offer Model}} \\
\includegraphics[scale=.55]{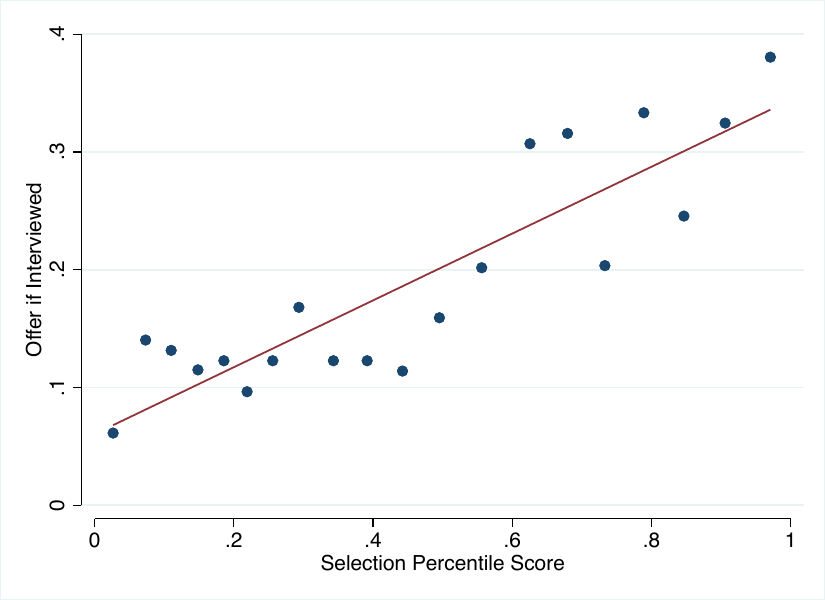} \\
\textsc{\footnotesize{D. UCB Offer Model}} \\
\includegraphics[scale=.55]{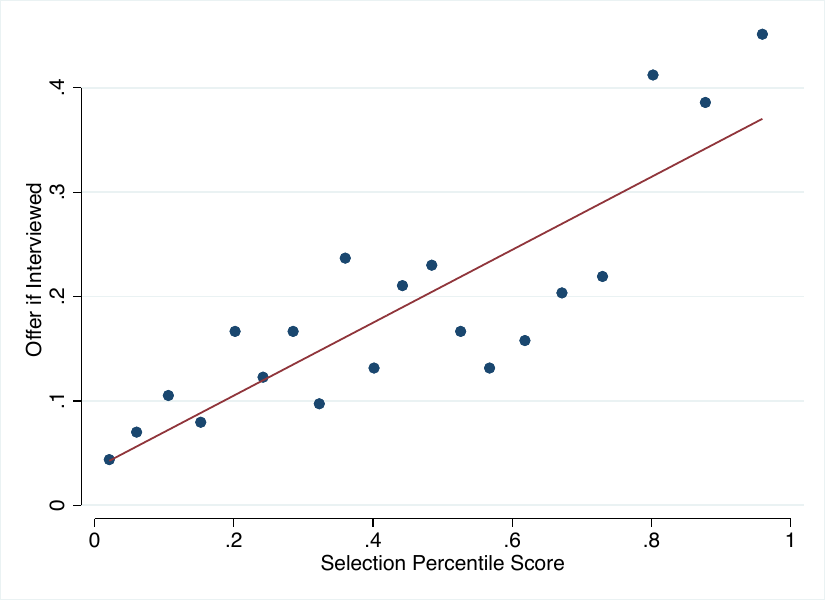} \\
\end{tabular}
}
\label{corr_quality_offer}
\end{center}
\begin{singlespace}
\begin{footnotesize} 
\justifying 
\noindent \textsc{Notes}: Each panel of this figure plots algorithm selection scores on the $x$-axis and the likelihood of an applicant receiving an offer if interviewed on the $y$-axis. Panel A shows the selection scores from an algorithm that predicts the firm's actual selection of which applicants to interview.  Panel B shows selection scores from the supervised learning algorithm described in Equation \eqref{SLrule} that predicts offer.  Panel C shows the selection scores from the UCB algorithm described in Equation \eqref{UCBrule} that predicts offer.  
\end{footnotesize}
\end{singlespace}
\end{figure}

\clearpage
\begin{figure}[ht!]
\begin{center}
\captionsetup{justification=centering}
\caption{\textsc{Figure \ref{iv_marginal_offered}: Characteristics of marginal interviewees---Offer Models}}

\makebox[\linewidth]{
\begin{tabular}{cc}
\textsc{\footnotesize{A. Offer Likelihood, SL}} & \textsc{\footnotesize{B.  Offer Likelihood, UCB}} \\
\includegraphics[scale=0.5]{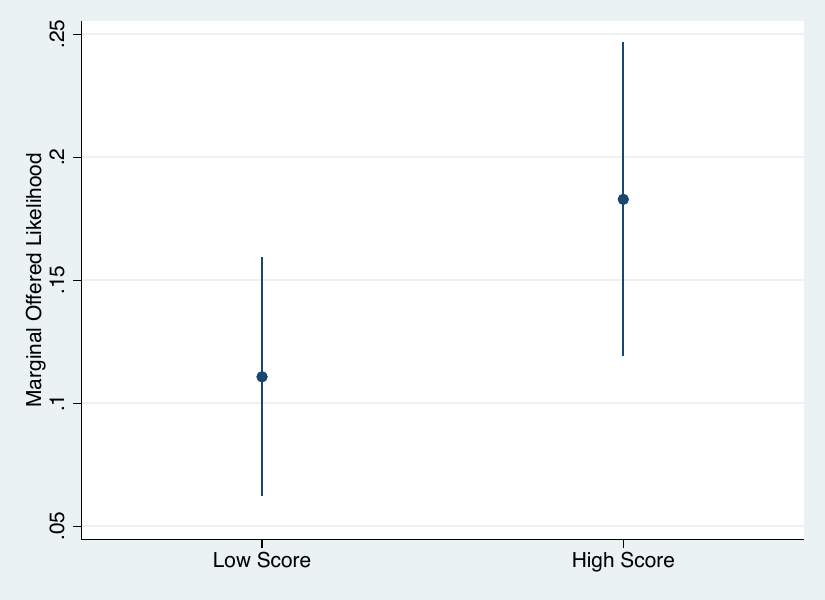} & \includegraphics[scale=0.5]{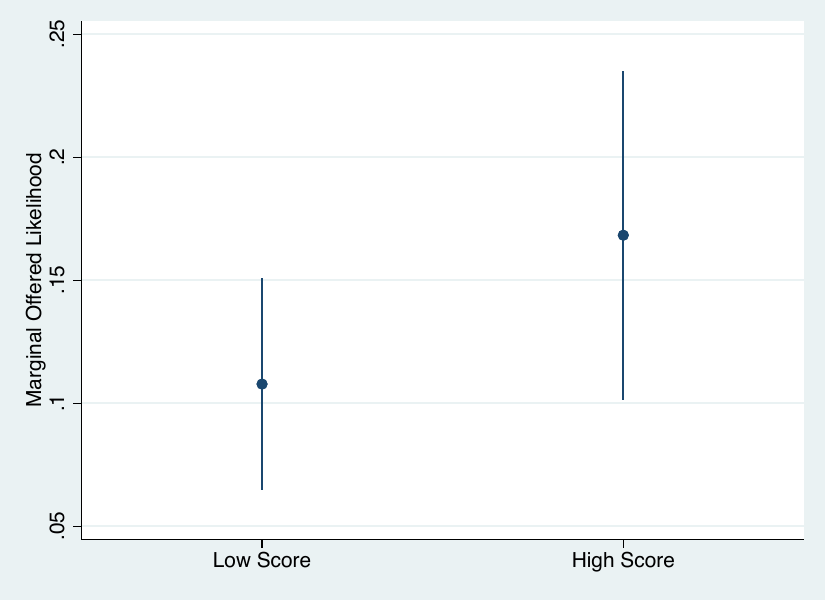}\\
\textsc{\footnotesize{C. Black/Hispanic, SL}} & \textsc{\footnotesize{D.  Black/Hispanic, UCB}} \\
\includegraphics[scale=0.5]{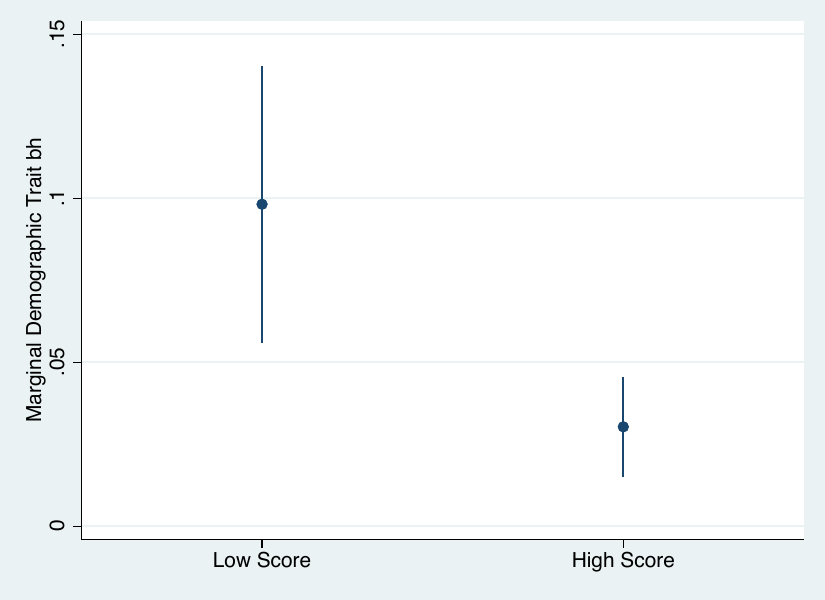} & \includegraphics[scale=0.5]{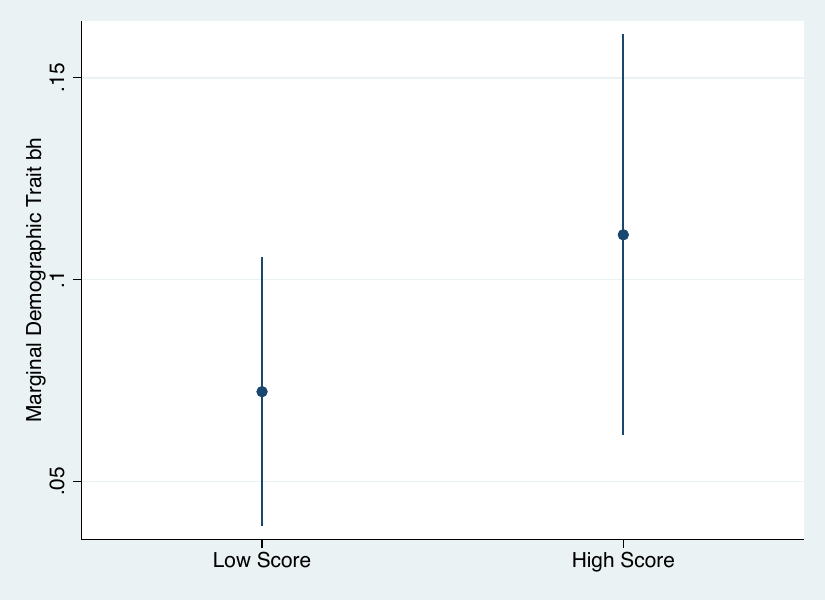}\\
\textsc{\footnotesize{E. Female, SL}} & \textsc{\footnotesize{F.  Female, UCB}} \\
\includegraphics[scale=0.5]{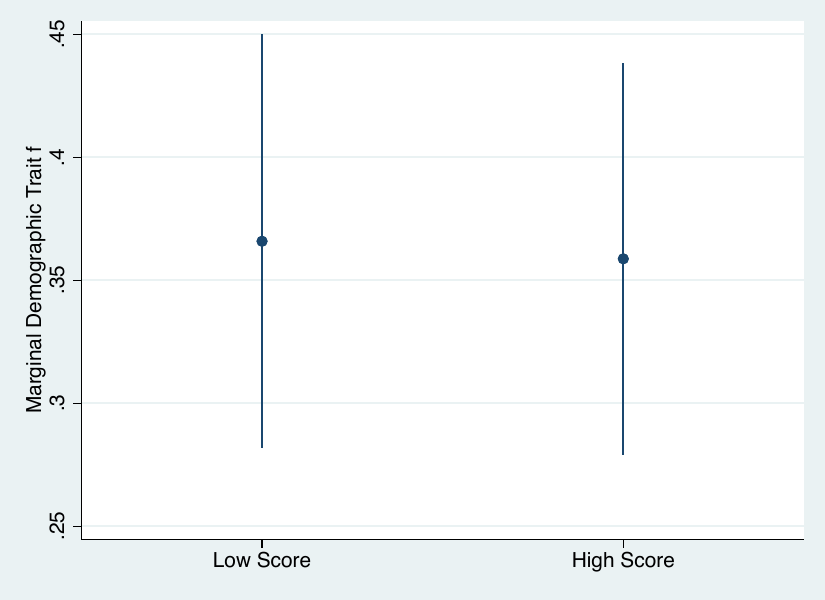} & \includegraphics[scale=0.5]{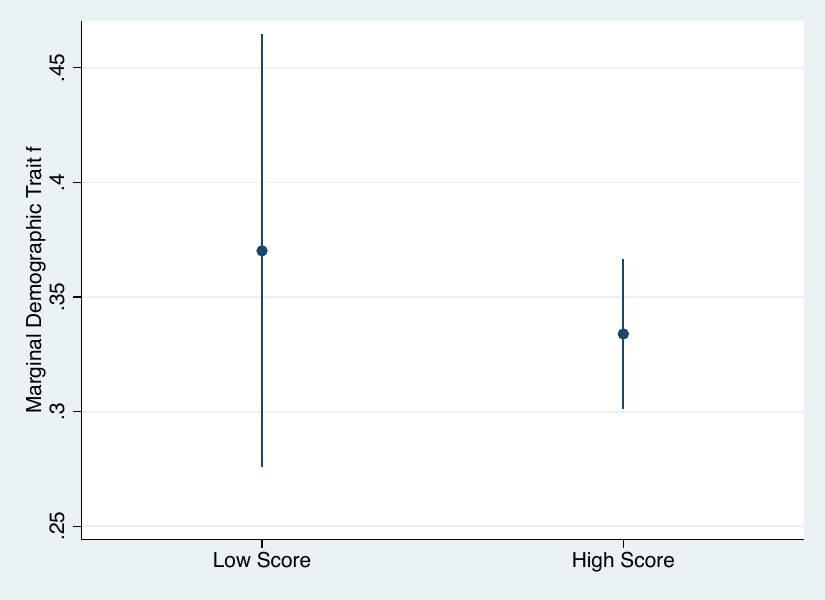}\\
\end{tabular}
}
\label{iv_marginal_offered}
\end{center}
\end{figure}
\begin{singlespace}
\begin{footnotesize}

\justifying 
\noindent \textsc{Notes}: Each panel in this figure shows the results of estimating the characteristics of applicants interviewed on the margin.  In each panel, these characteristics are estimated separately for applicants in the top and bottom half of the UCB algorithm's score.  Panels A, C, and E consider marginal applicants as defined by SL model scores.  Panels B, D, and F consider marginal applicants as defined by UCB model scores.  In Panels A and B, the $y$-axis is the average offer likelihood of marginally interviewed candidates; Panels C and D focus on the share of selected applicants who are Black or Hispanic; Panels E and F focus on the share of selected applicants who are female.  The confidence intervals shown in each panel are derived from robust standard errors clustered at the recruiter level.  
\end{footnotesize}
\end{singlespace}


\newpage
\clearpage
\section{Alternative Policies} \label{asec:altpolicies}

\subsection{Demographics Blinding}

\clearpage

\begin{figure}[ht!]
	\begin{center}
		\captionsetup{justification=centering}
		\caption{\textsc{Figure \ref{afig:blinding}: Change in Applicant Scores, Demographics Blind vs. Aware UCB model}}
		\makebox[\linewidth]{
			\begin{tabular}{c}
				\textsc{\footnotesize{A. Change in UCB Bonuses, by Demographics}}  \\
   	\includegraphics[scale=0.6]{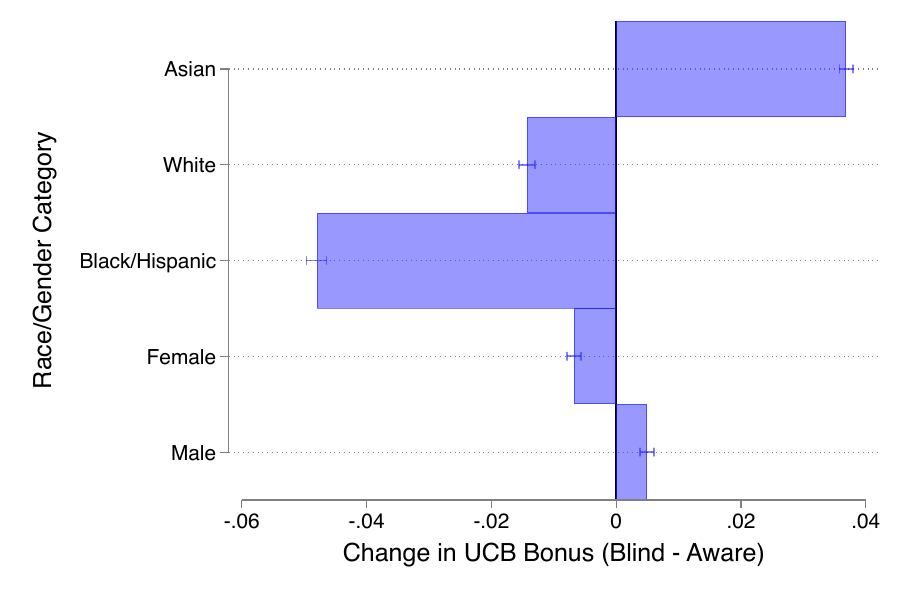} \\
				\textsc{\footnotesize{B. Returns to Attending a Top-Ranked School, by Demographic Blinding}}  \\
				\includegraphics[scale=0.6]{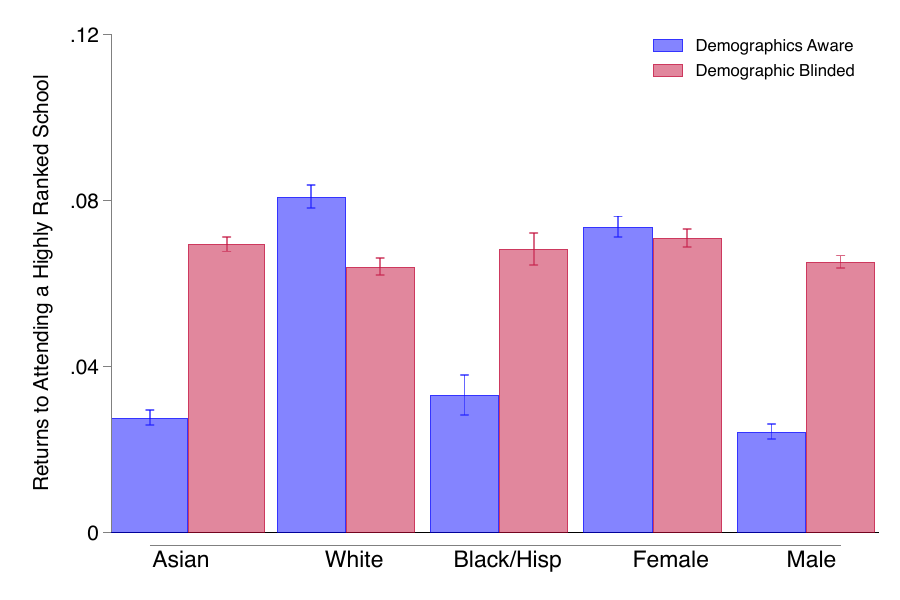} \\ 
			\end{tabular}
		}
		\label{afig:blinding}
	\end{center}
\begin{footnotesize}
	\begin{singlespace}		
		\justifying \noindent \textsc{Notes}: Panel A plots the coefficients and 95\% confidence intervals for aregression of applicant demographics on the change in UCB bonus, e.g. the difference in bonus between the demographics blind and demographics aware UCB models. A positive coefficient indicates that, on average, the demographic characteristic is associated with a positive change in bonuses. All regressions include controls for application year, management level, and job family.  Panel B plots the correlation between attending a highly ranked university and UCB model scores. ``Top-ranked'' is defined as attending a top 50-ranked US institution. The analysis is conducted separately for the demographically aware and demographics-blind UCB models.
	\end{singlespace}	
\end{footnotesize}
\end{figure}

\subsection{SL with Quota}

\clearpage
\begin{figure}[ht!]
	\begin{center}
		\captionsetup{justification=centering}
		\caption{\textsc{Figure \ref{afig_quotatiming}: Number of Black or Hispanic Candidates Selected, UCB versus SL with Quota}}
		\makebox[\linewidth]{\includegraphics[scale=.42]{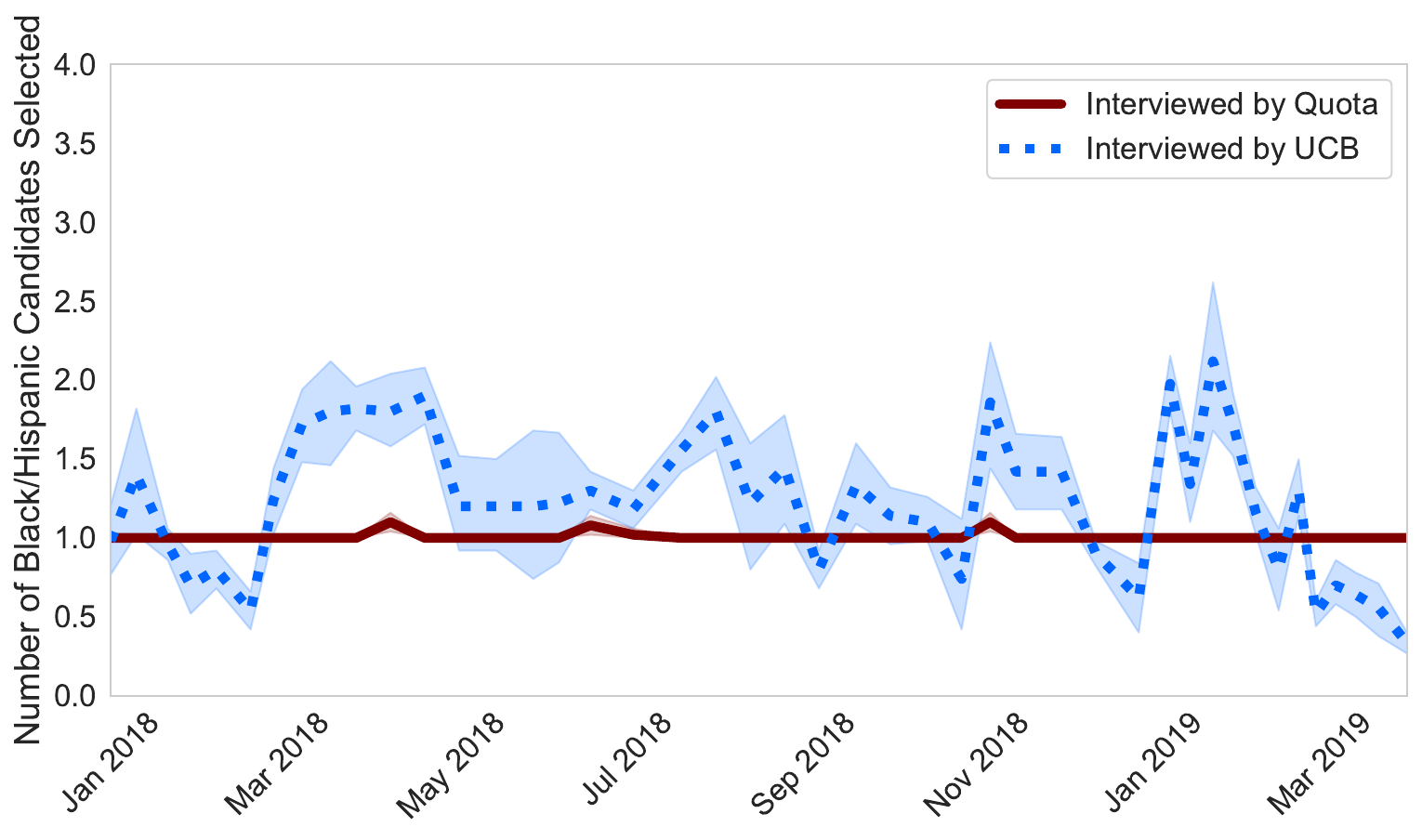}}
		\label{afig_quotatiming}
		
	\end{center}
\end{figure}
\begin{singlespace}
	\begin{footnotesize}
		\begin{singlespace}
			\vspace{-20pt}
			\justifying \noindent \textsc{Notes}:  This figure plots the number of Black or Hispanic applicants selected in each round for our SL with quota model. We use a supervised learning model to rank applicants each round and then select the top Black or Hispanic so the Black/Hispanic share of the interviewed pool matches the overall Black/Hispanic share of the applicant pool (13\%). We note the percentages may not always be exact because we must select discrete numbers of candidates, so the total share is not always exactly equal to the population share. We also plot the number of Black or Hispanic candidates selected by our UCB model. Both our SL and UCB models predicted hiring likelihood. In each round, both models are constrained to select the same number of candidates as human recruiters. For increased readability, we plot the rolling average over a five-period window.   
		\end{singlespace}
		
	\end{footnotesize}
\end{singlespace}


\newpage
\clearpage
\section{Time Dynamics} \label{asec:timing}

\subsection{Changes in Applicant Quality}

\newpage
\clearpage
\subsection{Simulations} \label{asec:simulations}

In our simulations, we take our existing applicants (e.g. keeping their covariates the same), but assign different values of hiring potential, $Y_{it}$.  For each simulation, we choose one racial group, $R$, to experience an increase (or decrease) in their hiring likelihood.  At the start of 2018, we assume that group $R$ applicants have the same average hiring likelihood as their true 2018 mean.  Because $Y_{it}$ is binary, we accomplish this by sampling from a binomial distribution with the given mean we are seeking to reproduce.
Over the course of 2018, we assume that their quality linearly increases from there so that, by the end of 2018, all incoming group $R$ candidates have $Y_{it}=1$.  In the meantime, we hold the quality of applicants from all other groups constant at their true 2018 mean.  

Importantly, we assign these values of hiring potential to \textit{all} applicants, regardless of whether they are interviewed in practice.  In this way, our simulation comes closer to a live-implementation in which we would be able to update our model with the hiring outcomes of all applicants selected by our models, not just the ones who were actually selected by human screeners in practice.  

To evaluate the capacity of our models to identify changes in applicant quality, we examine how they would assess a given group of applicants at different points in time (e.g. corresponding to different beliefs the model may have given the state of its training data). Specifically, we use the actual set of candidates who submitted applications from January 2019 to April 2019 and calculate their ML model scores at distinct points throughout 2018. By maintaining a constant evaluation cohort, we can identify score changes that result from the algorithm's learning and exploration over time, rather than from fluctuations in the applicant pool itself.

Figure \ref{learning_black} of the main text presents the results of one particular simulation, the one in which we assume Black candidates increase in quality.  Here, we present the full set of simulations: increases in quality for each race/ethnic group, as well as decreases in quality for each race/ethnic group.

\begin{figure}[ht!]
	\begin{center}
		\captionsetup{justification=centering}
		\caption{\textsc{Figure \ref{learning_good}: Increased quality simulations: demographics of selected candidates}}
		\makebox[\linewidth]{
			\begin{tabular}{cc}
				\textsc{\footnotesize{A. Black}} & \textsc{\footnotesize{B. Hispanic}}  \\
				\includegraphics[scale=0.25]{Data/Revision/AnalysisData/learning_simulation/lasso_fill_mean_slow_q440_good_candidate_black/learning_examples/V4_selection_candidate_black.png}  &  \includegraphics[scale=0.25]{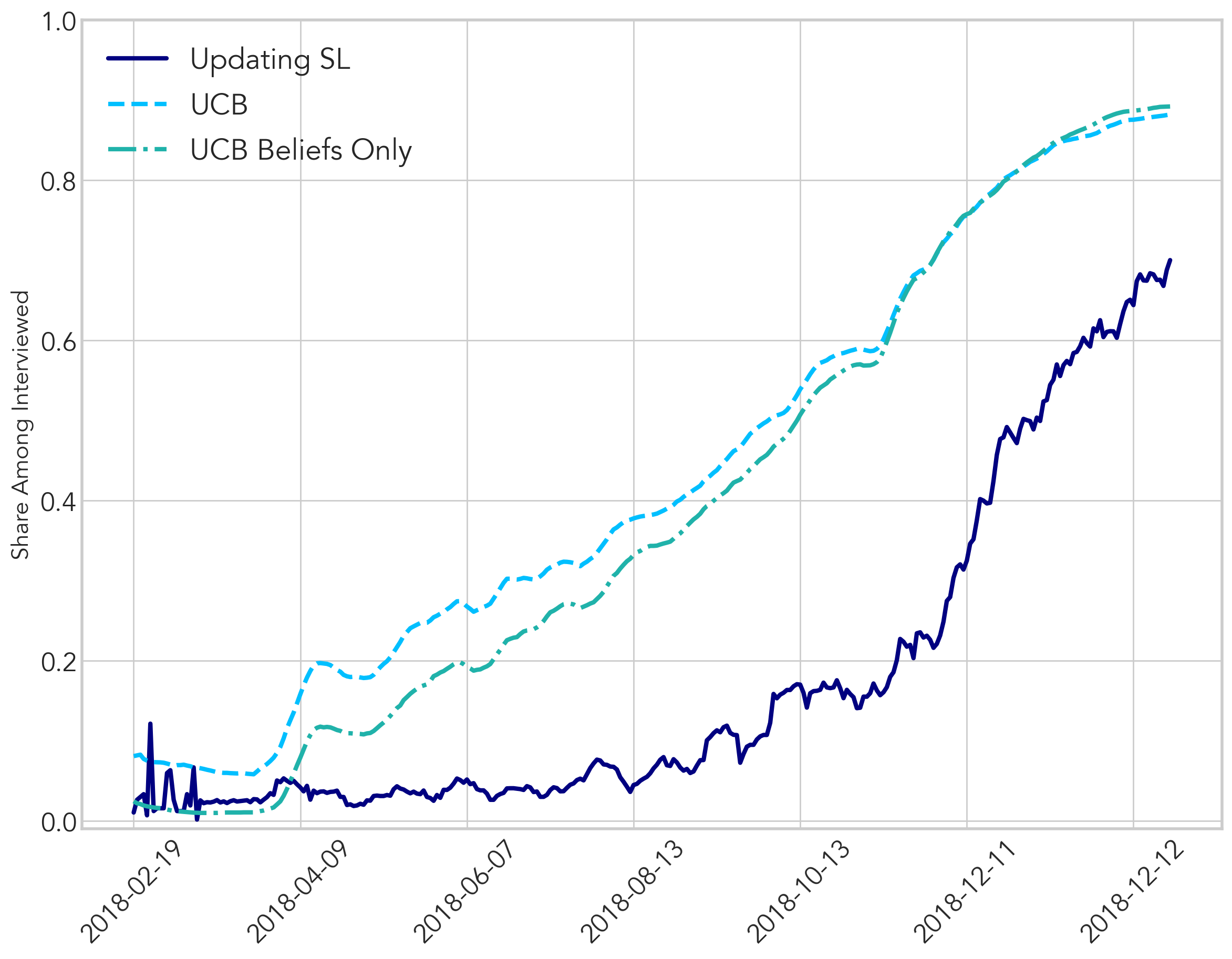} \\
				\textsc{\footnotesize{C. Asian}} & \textsc{\footnotesize{D. White}}  \\
				\includegraphics[scale=0.25]{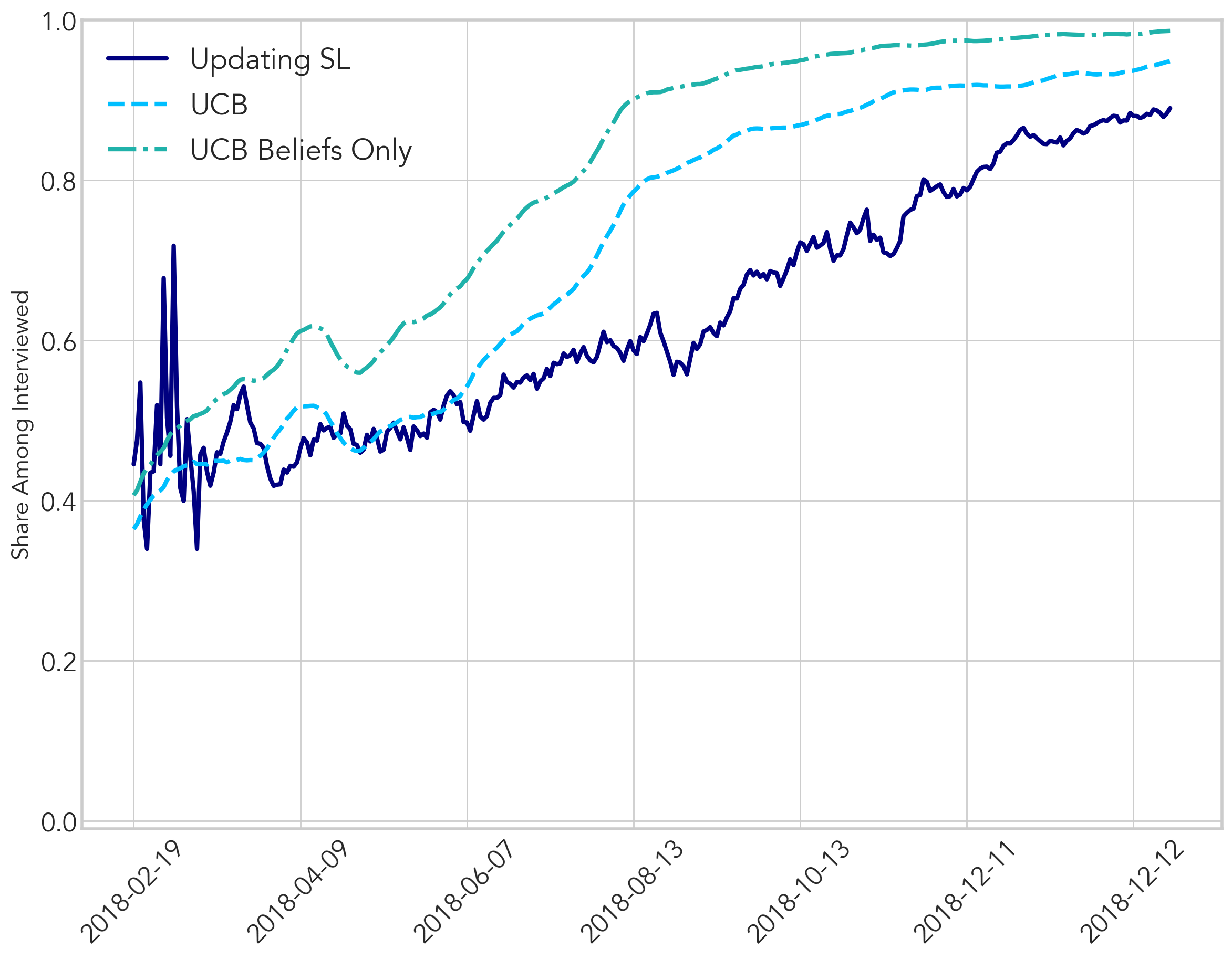} &  \includegraphics[scale=0.25]{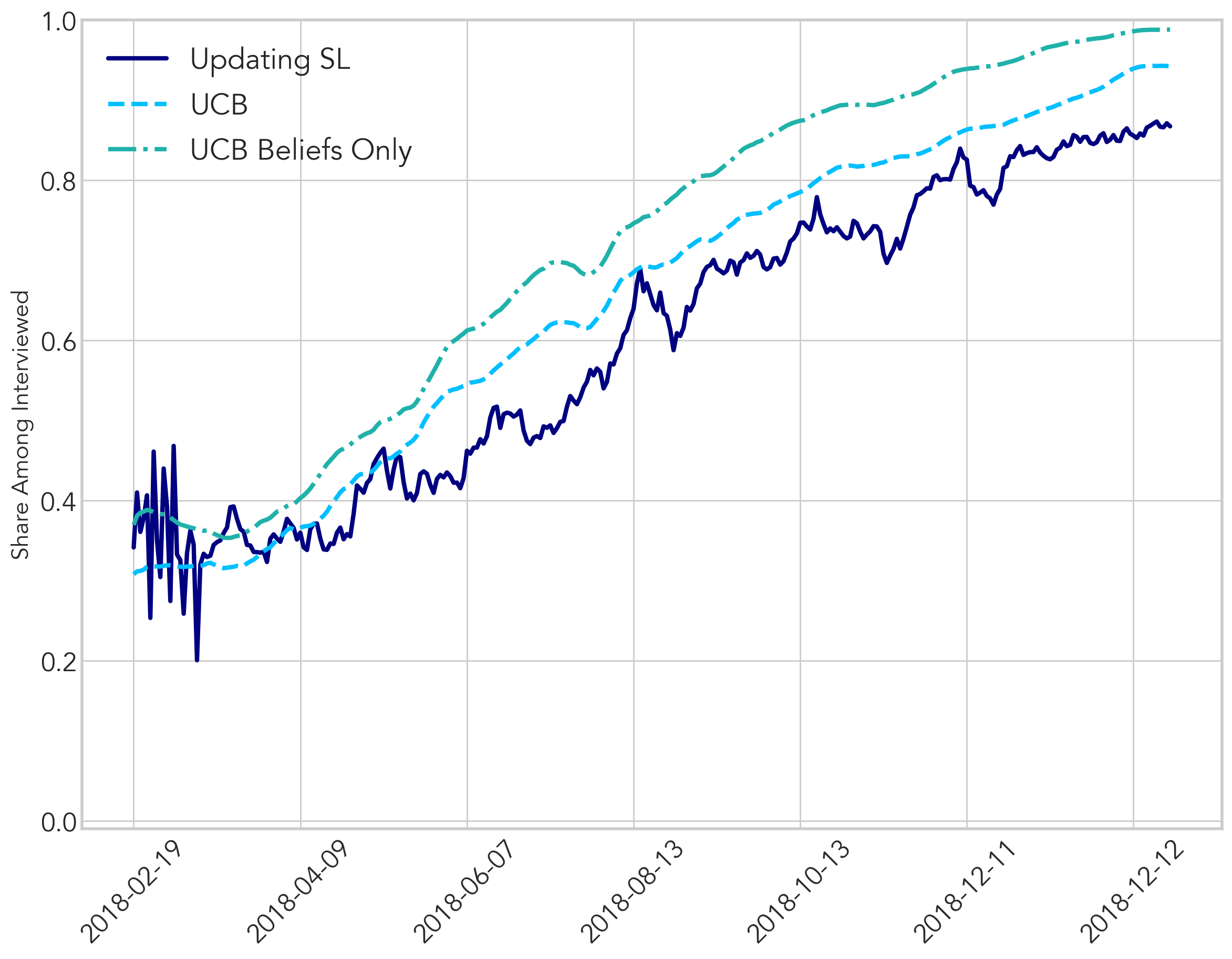} \\
			\end{tabular}
		}
		\label{learning_good}
	\end{center}
\end{figure}

\begin{figure}[ht!]
	\begin{center}
		\captionsetup{justification=centering}
		\caption{\textsc{Figure \ref{learning_good_quality}: Increased quality simulations: hiring yield among selected applicants}}
		\makebox[\linewidth]{
			\begin{tabular}{cc}
				\textsc{\footnotesize{A. Black}} & \textsc{\footnotesize{B. Hispanic}}  \\
				\includegraphics[scale=0.25]{Data/Revision/AnalysisData/learning_simulation/hired_race_aware_feb2023/quality_over_time_lasso_fill_mean_slow_q440_good_candidate_black.png}  &  \includegraphics[scale=0.25]{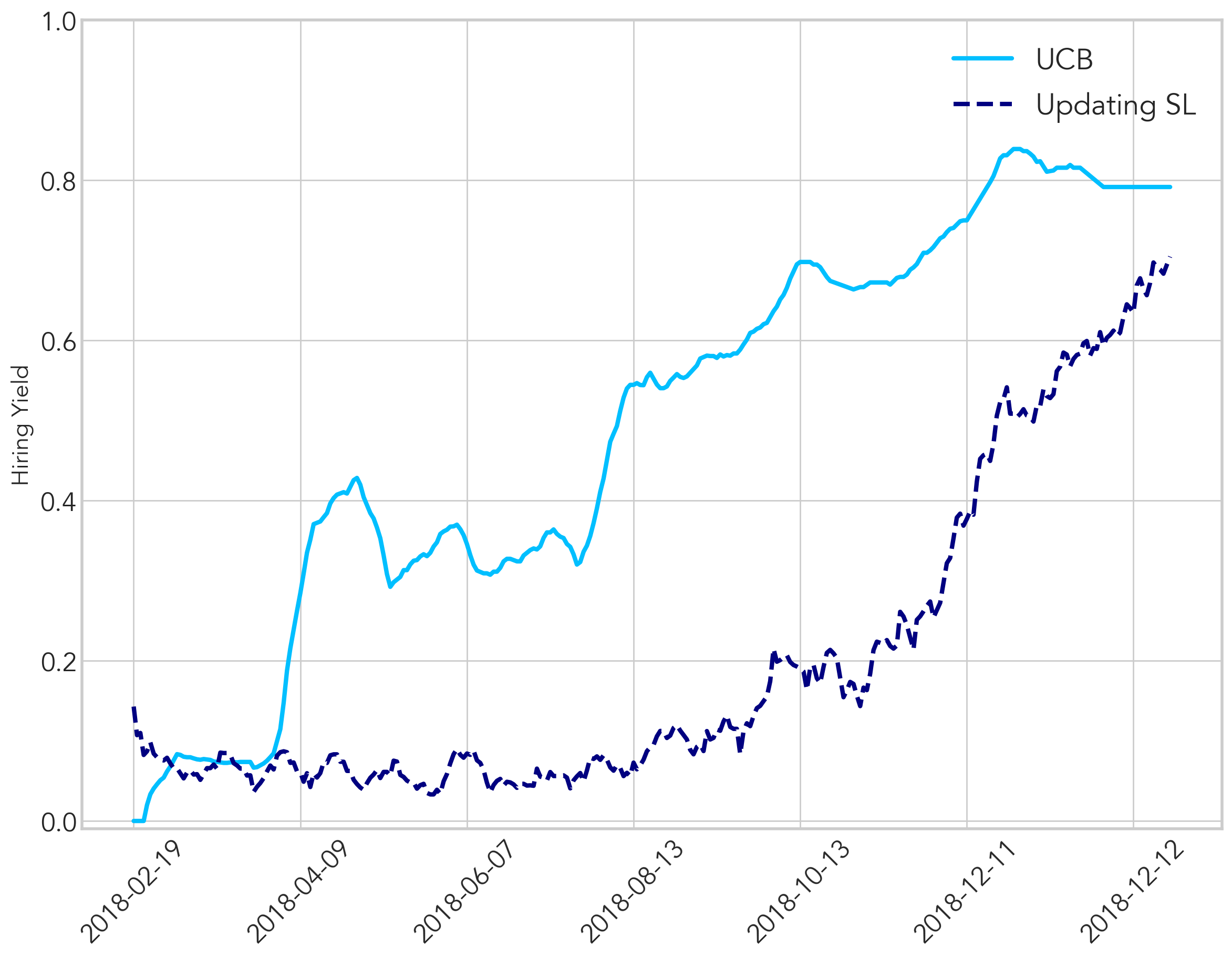} \\
				\textsc{\footnotesize{C. Asian}} & \textsc{\footnotesize{D. White}}  \\
				\includegraphics[scale=0.25]{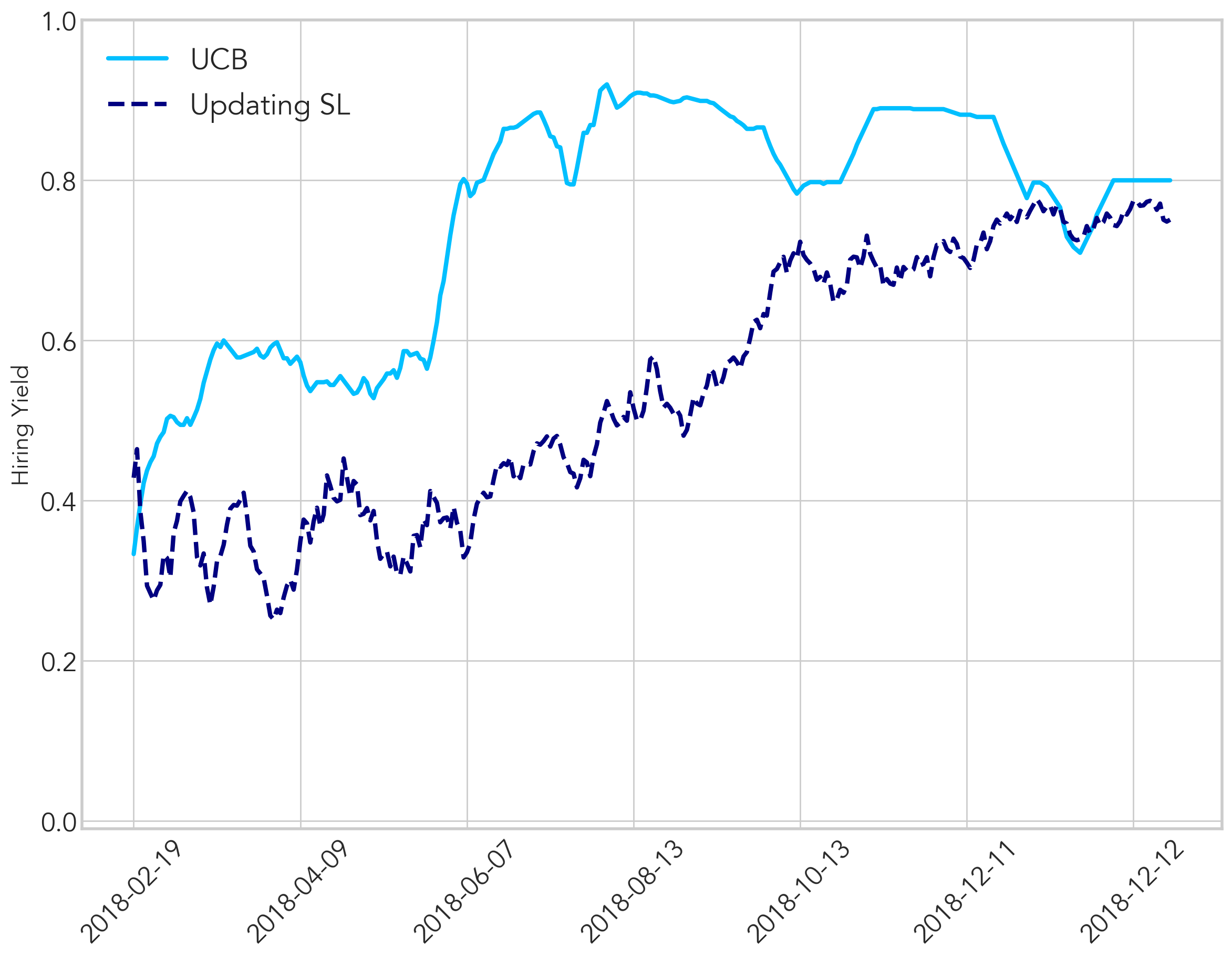} &  \includegraphics[scale=0.25]{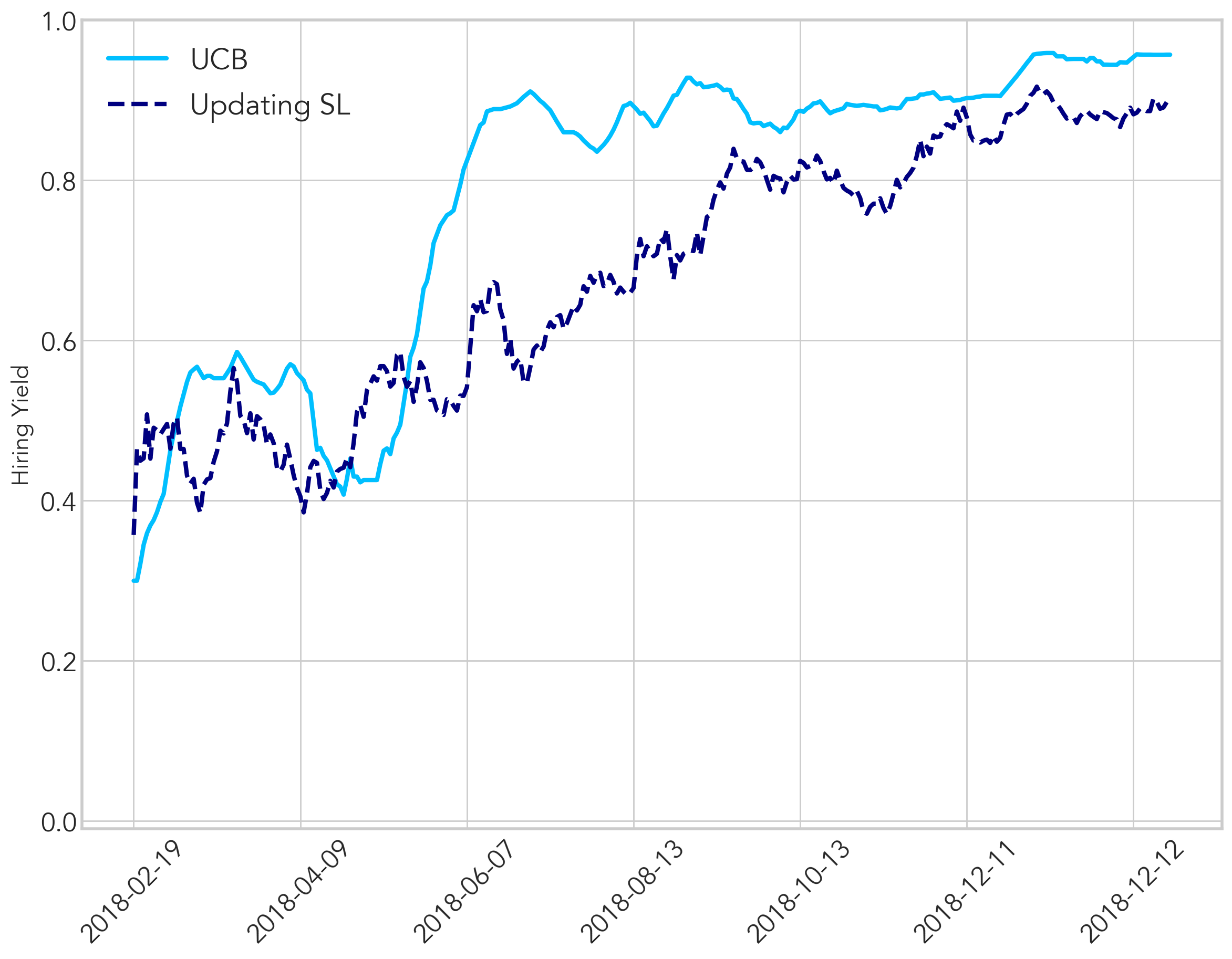} \\
			\end{tabular}
		}
		\label{learning_good_quality}
	\end{center}
\end{figure}

Appendix Figures \ref{learning_good} and \ref{learning_good_quality} present results for simulations in which applicant quality increases, with Appendix \ref{learning_good} focusing on changes in the demographics of candidates selected in each simulation and Appendix \ref{learning_good_quality} capturing changes in the hiring potential of selected candidates.  

Panel A of each figure reproduces the results for the simulation discussed in the main text.  Panel B of Appendix Figures \ref{learning_good} and \ref{learning_good_quality} present the results of the analogous simulation in which we increase the hiring potential of Hispanic, rather than Black, candidates.  Our results are broadly similar: the UCB model increases its selection of Hispanic applicants more quickly than an SL model, and the quality of its decisions, as measured by hiring yield increases more quickly as well.  In this case, however, the difference in learning speed is less stark than for the simulation increasing the quality of Black applicants because the baseline SL model selects a higher share of Hispanic applicants and so is able to detect the change in their quality more quickly.

Panels C and D of both figures plot outcomes for simulations in which the quality of Asian and White applicants increases, respectively.  Because these groups are already well represented in our training data, our results are slightly different.  Here, the SL model learns much more quickly about changes in the quality of Asian and White applicants because it selects a large number of candidates from these groups at baseline, making it easier to pick up on changes in their quality.  Another feature to note in Panels C and D of Figure \ref{learning_good} is that there is a large gap between the UCB model's beliefs and its selection choices: the UCB algorithm learns very quickly about increases in the quality of Asian and White applicants but does not initially select as many of these candidates.  This occurs because the UCB model is hesitant to exclusively select members of a large group (White or Asian), having seen very few Black and Hispanic applicants. In contrast, the UCB is much more willing to exclusively select Black or Hispanic applicants in the simulation results from Panels A and B because it already has more certainty about the quality of White or Asian applicants.

In Appendix Figures \ref{learning_bad} and \ref{learning_bad_quality}, we present analogous results from simulations in which we decrease quality.  When we do this for Black and Hispanic applicants, the UCB's beliefs fall very quickly.  However, because Black and Hispanic candidates continue to be so rare in the data, the UCB model continues to select a small number of these candidates, in order to continue exploring, even as their overall share among those selected trends down over time.  This is an example of how the UCB model trades off immediate gains in hiring yield for the option value of increased learning in the future.  When we do the same for White or Asian candidates, both the SL and UCB models reduce the share of such applicants that they select at approximately the same rate; this is because both models have already seen a large number of applicants from these groups.


\clearpage
\begin{singlespace}
	\begin{footnotesize}

\begin{singlespace}

\justifying \noindent \textsc{Notes}: This figure shows the demographics of applicants recommended for interviews under three different algorithmic selection strategies: SL, UCB, and the beliefs component of UCB (that is, the $\hat{E}_t[H|X; D^{UCB}_{t}]$ term in Equation \eqref{UCBrule}).  In each panel, the $y$-axis graphs a demographic group share of ``evaluation cohort'' (2019) applicants who would be selected under each simulation.  Panel A plots the Black share of interviewed applicants in a simulation in which the hiring potential of Black candidates increases linearly over the course of 2018, as described in Section \ref{sec:learning}.  Panel B plots the Hispanic share of interviewed applicants in a simulation in which the hiring potential of Hispanic candidates in 2018 increases.  Similarly, Panels C and D show the White and Asian shares from simulations in which the hiring potential of White and Asian applicants increases, respectively.  
\end{singlespace}

\end{footnotesize}
\end{singlespace}


\clearpage
\begin{singlespace}
	\begin{footnotesize}
		
		\begin{singlespace}
			\justifying \noindent \textsc{Notes}: This figure shows the hiring yield under three different algorithmic selection strategies: SL, UCB, and the beliefs component of UCB (that is, the $\hat{E}_t[H|X; D^{UCB}_{t}]$ term in Equation \eqref{UCBrule}).  In each panel, the $y$-axis graphs the share of interviewed applicants who are hired. Panel A plots the results from a simulation in which the hiring potential of Black candidates increases linearly over the course of 2018, as described in Section \ref{sec:learning}.  Panel B shows results from a simulation in which the hiring potential of Hispanic candidates in 2018 increases linearly.  Similarly, Panels C and D show results from simulations in which the hiring potential of White and Asian applicants increases, respectively.  
		\end{singlespace}
		
	\end{footnotesize}
\end{singlespace}

\clearpage
\begin{figure}[ht!]
\begin{center}
\captionsetup{justification=centering}
\caption{\textsc{Figure \ref{learning_bad}: Decreased quality simulations: demographics of selected candidates}}
\makebox[\linewidth]{
\begin{tabular}{cc}
\textsc{\footnotesize{A. Black}} & \textsc{\footnotesize{B. Hispanic}}  \\
\includegraphics[scale=0.25]{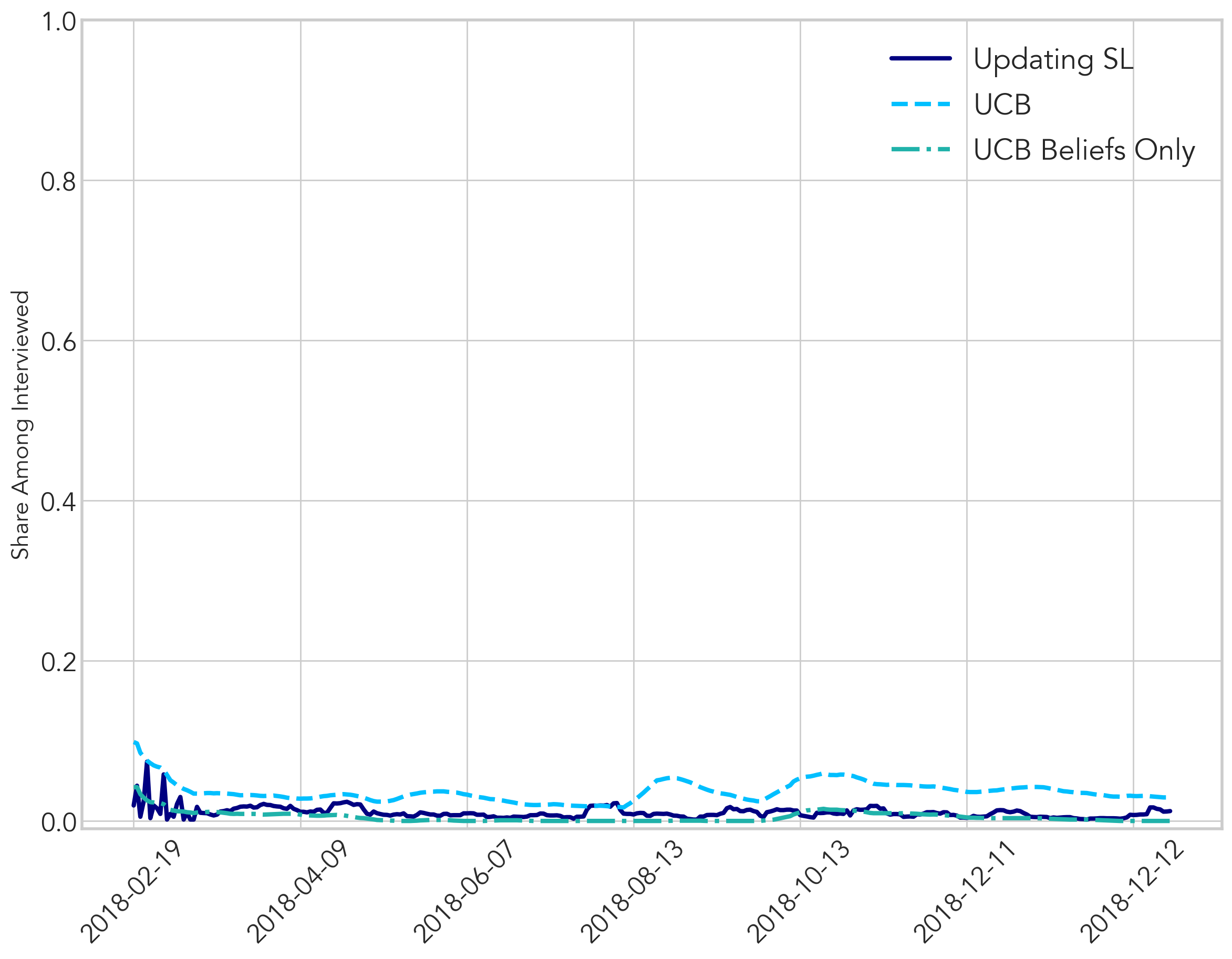}  &  \includegraphics[scale=0.25]{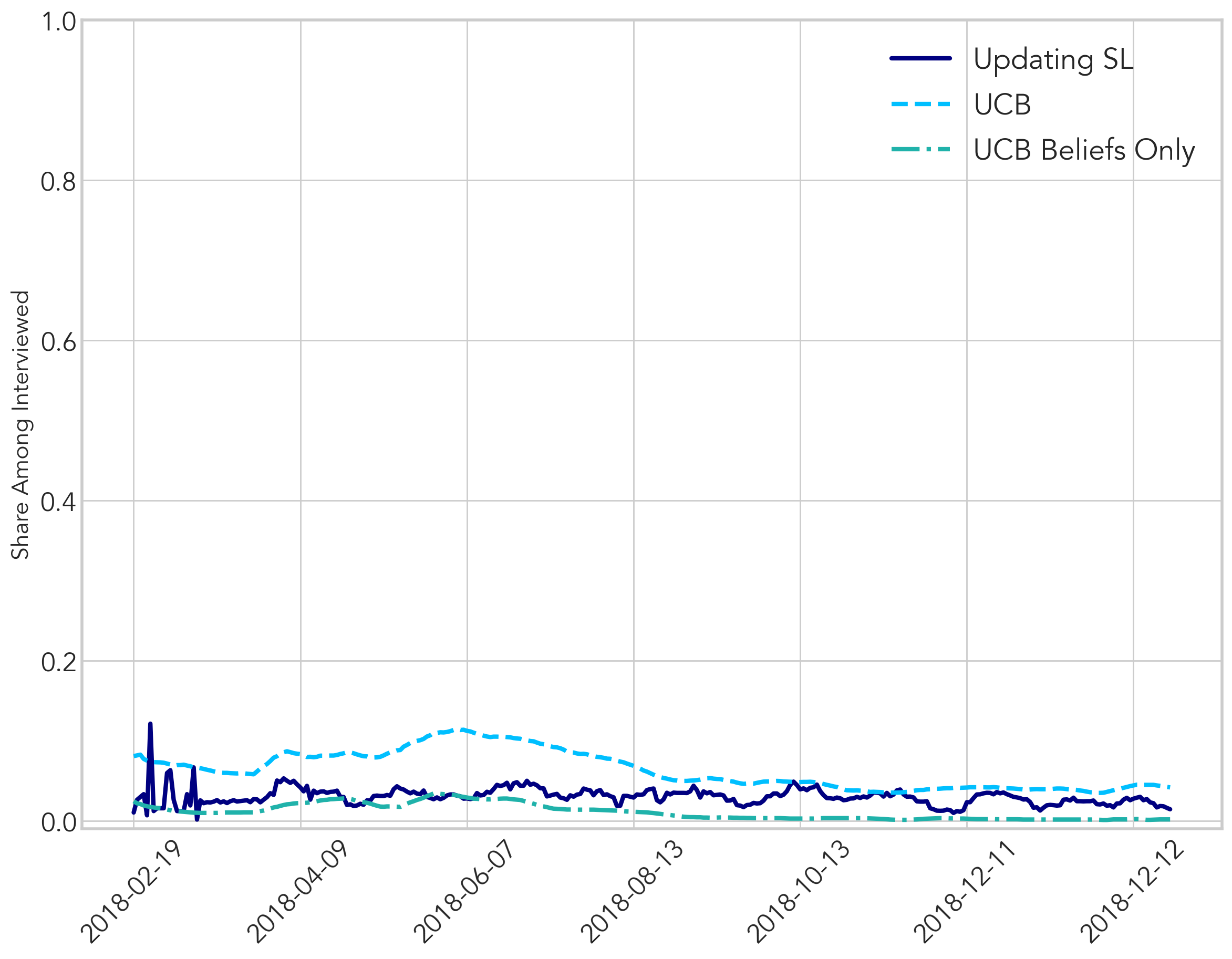} \\
\textsc{\footnotesize{C. Asian}} & \textsc{\footnotesize{D. White}}  \\
\includegraphics[scale=0.25]{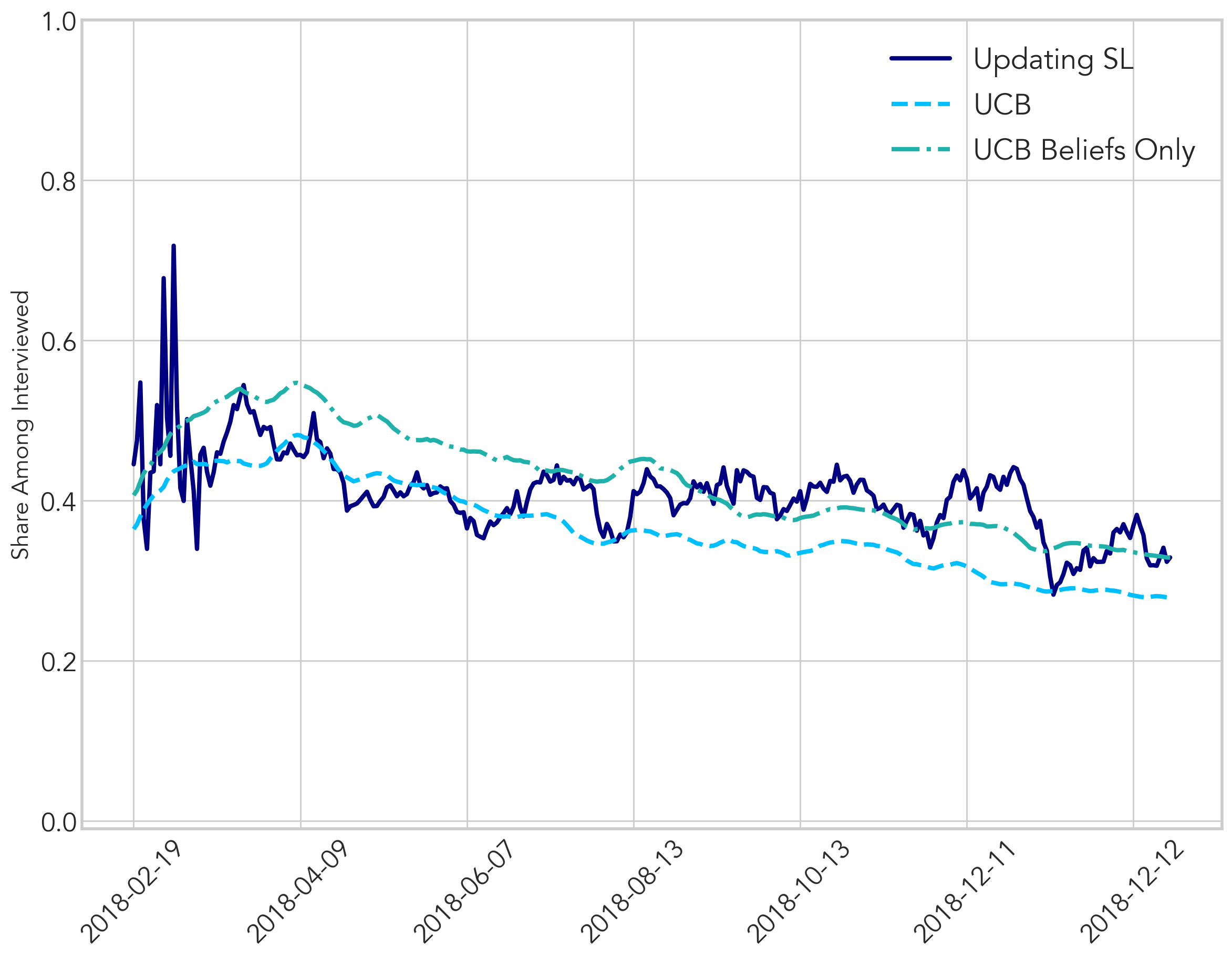} &  \includegraphics[scale=0.25]{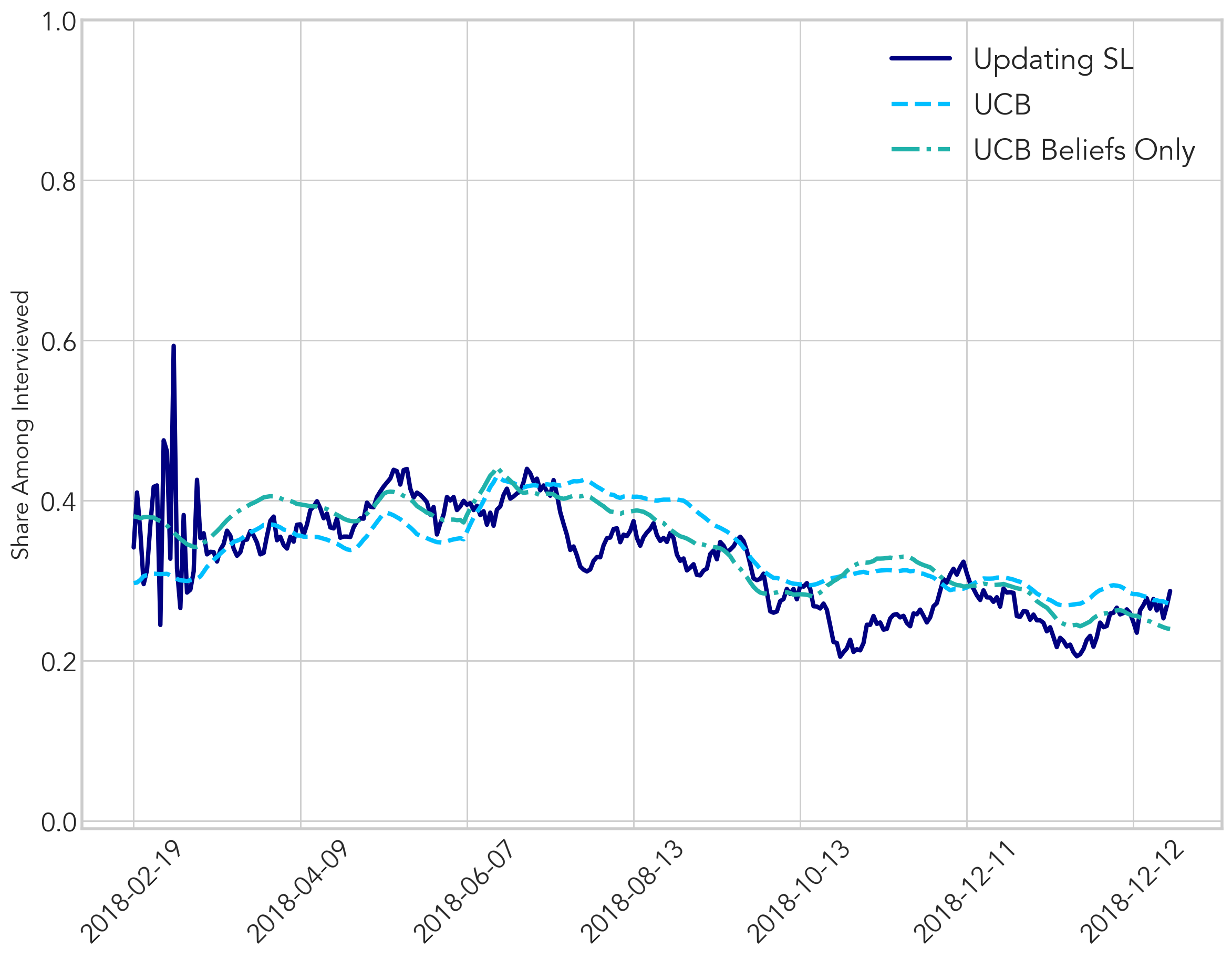} \\
\end{tabular}
}
\label{learning_bad}
\end{center}
\end{figure}
\begin{footnotesize}
\begin{singlespace}		
			\justifying \noindent \textsc{Notes}: 
This figure shows the demographics of applicants recommended for interviews under three different algorithmic selection strategies: SL, UCB, and the beliefs component of UCB (that is, the $\hat{E}_t[H|X; D^{UCB}_{t}]$ term in Equation \eqref{UCBrule}).  In each panel, the $y$-axis graphs a demographic group share of ``evaluation cohort'' (2019) applicants who would be selected under each simulation.  Panel A plots the Black share of interviewed applicants in a simulation in which the hiring potential of Black candidates decreases linearly over the course of 2018, as described in Section \ref{sec:learning}.  Panel B plots the Hispanic share of interviewed applicants in a simulation in which the hiring potential of Hispanic candidates in 2018 falls linearly.  Similarly, Panels C and D show the White and Asian shares from simulations in which the hiring potential of White and Asian applicants linearly decreases, respectively.  

					\end{singlespace}	
	\end{footnotesize}


\clearpage

\begin{figure}[ht!]
	\begin{center}
		\captionsetup{justification=centering}
		\caption{\textsc{Figure \ref{learning_bad_quality}: Decreased quality simulations: hiring yield of selected candidates}}
		\makebox[\linewidth]{
			\begin{tabular}{cc}
				\textsc{\footnotesize{A. Black}} & \textsc{\footnotesize{B. Hispanic}}  \\
				\includegraphics[scale=0.25]{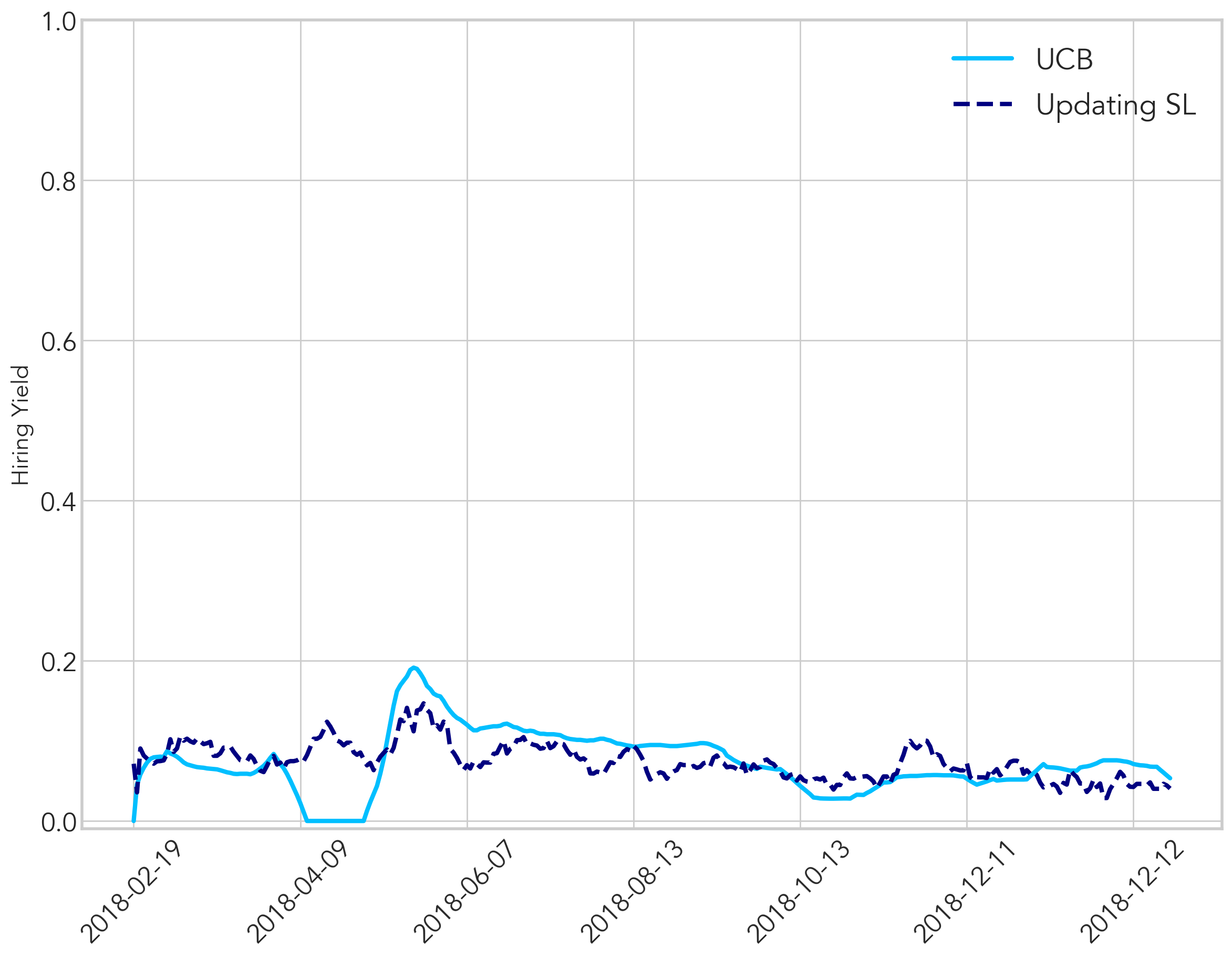}  &  \includegraphics[scale=0.25]{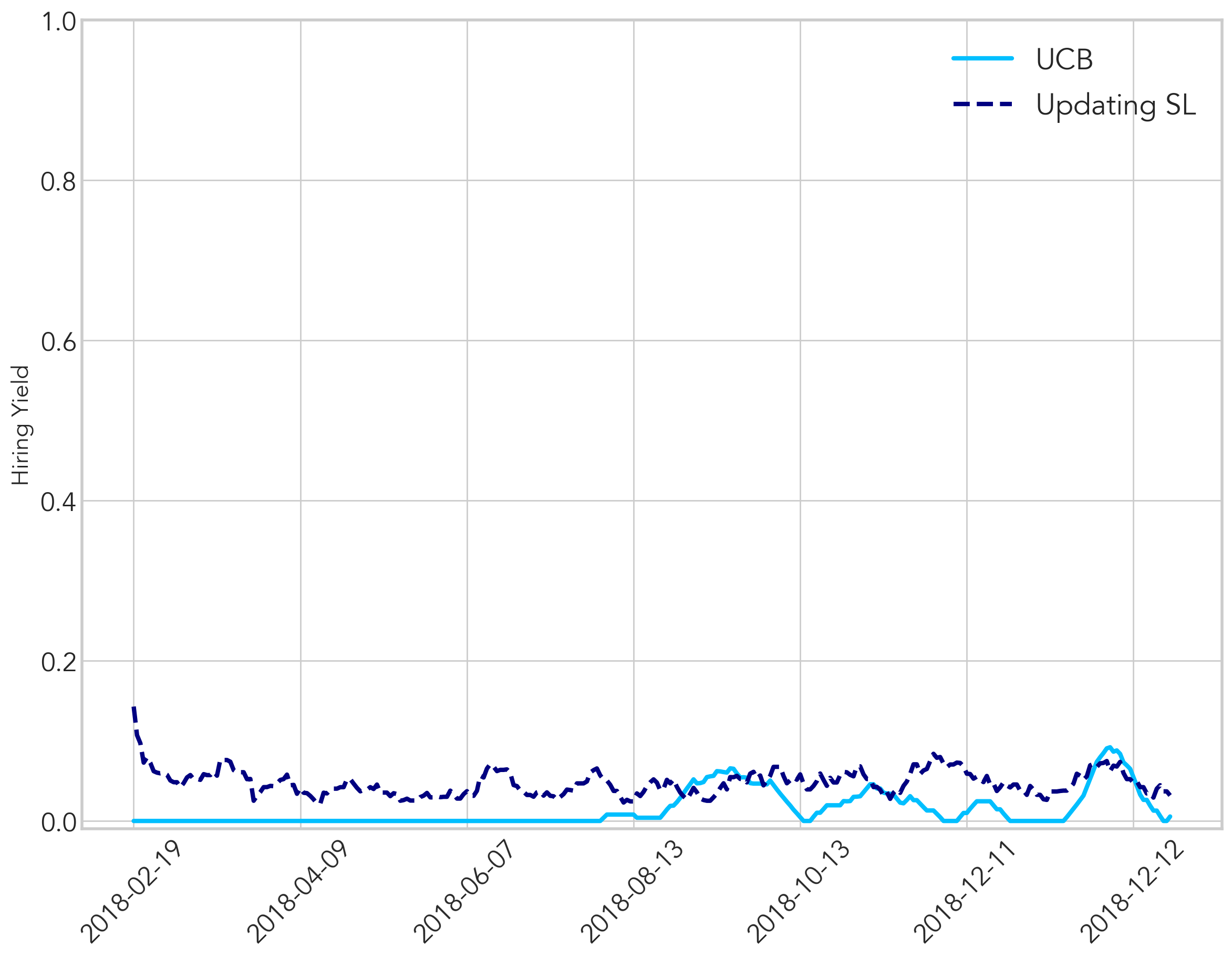} \\
				\textsc{\footnotesize{C. Asian}} & \textsc{\footnotesize{D. White}}  \\
				\includegraphics[scale=0.25]{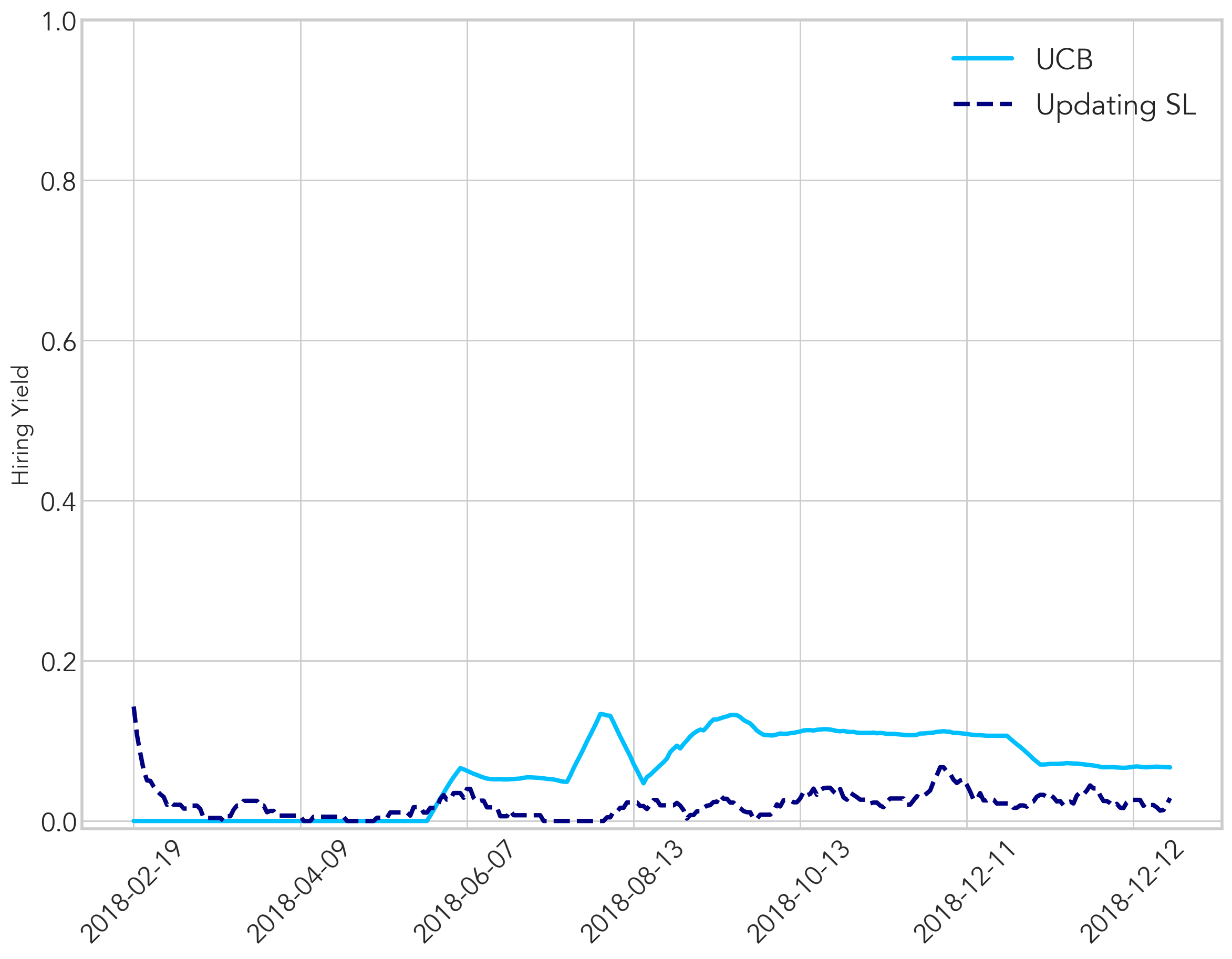} &  \includegraphics[scale=0.25]{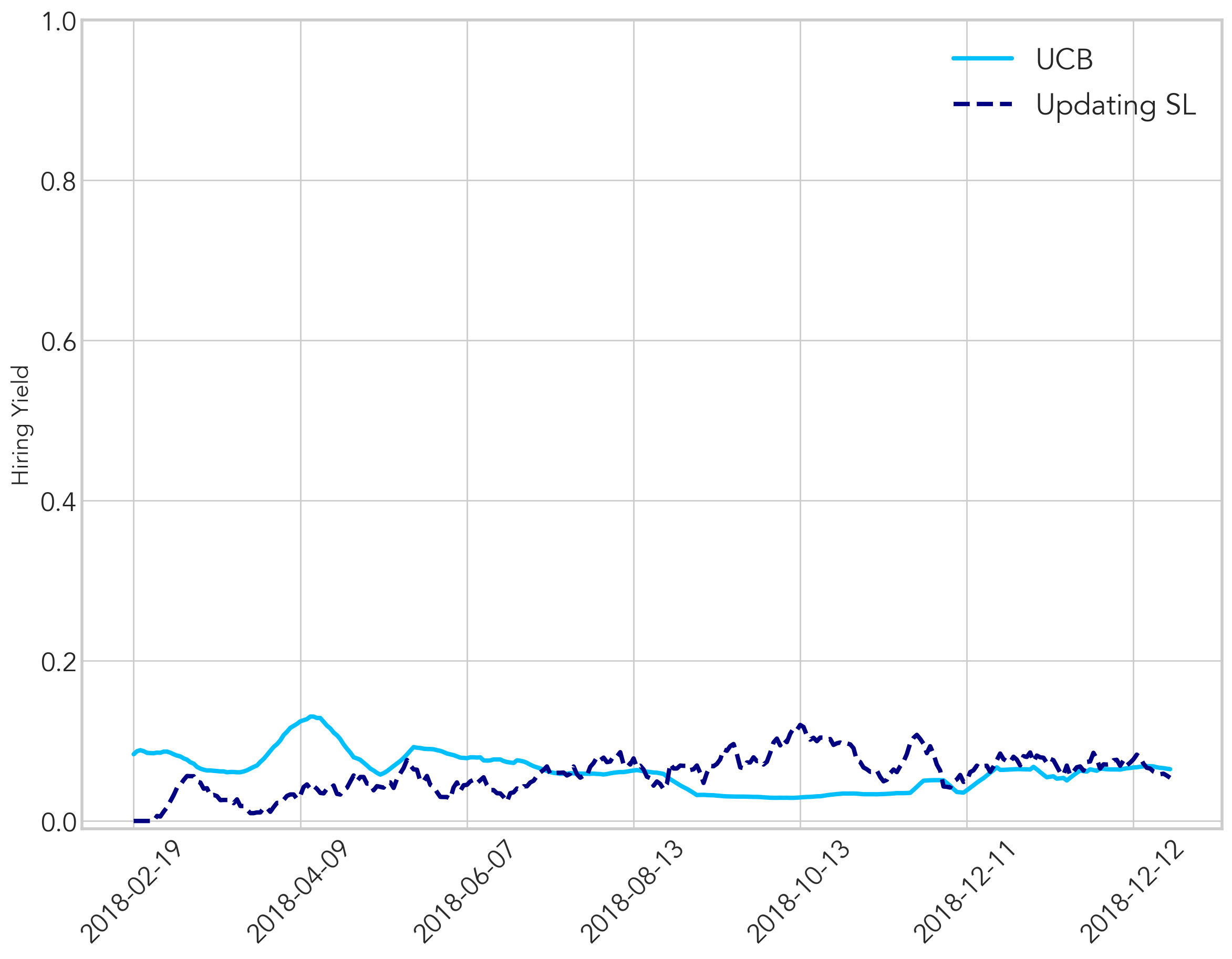} \\
			\end{tabular}
		}
		\label{learning_bad_quality}
	\end{center}
\end{figure}
\begin{singlespace}
	\begin{footnotesize}
		
		\begin{singlespace}
			\justifying \noindent \textsc{Notes}: This figure shows the hiring yield under three different algorithmic selection strategies: SL, UCB, and the beliefs component of UCB (that is, the $\hat{E}_t[H|X; D^{UCB}_{t}]$ term in Equation \eqref{UCBrule}).  In each panel, the $y$-axis graphs the share of interviewed applicants who are hired. Panel A plots the results from a simulation in which the hiring potential of Black candidates \textbf{decreases} linearly over the course of 2018, as described in Section \ref{sec:learning}.  Panel B shows results from a simulation in which the hiring potential of Hispanic candidates decreases linearly.  Similarly, Panels C and D show results from simulations in which the hiring potential of White and Asian applicants decreases, respectively.   
		\end{singlespace}
		
	\end{footnotesize}
\end{singlespace}

\end{appendix}
\end{document}

%% file: preambles/header.tex
    \usepackage[dvipsnames]{xcolor}
    \usepackage[T1]{fontenc}
	\usepackage[utf8]{inputenc}
	\usepackage[english]{babel}
	\usepackage{amsfonts}
	\usepackage{amssymb}
	    \usepackage{amsmath}
	
	\usepackage{enumerate}
	\usepackage{bbm}
	\usepackage{mathtools}
	\usepackage{subfig}
	\usepackage{booktabs}	
	\usepackage{setspace}
	\usepackage{graphicx}
	\usepackage{fullpage}
	
	\usepackage{soul}
	\usepackage{xspace}
	\usepackage{csquotes}
	\usepackage{placeins}
	\usepackage[footfont=normalsize,font=normalsize]{floatrow}
	\usepackage{floatrow}
	\usepackage{lscape}
	\usepackage{longtable}
	\usepackage{mathabx}
    \usepackage{accents}
    \usepackage{float}

    \usepackage{tikz}
    \usetikzlibrary{calc,intersections}

	\usepackage[]{natbib}
	\usepackage{bibunits}
	\defaultbibliography{AugI_Bibliography.bib}  
	\defaultbibliographystyle{aer}

\usepackage{geometry}
\usepackage{caption}
\usepackage{pgfplotstable}

\usepackage{pgfplotstable}
	
	\usepackage{hyperref}
	\hypersetup{
    colorlinks=true,
    linkcolor=Red,
    filecolor=magenta,      
    urlcolor=blue,
    citecolor = Blue
}
	\usepackage{cleveref}

    \usepackage{rotating}
	\usepackage[framemethod=tikz]{mdframed}
	\usepackage{todonotes}
	\presetkeys{todonotes}{color=black!5,inline}{} 

    \usepackage{tcolorbox}
    \newtcbox{\feedback}{nobeforeafter,colframe=black,colback=white,boxrule=0.5pt,arc=2pt,
      boxsep=0pt,left=2pt,right=2pt,top=2pt,bottom=2pt,tcbox raise base}
      
    \usepackage{amsthm}

    \theoremstyle{definition}

\usepackage{array}
\newcolumntype{L}[1]{>{\raggedright\let\newline\\\arraybackslash}m{#1}}
\newcolumntype{C}[1]{>{\centering\let\newline\\\arraybackslash\hspace{0pt}}m{#1}}
\newcolumntype{R}[1]{>{\raggedleft\let\newline\\\arraybackslash\hspace{0pt}}m{#1}}

\usepackage{ragged2e}
\newlength\ubwidth

\usepackage{algorithm}
\usepackage{algpseudocode}


%% file: preambles/notation.tex
\def\eea{\end{eqnarray*}}
\def\bea{\begin{eqnarray*}}
\renewcommand{\[}{\begin{equation}}
\renewcommand{\]}{\end{equation}}
\newcommand{\bi}{\begin{itemize}}
\newcommand{\ei}{\end{itemize}}

%% file: Output/Dec2022/summary.tex
\begin{tabular}{lccc} \toprule \toprule
Variable            &       Mean Training   & Mean Test & Mean Overall \\
\addlinespace \hline \addlinespace
Black &  0.087 &  0.087 &  0.087 \\
Hispanic &  0.040 &  0.043 &  0.042 \\
Asian &  0.573 &  0.591 &  0.581 \\
White &  0.300 &  0.279 &  0.290 \\
Male &  0.677 &  0.658 &  0.668 \\
Female &  0.323 &  0.342 &  0.332 \\
Referred &  0.140 &  0.114 &  0.129 \\
B.A. Degree &  0.232 &  0.242 &  0.237 \\
Associate Degree &  0.005 &  0.005 &  0.005 \\
Master's Degree &  0.612 &  0.643 &  0.626 \\
Ph.D. &  0.065 &  0.074 &  0.069 \\
Attended a U.S. College &  0.747 &  0.804 &  0.772 \\
Attended Elite U.S. College &  0.128 &  0.143 &  0.134 \\
Interviewed &  0.054 &  0.053 &  0.054 \\
Offered &  0.012 &  0.010 &  0.011 \\
Hired &  0.006 &  0.005 &  0.006 \\
Observations &       48,719 &       39,947 &       88,666 \\
\addlinespace \bottomrule \bottomrule
\end{tabular}

%% file: Output/Dec2022/IVvalidation.tex
{
\def\sym#1{\ifmmode^{#1}\else\(^{#1}\)\fi}
\begin{tabular}{l*{8}{c}}
\hline\hline
                         &\multicolumn{1}{c}{Interviewed}&\multicolumn{1}{c}{Black}&\multicolumn{1}{c}{Hispanic}&\multicolumn{1}{c}{Asian}&\multicolumn{1}{c}{White}&\multicolumn{1}{c}{Female}&\multicolumn{1}{c}{Ref.}&\multicolumn{1}{c}{MA}\\
                         &\multicolumn{1}{c}{(1)}   &\multicolumn{1}{c}{(2)}   &\multicolumn{1}{c}{(3)}   &\multicolumn{1}{c}{(4)}   &\multicolumn{1}{c}{(5)}   &\multicolumn{1}{c}{(6)}   &\multicolumn{1}{c}{(7)}   &\multicolumn{1}{c}{(8)}   \\
\hline
JK interview rate        &      0.0898***&    0.000158   &   -0.000433   &     0.00899   &    -0.00716   &    -0.00448   &     -0.0113   &     0.00910   \\
                         &   (0.00832)   &   (0.00470)   &   (0.00189)   &    (0.0122)   &   (0.00972)   &   (0.00557)   &    (0.0126)   &   (0.00961)   \\
\hline
Observations             &       37662   &       37662   &       37662   &       37662   &       37662   &       37662   &       37662   &       37662   \\
\hline\hline
\end{tabular}
}

%% file: Output/Dec2022/tab_IVhired.tex
{
\def\sym#1{\ifmmode^{#1}\else\(^{#1}\)\fi}
\begin{tabular}{l*{4}{c}}
\hline\hline
                    &\multicolumn{1}{c}{Low UCB}&\multicolumn{1}{c}{High UCB}&\multicolumn{1}{c}{Low SL}&\multicolumn{1}{c}{High SL}\\
                    &\multicolumn{1}{c}{(1)}   &\multicolumn{1}{c}{(2)}   &\multicolumn{1}{c}{(3)}   &\multicolumn{1}{c}{(4)}   \\
\hline  \addlinespace
Marginally Selected &      0.0629***&      0.0849***&      0.0564***&      0.0854***\\
                    &    (0.0123)   &    (0.0260)   &    (0.0130)   &    (0.0226)   \\
\hline
Observations        &       18710   &       18956   &       18862   &       18804   \\
\hline\hline
\end{tabular}
}

%% file: Output/Dec2022/tab_IVoffered.tex
{
\def\sym#1{\ifmmode^{#1}\else\(^{#1}\)\fi}
\begin{tabular}{l*{4}{c}}
\hline\hline
                    &\multicolumn{1}{c}{Low UCB}&\multicolumn{1}{c}{High UCB}&\multicolumn{1}{c}{Low SL}&\multicolumn{1}{c}{High SL}\\
                    &\multicolumn{1}{c}{(1)}   &\multicolumn{1}{c}{(2)}   &\multicolumn{1}{c}{(3)}   &\multicolumn{1}{c}{(4)}   \\
\hline  \addlinespace
Marginally Selected &       0.108***&       0.168***&       0.111***&       0.183***\\
                    &    (0.0220)   &    (0.0341)   &    (0.0248)   &    (0.0326)   \\
\hline
Observations        &       18538   &       19128   &       18417   &       19249   \\
\hline\hline
\end{tabular}
}

%% file: Output/Dec2022/tab_IVbh.tex
{
\def\sym#1{\ifmmode^{#1}\else\(^{#1}\)\fi}
\begin{tabular}{l*{4}{c}}
\hline\hline
                    &\multicolumn{1}{c}{Low UCB}&\multicolumn{1}{c}{High UCB}&\multicolumn{1}{c}{Low SL}&\multicolumn{1}{c}{High SL}\\
                    &\multicolumn{1}{c}{(1)}   &\multicolumn{1}{c}{(2)}   &\multicolumn{1}{c}{(3)}   &\multicolumn{1}{c}{(4)}   \\
\hline  \addlinespace
Marginally Selected &      0.0689***&      0.0982***&       0.139***&      0.0447***\\
                    &    (0.0145)   &    (0.0264)   &    (0.0190)   &    (0.0157)   \\
\hline
Observations        &       18710   &       18956   &       18862   &       18804   \\
\hline\hline
\end{tabular}
}

%% file: Output/Dec2022/tab_IVf.tex
{
\def\sym#1{\ifmmode^{#1}\else\(^{#1}\)\fi}
\begin{tabular}{l*{4}{c}}
\hline\hline
                    &\multicolumn{1}{c}{Low UCB}&\multicolumn{1}{c}{High UCB}&\multicolumn{1}{c}{Low SL}&\multicolumn{1}{c}{High SL}\\
                    &\multicolumn{1}{c}{(1)}   &\multicolumn{1}{c}{(2)}   &\multicolumn{1}{c}{(3)}   &\multicolumn{1}{c}{(4)}   \\
\hline  \addlinespace
Marginally Selected &       0.309***&       0.413***&       0.343***&       0.377***\\
                    &    (0.0181)   &    (0.0431)   &    (0.0442)   &    (0.0252)   \\
\hline
Observations        &       18710   &       18956   &       18862   &       18804   \\
\hline\hline
\end{tabular}
}

%% file: Output/Dec2022/tab_corr_onthejob_human.tex
{
\def\sym#1{\ifmmode^{#1}\else\(^{#1}\)\fi}
\begin{tabular}{l*{2}{c}}
\hline\hline
                    &\multicolumn{1}{c}{Top Rating}&\multicolumn{1}{c}{Promoted}\\
                    &\multicolumn{1}{c}{(1)}   &\multicolumn{1}{c}{(2)}   \\
\hline  \addlinespace
Human SL Score      &      -0.282** &     -0.0961   \\
                    &     (0.116)   &    (0.0782)   \\
\hline
Observations        &         180   &         233   \\
\hline\hline
\end{tabular}
}

%% file: Output/Dec2022/tab_corr_onthejob_ssup.tex
{
\def\sym#1{\ifmmode^{#1}\else\(^{#1}\)\fi}
\begin{tabular}{l*{4}{c}}
\hline\hline
                    &\multicolumn{2}{c}{Top Rating} &\multicolumn{2}{c}{Promoted}   \\
                    &\multicolumn{1}{c}{(1)}   &\multicolumn{1}{c}{(2)}   &\multicolumn{1}{c}{(3)}   &\multicolumn{1}{c}{(4)}   \\
\hline  \addlinespace
SL Hired            &      0.0791   &               &      0.0816   &               \\
                    &     (0.103)   &               &    (0.0641)   &               \\
[1em]
SL Offered          &               &       0.168** &               &     -0.0170   \\
                    &               &    (0.0800)   &               &    (0.0537)   \\
\hline
Observations        &         180   &         180   &         233   &         233   \\
\hline\hline
\end{tabular}
}

%% file: Output/Dec2022/tab_corr_onthejob_ucb.tex
{
\def\sym#1{\ifmmode^{#1}\else\(^{#1}\)\fi}
\begin{tabular}{l*{4}{c}}
\hline\hline
                    &\multicolumn{2}{c}{Top Rating} &\multicolumn{2}{c}{Promoted}   \\
                    &\multicolumn{1}{c}{(1)}   &\multicolumn{1}{c}{(2)}   &\multicolumn{1}{c}{(3)}   &\multicolumn{1}{c}{(4)}   \\
\hline  \addlinespace
UCB Hired           &      0.0377   &               &       0.161***&               \\
                    &     (0.106)   &               &    (0.0619)   &               \\
[1em]
UCB Offered         &               &       0.163*  &               &     -0.0245   \\
                    &               &    (0.0850)   &               &    (0.0576)   \\
\hline
Observations        &         180   &         180   &         233   &         233   \\
\hline\hline
\end{tabular}
}

%% file: Output/Dec2022/summary_features.tex
\begin{tabular}{lccc} \toprule \toprule
Variable            &       Mean Training   & Mean Test & Mean Overall \\
\addlinespace \hline \addlinespace
B.A. Degree &   0.23 &   0.24 &   0.24 \\
Graduate Degree &   0.68 &   0.72 &   0.69 \\
Has a Quantitative Background &   0.23 &   0.28 &   0.25 \\
Worked at a Fortune 500 Co. &   0.02 &   0.02 &   0.02 \\
Attended School in China &   0.07 &   0.08 &   0.07 \\
Attended School in India &   0.22 &   0.23 &   0.23 \\
Attended School in Europe &   0.04 &   0.05 &   0.05 \\
Attended Elite International School &   0.09 &   0.10 &   0.09 \\
Attended a U.S. College &   0.75 &   0.80 &   0.77 \\
Attended US News Top 25 Ranked College &   0.13 &   0.14 &   0.13 \\
Attended US News Top 50 Ranked College &   0.26 &   0.28 &   0.27 \\
Military Experience &   0.04 &   0.04 &   0.04 \\
Referred Applicant &   0.14 &   0.11 &   0.13 \\
Service Sector Experience &   0.00 &   0.00 &   0.00 \\
Number of Work Histories &   3.78 &   4.01 &   3.89 \\
Number of Degrees &   1.69 &   1.73 &   1.71 \\
Major Description Business Management &   0.19 &   0.16 &   0.18 \\
Major Description Computer Science &   0.15 &   0.14 &   0.15 \\
Major Description Finance/Economics &   0.14 &   0.13 &   0.14 \\
Major Description Engineering &   0.06 &   0.05 &   0.06 \\
Observations &       48,719 &       39,947 &       88,666 \\
\addlinespace \bottomrule \bottomrule
\end{tabular}

%% file: Output/Dec2022/tab_corr_H1_scores.tex
{
\def\sym#1{\ifmmode^{#1}\else\(^{#1}\)\fi}
\begin{tabular}{l*{6}{c}}
\hline\hline
                    &\multicolumn{3}{c}{Hired}                      &\multicolumn{3}{c}{Offered}                    \\
                    &\multicolumn{1}{c}{(1)}   &\multicolumn{1}{c}{(2)}   &\multicolumn{1}{c}{(3)}   &\multicolumn{1}{c}{(4)}   &\multicolumn{1}{c}{(5)}   &\multicolumn{1}{c}{(6)}   \\
\hline  \addlinespace
Human               &     -0.0652** &               &               &      -0.125***&               &               \\
                    &    (0.0280)   &               &               &    (0.0363)   &               &               \\
[1em]
SL Hired            &               &       0.171***&               &               &               &               \\
                    &               &    (0.0267)   &               &               &               &               \\
[1em]
UCB Hired           &               &               &       0.205***&               &               &               \\
                    &               &               &    (0.0261)   &               &               &               \\
[1em]
SL Offered          &               &               &               &               &       0.284***&               \\
                    &               &               &               &               &    (0.0294)   &               \\
[1em]
UCB Offered         &               &               &               &               &               &       0.349***\\
                    &               &               &               &               &               &    (0.0311)   \\
\hline
Observations        &        2275   &        2275   &        2275   &        2275   &        2275   &        2275   \\
Mean of Hired: .102 &               &               &               &               &               &               \\
Mean of Offered: .189&               &               &               &               &               &               \\
\hline\hline
\end{tabular}
}

%% file: Output/InstrumentTestsNov2021/IV_monotonicity_race_gender.tex
{
\def\sym#1{\ifmmode^{#1}\else\(^{#1}\)\fi}
\begin{tabular}{l*{9}{c}}
\toprule
                &\multicolumn{1}{c}{(1)}&\multicolumn{1}{c}{(2)}&\multicolumn{1}{c}{(3)}&\multicolumn{1}{c}{(4)}&\multicolumn{1}{c}{(5)}&\multicolumn{1}{c}{(6)}&\multicolumn{1}{c}{(7)}&\multicolumn{1}{c}{(8)}&\multicolumn{1}{c}{(9)}\\
                &\multicolumn{1}{c}{All}&\multicolumn{1}{c}{White}&\multicolumn{1}{c}{Black}&\multicolumn{1}{c}{Asian}&\multicolumn{1}{c}{Hispanic}&\multicolumn{1}{c}{Female}&\multicolumn{1}{c}{Male}&\multicolumn{1}{c}{MA}&\multicolumn{1}{c}{Ref.}\\
\hline  \addlinespace
Interviewer &     0.08***&     0.08***&     0.06***&     0.09***&     0.04   &     0.08***&     0.08***&     0.08***&     0.10***\\
Leniency         &   (0.01)   &   (0.01)   &   (0.01)   &   (0.01)   &   (0.02)   &   (0.01)   &   (0.01)   &   (0.01)   &   (0.01)   \\
\addlinespace
Observations    &    23803   &     7055   &     2150   &    13565   &     1097   &     8825   &    16575   &    16384   &     3245   \\
\bottomrule
\multicolumn{10}{l}{\footnotesize Standard errors in parentheses}\\
\multicolumn{10}{l}{\footnotesize * p<0.10, ** p<0.05, *** p<0.01}\\
\end{tabular}
}

%% file: Output/InstrumentTestsNov2021/leniency_correlations_randomsplit.tex
\begin{tabular}{l*{1}{c}}
\hline\hline
            &           Within-Person Correlation\\
            &        Strict and Lenient Selection Scores\\
\hline
All        &       0.688\\
White       &       0.836\\
Black       &       0.860\\
Asian       &       0.702\\
Hispanic    &       0.760\\
Female      &       0.703\\
Male        &       0.686\\
MA          &       0.686\\
Referral     &       0.689\\
\hline\hline
\end{tabular}